\documentclass[12pt,preprint]{aastex}

\def\lsim {$\rlap{\raise.4ex\hbox{$<$}}\lower.55ex\hbox{$\sim$}\,$}
\def\gsim {$\rlap{\raise.4ex\hbox{$>$}}\lower.55ex\hbox{$\sim$}\,$}
\shorttitle{Gas Motions in UCHII Regions} \shortauthors{Zhu et al.}

\begin{document}

\title{[Ne~II] Observations of Gas Motions in Compact and Ultracompact H~II Regions}

\author{Qingfeng Zhu\altaffilmark{1}}
\affil{Center for Imaging Science, Rochester Institute of Technology, Rochester, NY 14623}
\email{zhuqf@cis.rit.edu}

\and

\author{John H. Lacy\altaffilmark{1}, Daniel T. Jaffe\altaffilmark{1}}
\affil{Department of Astronomy, University of Texas, Austin, TX
78712}

\email{lacy@astro.as.utexas.edu, dtj@astro.as.utexas.edu}

\and

\author{Matthew J. Richter\altaffilmark{1}}
\affil{Department of Physics, University of California, Davis, CA 95616-8677}

\email{richter@physics.ucdavis.edu}

\and

\author{Thomas K. Greathouse\altaffilmark{1}}
\affil{Division \#15, Southwest Research Institute, San Antonio, TX
78228} \email{tgreathouse@swri.edu}

\altaffiltext{1}{Visiting Astronomer at the Infrared Telescope
Facility, which is operated by the University of Hawaii under
Cooperative Agreement no. NCC 5-538 with the National Aeronautics
and Space Administration, Office of Space Science, Planetary
Astronomy Program.}

\clearpage

\begin{abstract}
We present high spatial and spectral resolution observations of
sixteen Galactic compact and ultracompact H~II regions in the
[Ne~II] 12.8~$\mu$m fine structure line. The small thermal width of
the neon line and the high dynamic range of the maps provide an
unprecedented view of the kinematics of compact and ultracompact
H~II regions. These observations solidify an emerging picture of the
structure of ultracompact H~II regions suggested in our earlier
studies of G29.96 and Mon R2; systematic surface flows, rather than
turbulence or bulk expansion, dominate the gas motions in the H~II
regions. The observations show that almost all of the sources have
significant (5-20~km~s$^{-1}$) velocity gradients and that most of
the sources are limb-brightened. In many cases, the velocity pattern
implies tangential flow along a dense shell of ionized gas. None of
the observed sources clearly fits into the categories of filled
expanding spheres, expanding shells, filled blister flows, or
cometary H~II regions formed by rapidly moving stars. Instead, the
kinematics and morphologies of most of the sources lead to a picture
of H~II regions confined to the edges of cavities created by stellar
wind ram pressure and flowing along the cavity surfaces. In sources
where the radio continuum and [Ne~II] morphologies agree, the
majority of the ionic emission is blue-shifted relative to nearby
molecular gas. This is consistent with sources lying on the near
side of their natal clouds being less affected by extinction and
with gas motions being predominantly outward, as is expected for
pressure-driven flows.
\end{abstract}

\keywords{interstellar medium - ultracompact H~II regions - fine structure line}

\section{Introduction}
The short Kelvin-Helmholtz timescales of massive young stellar
objects (YSOs) allow these objects to begin emitting ionizing
photons and to excite nearby material while still embedded in their
natal clouds \citep{lad99}. One of the earliest observable
manifestations of newly formed OB stars is the ultracompact HII
(UCHII) region. These embedded HII regions are initially small ($D
{\leq} 0.1~pc$) and dense ($n_{e} {\sim} 10^4~cm^{-3}$) and have a
typical emission measure (EM) $\sim$ 10$^7$~pc~cm$^{-6}$
\citep{chu02}. As HII regions evolve their sizes increase, but they
are still considered to be UCHII regions as long as their emission
measures (or surface brightnesses) remain high.

Since most stars with M $\ge$ 8~M$_{\odot}$ should form UCHII
regions shortly after their birth, studies of UCHII regions can help
us learn about physical conditions in high-mass star forming regions
and the evolution of the physical conditions as the massive stars
turn on. A large percentage, $\ge$ 50$\%$, of UCHII regions have
cometary, core-halo or irregular morphologies
\citep{wooC89b,kurCW94,walBHR98,depWDMD05}, arguing against the
classical expanding sphere model of HII regions \citep{savG55}. In
addition, statistical arguments based on the Galactic population of
UCHII regions lead to estimates of the lifetime of these regions
that are much longer than expected from the expanding sphere model
\citep{wooC89a,hoaKLKH07}. Many models have been proposed for UCHII
region formation \citep[see the review by][and references
therein]{chu99}, with solving the lifetime problem a major goal.
However, the age of an individual UCHII region is hard to determine
through observations. A number of theoretical models have been
geared toward explaining both these long apparent lifetimes and the
observed morphologies of UCHII regions
\citep{holJLS94,dysWR95,redWD96,wilDR96,redWD98}.

Observations of the morphology and gas kinematics of UCHII regions
can help us to constrain models for the structure and evolution of
the regions much more definitively than observations of the
morphology alone. The morphological type is not a unique indicator
of the UCHII region physics because many models can produce similar
morphologies \citep{ten79, yorTB83, macVW91, vanM92, artH06}. While
it is difficult to choose between different models based on the
morphologies of HII regions, their kinematic signatures are very
distinct. To understand the structure and evolution of the regions
around the youngest massive stars, it is therefore necessary to
observe gas kinematics inside UCHII regions with the highest
spectral and spatial resolution. The large extinction associated
with molecular clouds around UCHII regions, however, prevents direct
observation at optical wavelengths and even at near-infrared
wavelengths \citep{dopFCSCT06}. Thus, observations at longer
wavelengths become crucial to UCHII region studies.

Radio interferometric observations can provide data with high
angular and spectral resolution. At high enough frequencies, where
non-LTE effects become less important, hydrogen radio recombination
line (RRL) emission has the same dependence on the electron density
as thermal free-free radio continuum emission, and the emissivities
of these lines are well known. RRL observations can provide
information on gas density and kinematics, and, to a limited degree,
on gas temperature. However, thermal broadening of hydrogen RRLs due
to the low mass of hydrogen atoms degrades the spectral resolution
of radio spectroscopy. The thermal line width of the RRLs (or any
hydrogen recombination lines like Br$\alpha$ and Br$\gamma$, for
that matter) in an ionized region (T$_e {\simeq} $10$^4$~K) is
$\simeq$ 20~km~s$^{-1}$, comparable to the velocities of bulk flows
in many sources. Furthermore, the incomplete uv sampling of radio
interferometric survey can both overemphasize compact structures
and, more generally limit the spatial dynamic range of the
observations.

Mid-IR fine structure line observations offer an alternative to
radio and IR recombination line observations. Fine structure lines
have small thermal line widths because they are emitted by heavy
elements. The importance of this feature of the fine structure lines
has been illustrated previously by high spectral resolution maps of
[Ne~II] emission in Mon R2 and G29.96-0.02 which show single lines
or multiple components of lines with widths significantly smaller
than the thermal widths of the hydrogen recombination lines
\citep{jafZLR03,zhuLJRG05}. In some cases, the broad, largely
Gaussian hydrogen recombination lines are, in fact, superpositions
of several much narrower kinematic components (see especially Fig. 3
of \cite{jafZLR03}). In addition, although the emitting species are
several orders of magnitude less abundant than hydrogen, the fine
structure lines are brighter than recombination lines because
collisional excitation cross-sections are much larger than the
relevant recombination cross-sections. Typical electron densities
over all compact and most ultracompact H~II regions are lower than
the critical density of the [Ne~II] 12.8~$\mu$m transition (${\sim}
5 \times 10^5~cm^{-3}$), so this line shares the property of the
recombination lines of having a line strength proportional to
emission measure. We should mention that turbulence can also have
significant effects on line widths. Moreover, turbulence affects
both RRL and fine structure line observations and prevents
detections of velocity variations (due to gas bulk motions) with
scales smaller than the intrinsic line width combining thermal and
turbulent effects along any line of sight direction.

Our previous [Ne~II] line observations of a few UCHII regions have
demonstrated the effectiveness of using high spectral and spatial
resolution maps of fine structure lines to study such regions
\citep{jafZLR03,zhuLJRG05}. In these studies, we demonstrated that
the kinematic patterns in the ionized gas in Mon R2 and G29.96-0.02
are consistent with tangential flows along the surfaces of thin,
more-or-less parabolic shells. The goal of the current paper is to
use [Ne~II] mapping as a tool to examine a larger sample of compact
and ultracompact HII regions with a range of shapes and sizes in
order to look for a common thread in the kinematics that might lead
to a better understanding of the physics and evolution of these
sources. We present scan-mapped spectroscopic observations of
fifteen Galactic compact and ultracompact HII regions. This survey
is the first mid-infrared high-spectral resolution study of gas
kinematics in a large sample of UCHII regions.

In \S~{\bf \ref{secobservation}}, we describe our observational
method. We introduce our data reduction procedure in \S~{\bf
\ref{secdatareduction}}. In \S~{\bf \ref{secresult}}, we investigate
the morphology and kinematics of individual objects. Finally, the
conclusions drawn from our observations are presented in \S~{\bf
\ref{secdiscussion}}.

\section{Observations}
\label{secobservation} Candidates in our survey were selected based
on their radio continuum flux density levels \citep{wooC89b},
biasing our list toward radio-bright compact and ultracompact HII
regions. Observations were performed with the high spectral
resolution, mid-IR (5-25~$\mu$m) cross dispersed spectrograph TEXES
\citep[Texas Echelon Cross Echelle Spectrograph,][]{lacRGJZ02} on
the 3.0 meter NASA IRTF on Mauna Kea in Hawaii.

We list our observational parameters in Table~\ref{obspara}. Our
spatial and spectral resolution were approximately 1.8$''$ and
4~km~s$^{-1}$. For each object, a north-south oriented
slit was used to scan the object from west to east. We chose the
step size along the scans so that there were at least two steps per
slit width and chose the scan lengths to cover the emission seen in
radio continuum maps available in the literature. Before each scan,
we offset the telescope to the west several arcseconds beyond the
edge of the maps shown in the figures. Several extra steps were
taken at the beginning and the end of each scan in order to measure
the sky background. The lengths of the scans and the initial offsets
were adjusted after examining a preliminary scan with our quick-look
program \citep{lacRGJZ02}. We made multiple overlapping scans,
offset north-south, to cover the sources. TEXES takes ambient
temperature blackbody and sky flats at the beginning of each scan
for flux calibration.

\section{Data Reduction}
\label{secdatareduction} We use a custom Fortran reduction program
\citep{lacRGJZ02} to correct optical distortions, flat field, remove
cosmic ray spikes, and fix bad pixels. The same program also does
wavelength and flux calibrations. We use atmospheric lines for
wavelength calibration, with wavelengths obtained from HITRAN
\citep{rotRGMEFP98}. We use an IDL script to subtract sky background
and combine multiple scans. The sky emission at each slit position
is linearly interpolated from the sky frames at the beginning and
the end of each scan. Sometimes, we start scans too late or end
scans too early. Extended line emission is present in our sky frames
and is subtracted off from other frames. This will result in an
underestimate of line flux from targets. Multiple scans are
cross-correlated, shifted and added to create the datacubes.
Normally, no shift is needed along the spectral direction. One
exception was the source W3, whose datacube was obtained by
combining data from two nights and where the maps were registered
using the compact continuum source IRS5 \citep{wynBN72}. Both the
Earth's motion and the spectral settings of the instrument on two
different nights affect the wavelength calibration. The continuum
flux level at each pixel in the map is obtained from spectral points
off the line and is subtracted from the spectra. Finally, we
resample the datacubes so that each pixel has a size of
$0.3''{\times}0.3''$ on the sky. The coordinates of the resulting
continuum maps and [Ne~II] line maps are relative to the peak
positions in the integrated [Ne~II] line maps. When radio continuum
maps are available, they are cross-correlated with the [Ne~II] maps
so that both maps can be plotted on the same scale. The coordinates
in Table 1 were taken from the literature and served as reference
points during our line observations.

\section{Results}
\label{secresult}

In this section we discuss highlights of individual objects and
features common to several sources. [Ne~II] line emission datacubes
for all objects are available online. {\bf Note: the data cubes will
be available through the ApJ web site.}

\subsection{G29.96~-0.02}

G29.96~-0.02 (G29.96 hereafter) is a cometary UCHII region at a
distance of $\sim7.4$~kpc, is $\sim$20$''$$\times$15$''$ in size,
and is one of the most luminous UCHII regions in the sky at 2 cm
wavelength. The ionized region is considered as an archetype of the
UCHII regions caused by bow shocks \citep{macVW91}, although some
past observations of the gas velocities along the symmetry axis of
this source argue against this interpretation, implying that the
source is a blister flow \citep{lumH96} or a blister flow modified
by a stellar wind \citep{lumH99}. We presented [Ne~II] line mapping
observations and compared gas kinematics in the region with a bow
shock model in \cite{zhuLJRG05}. We include here a summary of the
model and results for G29.96 from the earlier paper as a template
for understanding the kinematics of some of the other sources in the
sample. The bow shock model predicts that an approximately
paraboloidal shell of swept-up material will form in front of a star
moving supersonically inside a molecular cloud. The shell is
supported by the ram pressure of the stellar wind and the ambient
medium. Swept-up material, including ionized and neutral gas, is
forced to flow along the surface of the shell. With different
viewing angles for the shell, the model can reproduce many of the
observed features in ionic line observations of G29.96 as well as
several other UCHII regions in our sample, including morphologies
and the distinctive shapes that parabolic flows have in
position-velocity (p-v) diagrams.

\begin{figure}
\epsscale{1.0} \plotone{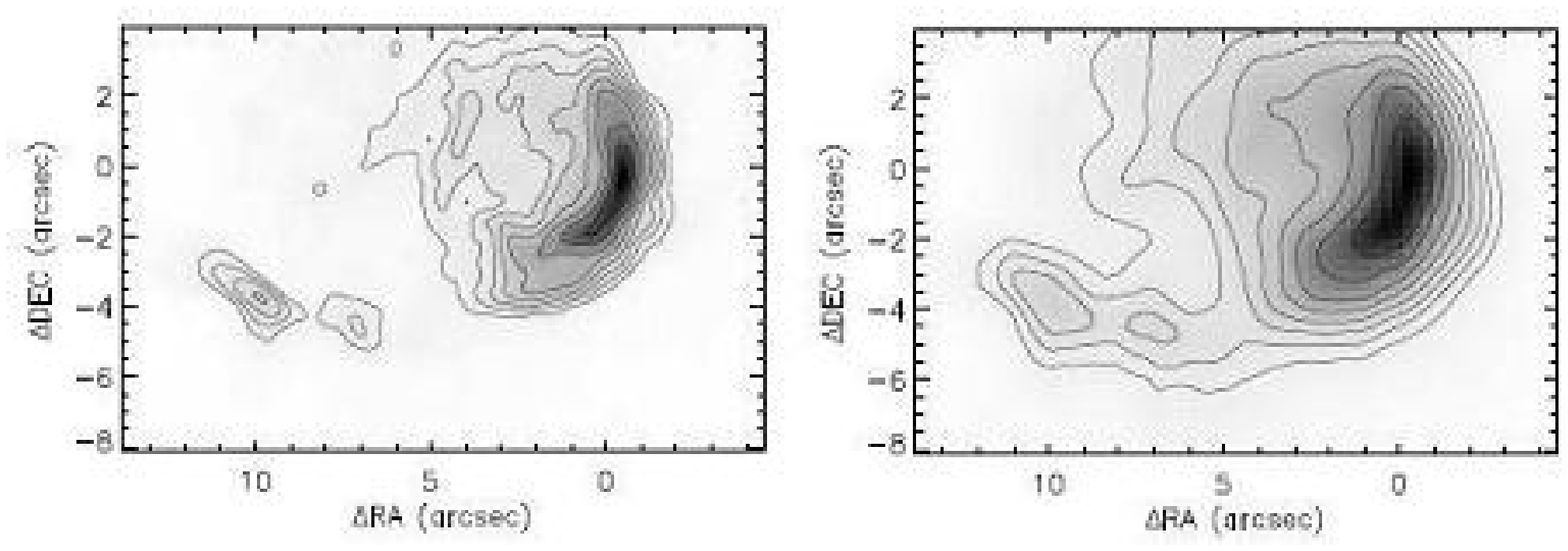} \caption{2~cm continuum map
\citep[left,][restoring beam: $0.''56 \times 0.''49$]{feyGCV95} and
integrated [Ne~II] line map (right) of G29.96~-0.02. The (0,0)
position marks the peak emission in the [Ne~II] map. The line map
has been cross-correlated with the continuum map and shifted to
match. Contours are drawn at $70\%$, $50\%$, $35\%$, $25\%$,
$17.5\%$, $12.5\%$, $9\%$ and $6\%$ of the peak value in each map.
The peak [Ne~II] surface brightness is
1.0~ergs~cm$^{-2}$~s$^{-1}$~sr$^{-1}$. The total [Ne~II] flux is
6.3$\times$10$^{-10}$~ergs~cm$^{-2}$~s$^{-1}$. \label{g2996mapneii}}
\end{figure}

\begin{figure}
\epsscale{1.0} \plotone{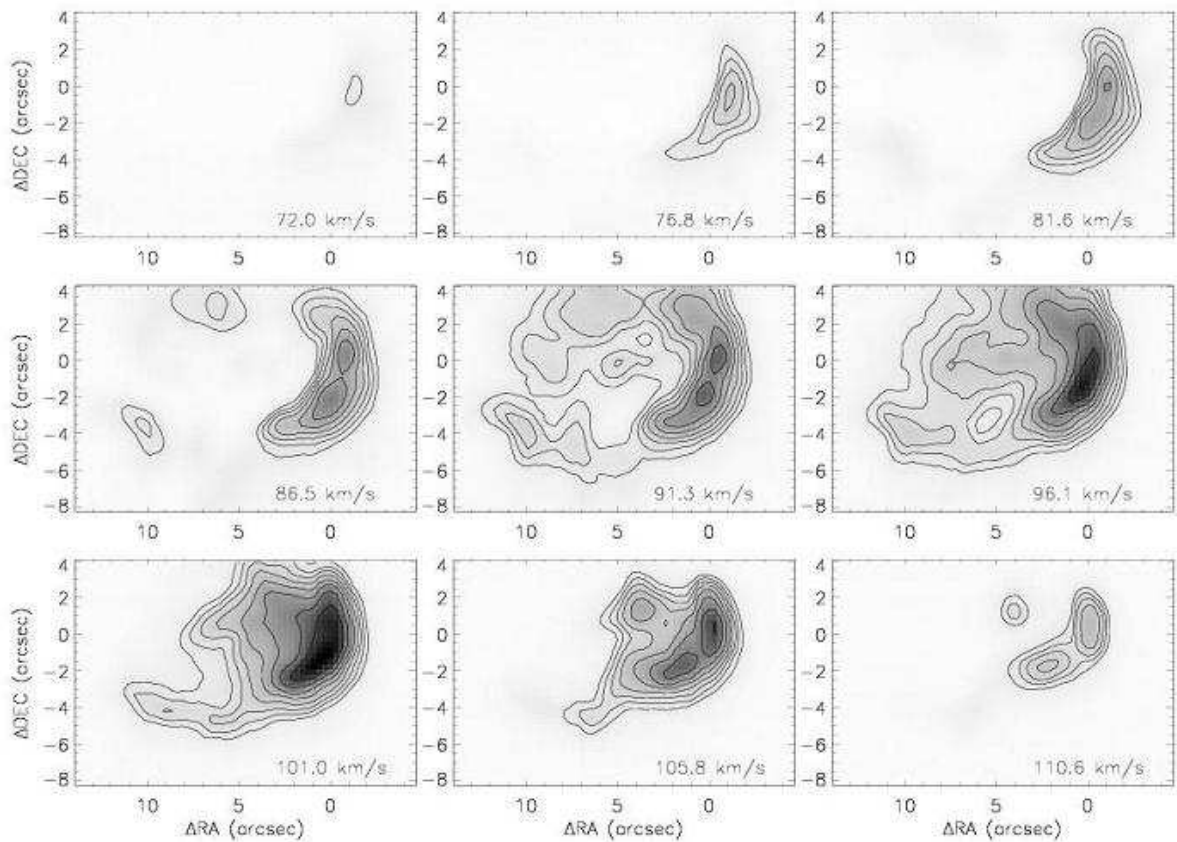} \caption{Channel maps of [Ne~II]
line observations for G29.96~-0.02.
The individual maps sample a single velocity slice in the
data cube and therefore represent a width equal to the
spectral resolution, or about 4 km s$^{-1}$.
Contours are drawn at
$70\%$, $50\%$, $35\%$, $25\%$, $17.5\%$,
$12.5\%$, $9\%$ and $6\%$ of the
peak value of all channels. The peak value is
50.0~ergs~cm$^{-2}$~s$^{-1}$~sr$^{-1}$~(cm$^{-1}$)$^{-1}$.
The molecular cloud velocity is V$_{LSR}$ = 98 km s$^{-1}$.
\label{g2996nechanmapneii}}
\end{figure}

\begin{figure}
\epsscale{1.0} \plotone{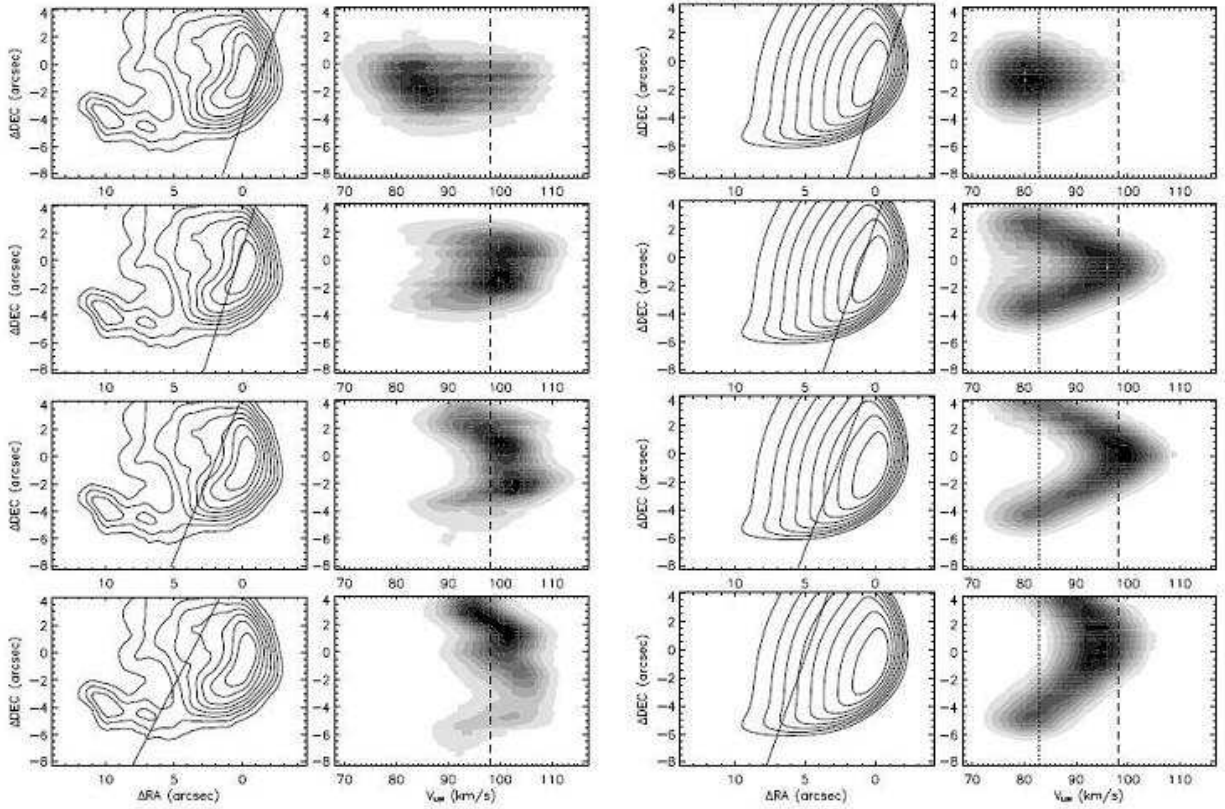} \caption{Position-Velocity (P-V)
diagrams of G29.96~-0.02 [Ne~II] observations (column 2) and a bow
shock model \citep[][column 4]{zhuLJRG05}. The straight lines
through the observed (column 1) and model (column 3) integrated line
flux maps show the position of each cut. Dashed lines in p-v
diagrams show the observed ambient molecular material velocity.
Dotted lines in the model p-v diagrams show the ambient material
velocity in the model. The contour maps are drawn as in
Fig.\ref{g2996mapneii} with the peak value
$\sim$1.0~ergs~cm$^{-2}$~s$^{-1}$~sr$^{-1}$ for the line maps of
observations. \label{g2996pv1}}
\end{figure}

\begin{figure}
\epsscale{0.8} \plotone{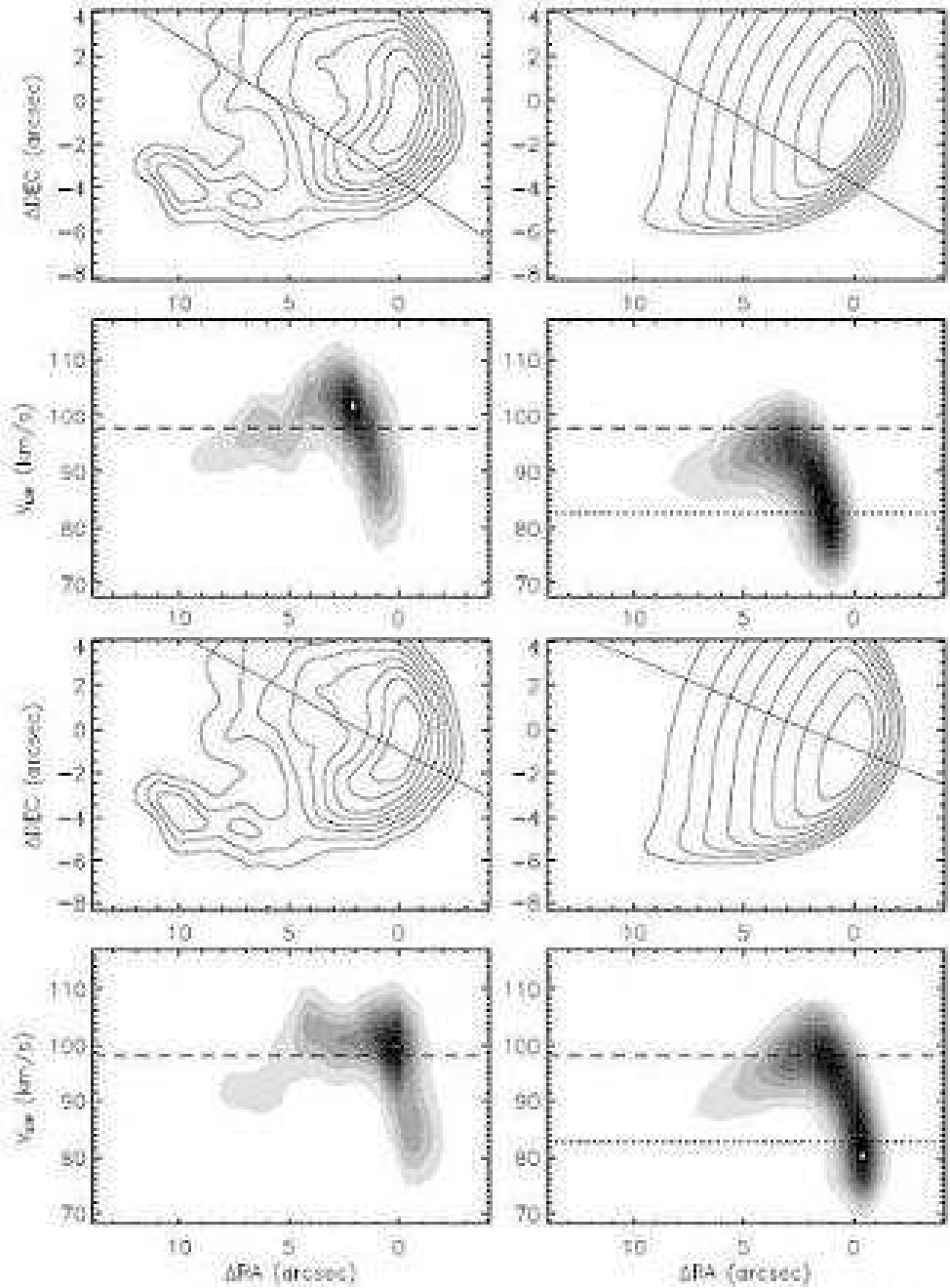} \caption{Position-velocity diagrams
of G29.96~-0.02 [Ne~II] observations (left) and the \cite{zhuLJRG05}
bow shock model (right). The straight line in the integrated line
map above each position-velocity diagram shows the location of the
cut. The dashed lines in the p-v diagrams show the ambient molecular
material velocity. Dotted lines in the model p-v diagrams show the
ambient material velocity in the model. The contours are drawn as in
Fig.\ref{g2996mapneii} with the peak value
$\sim$1.0~ergs~cm$^{-2}$~s$^{-1}$~sr$^{-1}$ for the line maps of
observations. \label{g2996pv2}}
\end{figure}

Figure~\ref{g2996mapneii} shows a 2~cm radio continuum map from
\cite{feyGCV95} and our spectrally integrated [Ne~II] line map of
G29.96. The two maps closely resemble each other, indicating that
most of the [Ne~II] emission arises from gas with n$<$n$_{crit}$ and
that the extinction to the gas is either rather uniform or small at
12.8~$\mu$m. In fact, \cite{morSMPDBCR02} observed that the electron
density in the region is sub-critical (N$_e$$\sim$680 cm$^{-3}$ -
5.7$\times$10$^4$ cm$^{-3}$) and \cite{marPMTCR02} reported low
infrared extinction (A$_K$$\simeq$1.6) toward the region. Our
findings are consistent with these observations. Both maps show an
emission arc open to the east with extended emission in the eastern
part of the region. A filament extends from the southern edge of the
nebula to the east, forming a ``tail'' of the cometary HII region.
Channel maps at nine equally spaced velocities are shown in
Figure~\ref{g2996nechanmapneii}. These channel maps show a
morphology change from a thin crescent in the most blue-shifted
channels to a fan-like region in red-shifted channels.
Figures~\ref{g2996pv1} and \ref{g2996pv2} show position-velocity
(p-v) diagrams of several representative cross cuts made through the
observed and model data cubes. The observations and the model
display similar p-v structures (as shown in our previous paper, a
characteristic ``$\Lambda$'' shape for cuts perpendicular to the
nebular symmetry axis and a ``7'' shape for cuts parallel to the
axis). Similar changes in structure from cut to cut can also be
seen. The similarities in line emission morphology and gas
kinematics between the observations and the model indicate that the
ionized gas in G29.96 flows along a thin shell at
$\sim20$~km~s$^{-1}$, as it would if it were produced by a bow
shock. We use these distinctive patterns in the p-v diagrams as
indicators of the presence of surface flows in other sources where
we have not carried out detailed model fits. However, one clear
discrepancy between the bow shock model that gives the correct shape
to the contours in the model p-v diagrams in Figures~\ref{g2996pv1}
and \ref{g2996pv2} and the observed sources is the
$\sim$15~km~s$^{-1}$ difference between the observed and modeled
velocity of the ambient neutral material. We discuss in
\S~\ref{secdiscussion} modifications of the bow-shock model to
explain this discrepancy.

\subsection{G5.89~-0.39}

\begin{figure}
\epsscale{1.0} \plotone{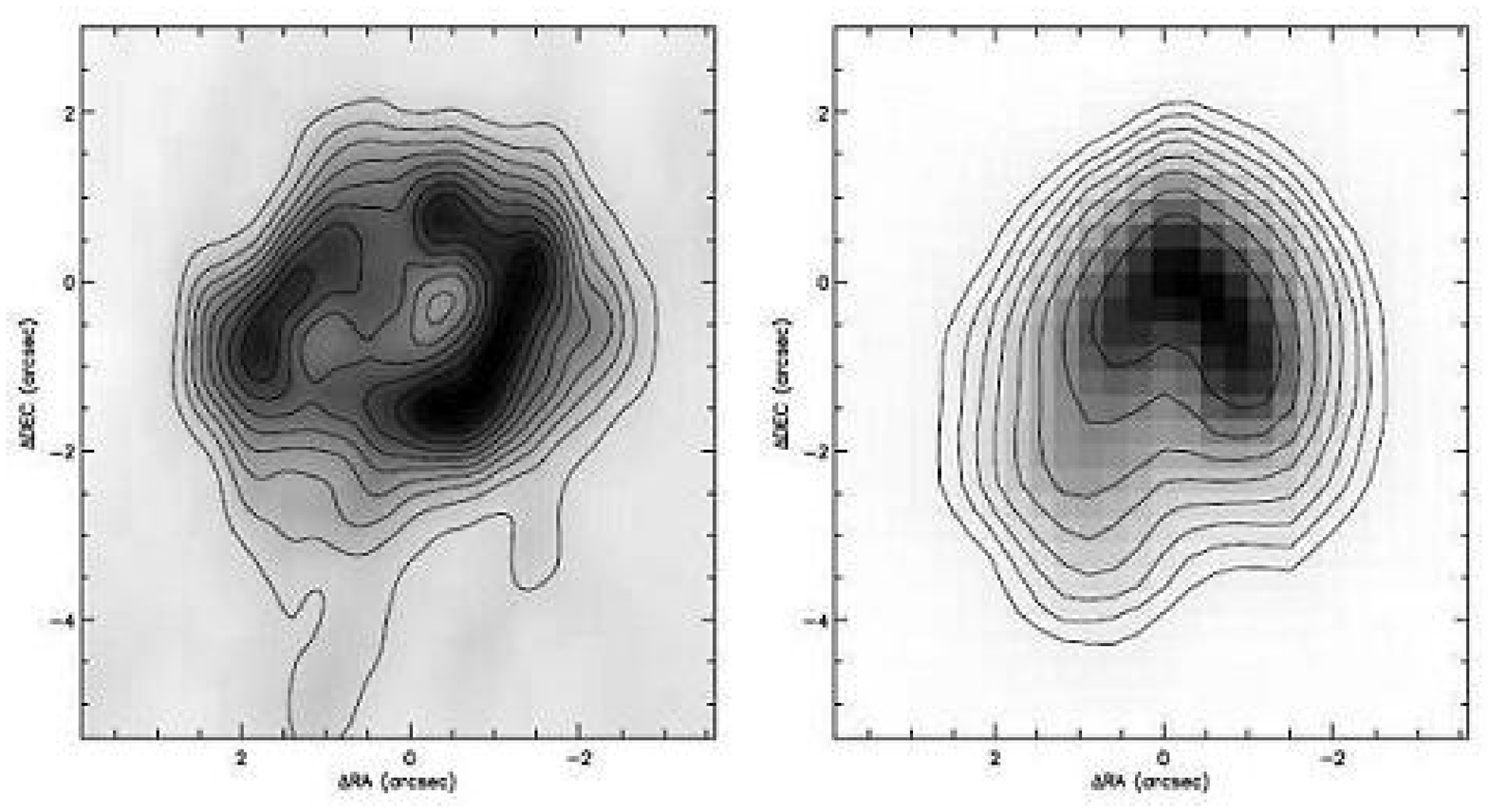} \caption{6~cm continuum map
\citep[left,][restoring beam: $0.''38 \times 0.''38$]{wooC89b} and
integrated [Ne~II] line map (right) of G5.89~-0.39. The [Ne~II] line
map has been cross-correlated with the radio map and the (0,0)
position has been shifted to the best match. The (0,0) position is
the location of the emission peak in the line map. Contours are
drawn at $95\%$, $90\%$, $80\%$, $70\%$, $60\%$, $50\%$, $40\%$
,$30\%$, $20\%$ and $10\%$ of peak value for the continuum map and
at $70\%$, $50\%$, $35\%$, $25\%$, $17.5\%$, $12.5\%$, $9\%$ and
$6\%$ of the peak value for the line map. The peak [Ne~II] surface
brightness is 1.08~ergs~cm$^{-2}$~s$^{-1}$~sr$^{-1}$. The total
[Ne~II] flux is 2.26$\times$10$^{-10}$~ergs~cm$^{-2}$~s$^{-1}$.
\label{g589map}}
\end{figure}

G5.89~-0.39, also called W28 A2(1), is 2.6~kpc from the Sun
\citep{dowWBW80,kimK03}. \cite{wooC89b} measured continuum images of
the source at both 2~cm (Figure~\ref{g589map}, left) and 6~cm. A
shell-like morphology with significantly limb-brightened walls
surrounding the slightly elongated central cavity can be recognized.
The G5.89 region also hosts a source of extremely broad (FWHM
$\sim$140~km~s$^{-1}$) CO J=1-0 emission \citep{sheC96,harF88}.
\cite{acoCW98} have used multi-epoch radio continuum observations to
argue that the ionized shell is expanding at $\sim$35~km~s$^{-1}$.
Radio recombination line observations of \cite{rodRG02} show a
complex velocity structure across the region, which cannot be
explained with a consistent dynamical model. Mid-IR dust continuum
observations show that G5.89 deviates from spherical symmetry and
resembles a shell partially open to the south, with significant SW
and SE protrusions from the geometric center, implying an
anisotropic expansion of the ionization front due to density
variations in the surrounding molecular medium \citep{balAJKM92}.

\begin{figure}
\epsscale{1.0} \plotone{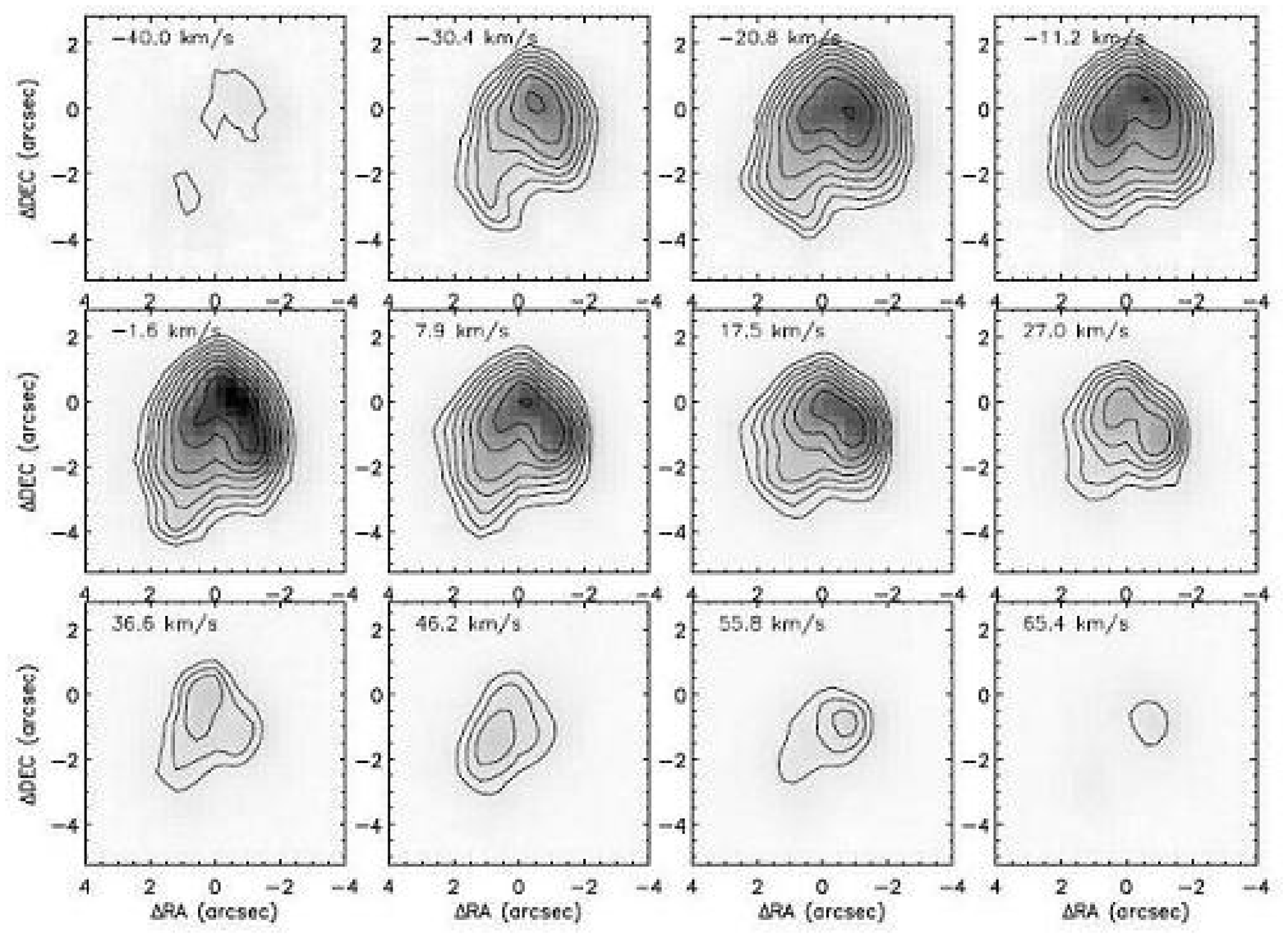} \caption{Channel maps of G5.89~-0.39
[Ne~II] line observations. Contours are drawn at $70\%$, $50\%$,
$35\%$, $25\%$, $17.5\%$, $12.5\%$, $9\%$ and $6\%$ of the peak
value of all channels. The peak value is
9.15~ergs~cm$^{-2}$~s$^{-1}$~sr$^{-1}$~(cm$^{-1}$)$^{-1}$. The
molecular cloud velocity is V$_{LSR}$ = 9 km s$^{-1}$.
\label{g589chanmapneii}}
\end{figure}

Our [Ne~II]~12.8~$\mu$m line map (Figure~\ref{g589map}, right)
shows a morphology similar to that of the mid-infrared continuum,
with a bright arc along the north side of the source.
With our $\sim 1.5''$ resolution, the kinematic signature of an
expanding shell is not apparent, although there is some evidence
that the line is broadest toward the center of the source
(Figure~\ref{g589chanmapneii}).
The majority of the emission is blue-shifted
by $\sim$9~km~s$^{-1}$ with respect to the molecular cloud
\citep{hatTM98}, and the extent of the
region becomes relatively smaller toward the red end of the
spectrum.

\subsection{G11.94~-0.62}
\begin{figure}
\epsscale{1.0} \plotone{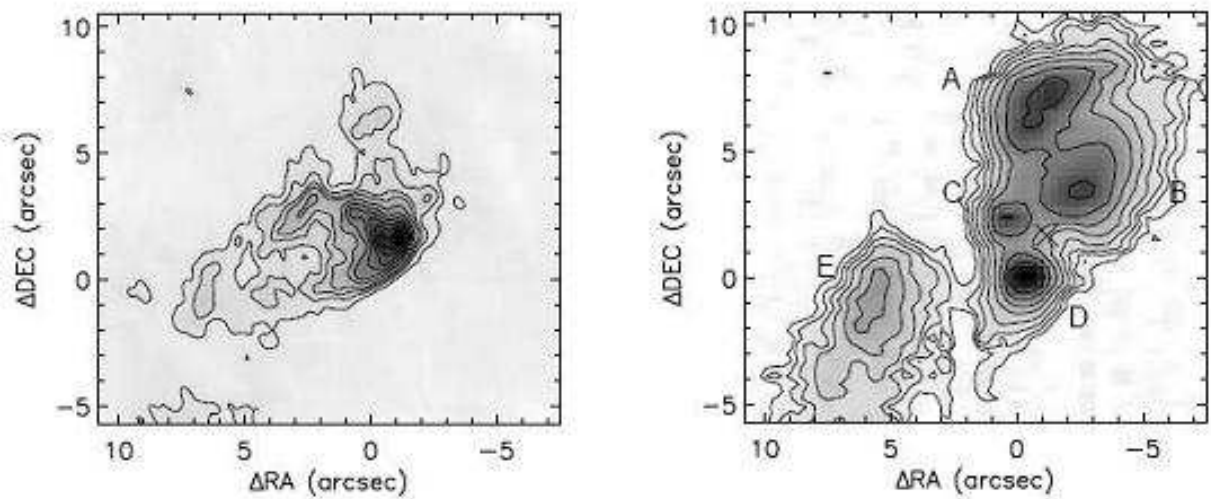} \caption{2 cm continuum map
\citep[left,][restoring beam: $0.''41 \times 0.''41$]{wooC89b} and
integrated [Ne~II] line flux map (right) of G11.94~-0.62. The (0,0)
is the location of the peak [Ne~II] line emission. The radio
continuum peak, marked with an ``X'', is offset 1.17$''$ west and
1.67$''$ north from the line emission peak according to
\cite{debRTP03}. Contours are drawn at $95\%$, $90\%$, $80\%$,
$70\%$, $60\%$, $50\%$, $40\%$ ,$30\%$, $20\%$,$15\%$, $10\%$ and
$5\%$ of the peak flux density for the continuum map and at $70\%$,
$50\%$, $35\%$, $25\%$, $17.5\%$, $12.5\%$, $9\%$ and $6\%$ of the
peak value for the line map. Five separate components (A, B, C, D
and E) are indicated in the line map. The peak [Ne~II] surface
brightness is 0.073~ergs~cm$^{-2}$~s$^{-1}$~sr$^{-1}$. The total
[Ne~II] flux is 6.3$\times$10$^{-11}$~ergs~cm$^{-2}$~s$^{-1}$.
\label{g1194map}}
\end{figure}

Located at a distance of 4.2~kpc \citep{chuWC90}, G11.94~-0.62
was categorized by \cite{wooC89b} as a cometary
UCHII region. In their radio continuum map (Figure~\ref{g1194map},
left), the region shows an emission arc with its apex pointing to
the west and diffuse emission extending to the east and north. A
broad $^{12}$CO(J=1-0) line was observed toward the object
\citep{sheC96}. The CO line width (full width at zero intensity) of
over 28~km~s$^{-1}$ suggests the presence of high velocity outflows.
\cite{debRTP03} imaged the region at 11.7~$\mu$m and 20.8~$\mu$m.
The region does not show a cometary shape in their mid-IR images
and, although their pointing has an astrometric accuracy about
1$''$, the radio peak does not match any structure in the mid-IR
images. The overall appearance of the region at mid-IR and radio
continuum wavelengths does not match well.

Our [Ne~II]~12.8~$\mu$m line map (Figure~\ref{g1194map}, right)
matches the mid-IR continuum observations \citep{debRTP03} very
well. By registering our map with the continuum map in
\cite{debRTP03}, we determine the position of the radio continuum
peak in our map, marked with an `X' at (-1.17$''$, 1.67$''$). Five
separate emission peaks (designated as A-E) can be seen in our
integrated [Ne~II] map. They all have corresponding 11.7$\mu$m and
20.8$\mu$m emission peaks in \cite{debRTP03}. Linear sizes of these
clumps range from $\sim$2$''$ in C to over 7$''$ in A and E. Like
the mid-IR continuum, the [Ne~II] morphology is a poor match to that
of the radio continuum.

\begin{figure}
\epsscale{1.0} \plotone{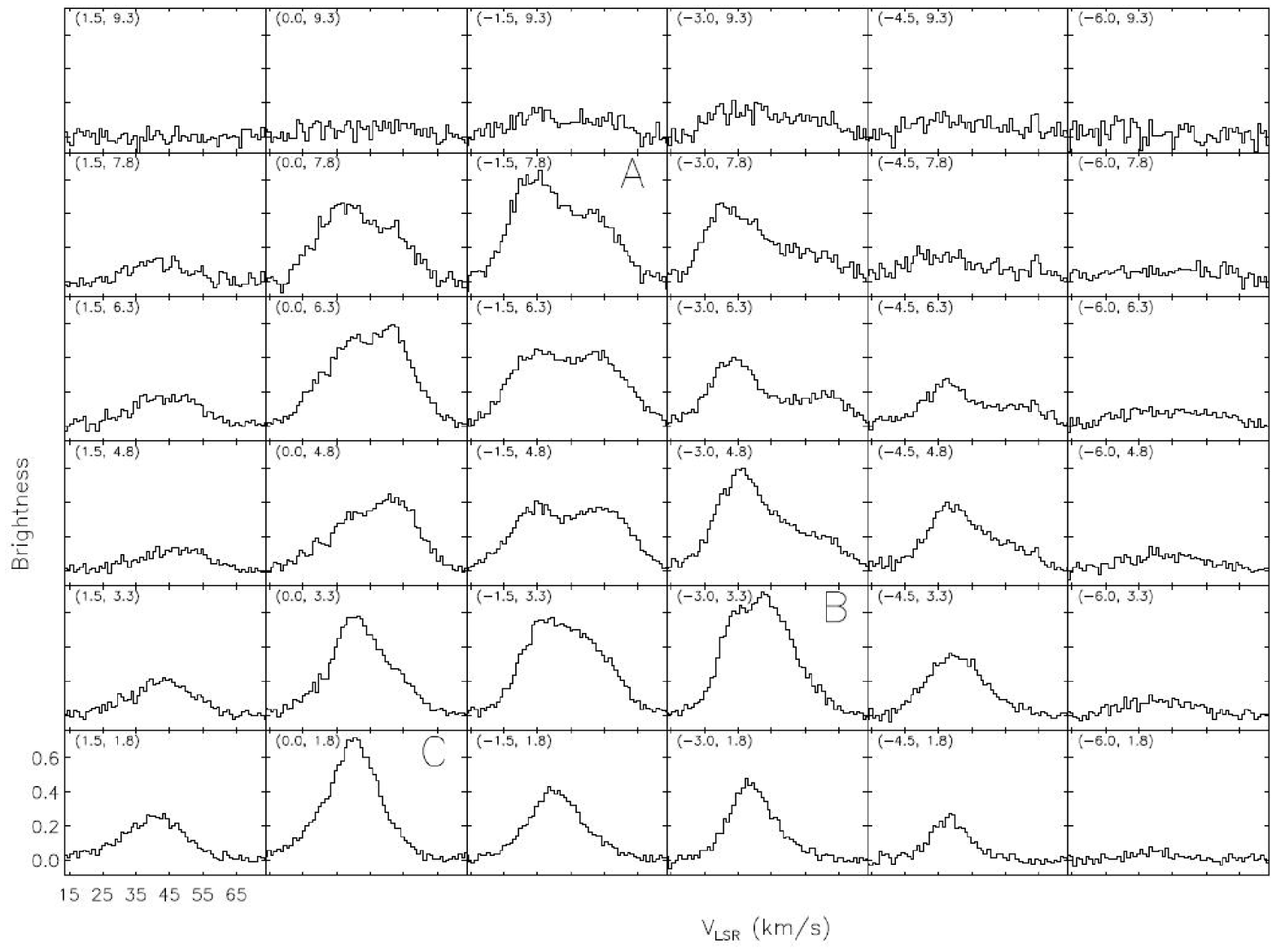} \caption{[Ne~II] line profiles
across G11.94~-0.62 (A, B and C). Spectra are averages over 1.5$''$
$\times 1.5''$. Central coordinates of the averaging areas are shown
in parentheses. The brightness has units of
erg~cm$^{-2}$~s$^{-1}$~sr$^{-1}$~(cm$^{-1}$)$^{-1}$. The pixel size
along the spectral direction is 0.0025~cm$^{-1}$
($\sim$0.95~km~s$^{-1}$). The molecular cloud velocity is V$_{LSR}$
= 39 km s$^{-1}$. \label{g1194ABCspec}}
\end{figure}

The spectra of the clumps C, D and E each can be fit with a single
Gaussian profile with a center velocity very close to the molecular
cloud velocity \citep[V$_{LSR} \simeq $39~km~s$^{-1}$,][]{chuWC90}.
Figure~\ref{g1194ABCspec} shows line profiles in the region of A, B
and C. Over this part of the source, the lines are broad, with most
of the additional emission to the red of the cloud velocity. The
profiles in Figure~\ref{g1194ABCspec} show that most of the [Ne~II]
emission occurs over the range V$_{LSR}$ = 30-60 km s$^{-1}$.
Complex morphology, probably induced by dust extinction, prevents a
detailed analysis of the kinematics in the region.

\subsection{G30.54~+0.02}
\begin{figure}
\epsscale{1.0} \plotone{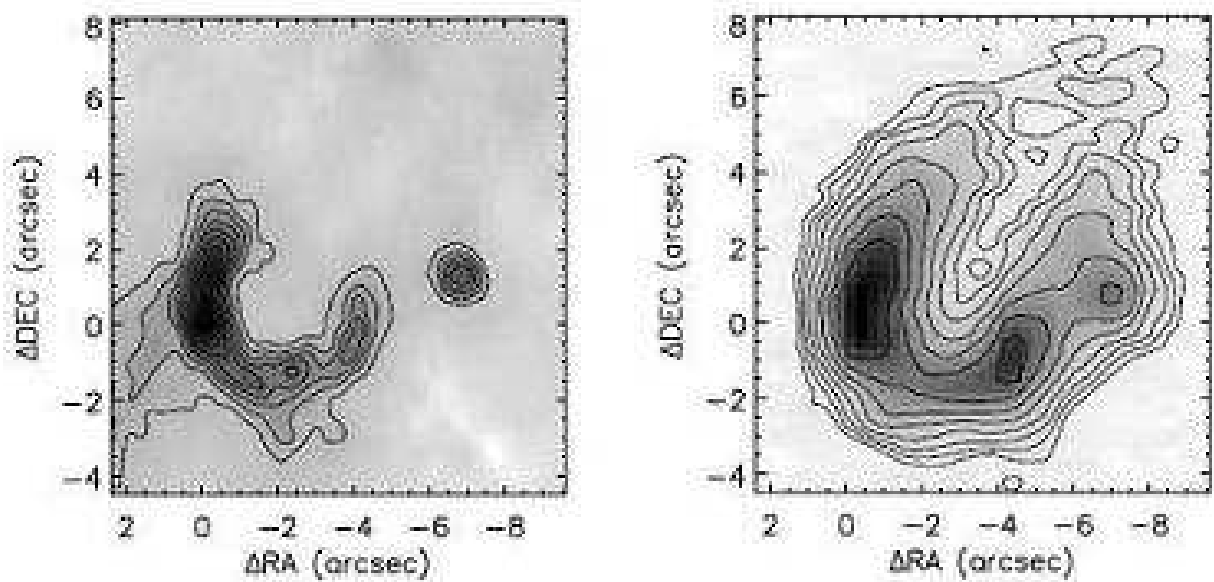} \caption{2~cm continuum map
\citep[left,][restoring beam: $0.''41 \times 0.''41$]{wooC89b} and
integrated [Ne~II] line map (right) of G30.54~+0.02. The [Ne~II]
line map is cross-correlated with the continuum map and shifted to
the best match. The (0,0) position marks the peak emission in the
line map. Contours are drawn at $95\%$, $90\%$, $80\%$, $70\%$,
$60\%$, $50\%$, $40\%$ ,$30\%$, $20\%$ and $10\%$ of the peak value
for the continuum map and at $70\%$, $50\%$, $35\%$, $25\%$,
$17.5\%$, $12.5\%$, $9\%$ and $6\%$ of the peak value for the line
map The peak [Ne~II] surface brightness is
0.086~ergs~cm$^{-2}$~s$^{-1}$~sr$^{-1}$. The total [Ne~II] flux is
4.8$\times$10$^{-11}$~ergs~cm$^{-2}$~s$^{-1}$. \label{g3054map}}
\end{figure}

G30.54~+0.02 has a limb-brightened, or ''horseshoe'' shape in both the radio
continuum map \citep{wooC89b} and our [Ne~II] line
map (Figure~\ref{g3054map}).
The more concentrated emission with a bigger opening in the
radio continuum map probably is the result of the higher resolution
of the interferometer observations and of the uneven sensitivity of the
snapshot mode aperture synthesis to compact and extended emission.
Both radio and [Ne~II] line observations show that
the emission arc breaks into at least three components.

\begin{figure}
\epsscale{1.0} \plotone{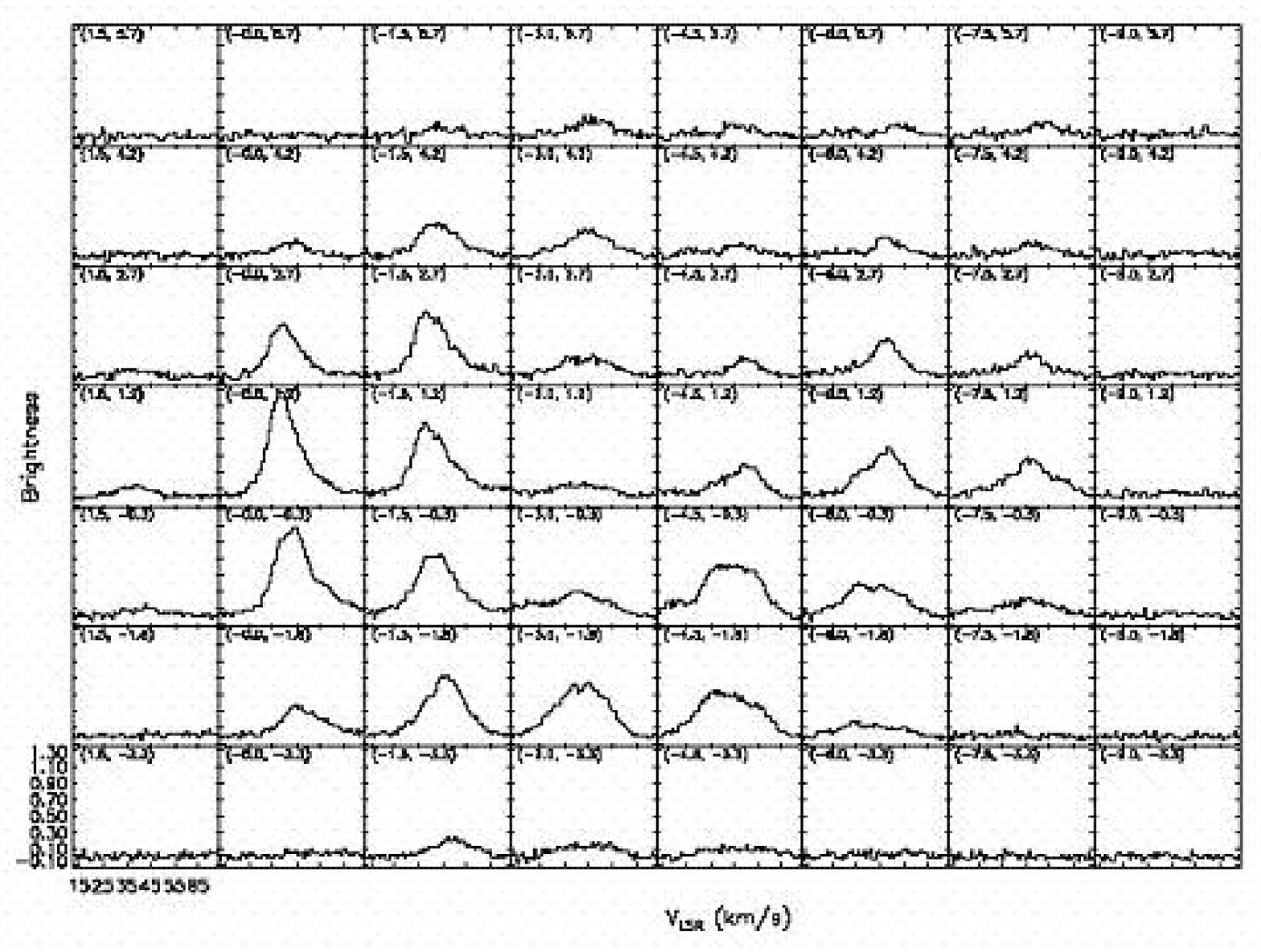} \caption{[Ne~II] line profiles for
G30.54~+0.02. Spectra are averages over $1.5'' \times 1.5''$.
Central coordinates of the averaging areas are shown in parentheses.
The brightness has units of
erg~cm$^{-2}$~s$^{-1}$~sr$^{-1}$~(cm$^{-1}$)$^{-1}$. The pixel size
along the spectral direction is 0.0025~cm$^{-1}$. The molecular
cloud velocity is V$_{LSR}$ = 48 km s$^{-1}$. \label{g3054profile}}
\end{figure}

\begin{figure}
\epsscale{0.6} \plotone{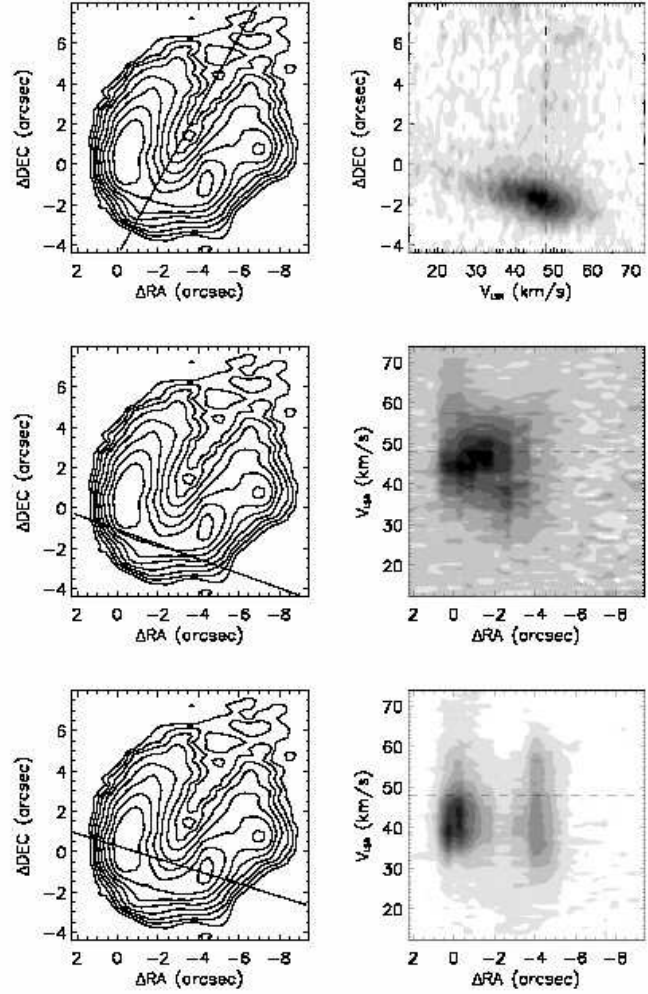} \caption{Position-velocity diagrams
of G30.54~+0.02. Contours in line flux maps are drawn at $70\%$,
$50\%$, $35\%$, $25\%$, $17.5\%$, $12.5\%$, $9\%$ and $6\%$ of the
peak value (0.086~ergs~cm$^{-2}$~s$^{-1}$~sr$^{-1}$). Dashed lines
in p-v diagrams show the ambient molecular material velocity.
\label{g3054pv}}
\end{figure}

Figure~\ref{g3054profile} shows [Ne~II] profiles across the region
while Figure~\ref{g3054pv} displays position-velocity diagrams on
selected cuts along and across the symmetry axis. Large line widths
(20-30~km~s$^{-1}$ FWHM), together with smooth profiles imply that
turbulence or substantial small-scale motions are present and may
mask the bulk motion of the gas. Nevertheless, a change of velocity
pattern within the structure is still apparent.

The velocity pattern along the bright arc is consistent with
significant tangential flows of the type seen in G29.96, but here
mostly in the plane of the sky. At the vertex of the horseshoe,
there is a single broad line with a velocity centroid close to the
molecular cloud velocity (Figure~\ref{g3054pv}). On the arm of the
horseshoe to the east of the vertex the line peak shifts $\sim$
7~km~s$^{-1}$ to the blue of the molecular cloud velocity while on
the western arm the molecular line and [Ne~II] peak velocities
agree. [Ne~II] emission is present over at least 40-50~km~s$^{-1}$
at all positions (Figures~\ref{g3054profile} and \ref{g3054pv}).

Along the symmetry axis and away from the bright arc, the match to a
tangential flow is less clear. Rather than breaking up into two
narrow lines, the emission on the axis and behind the vertex is at a
single velocity close to that of the molecular cloud. A small
velocity gradient across the apex may indicate an inclination to the
plane of sky.

\subsection{G33.92~+0.11}
G33.92~+0.11 is particularly remarkable because
of the apparent disconnect between its nondescript morphology and
its distinctive position-velocity structure. It was first
categorized as a core-halo UCHII region by \cite{wooC89b} and later
as a shell UCHII region by \cite{feyCGNJ92}. \cite{feyCGNJ92} showed
that the 20 cm extended emission is roughly shell-like, although the
cavity in their 20~cm image is offset to the southeast of the
emission peak, and the rest of the region has a morphology close to
a core-halo region with cometary extended emission. C$^{18}$O
observations showed two emission cores, and one of them is
associated with the peak of the UCHII region \citep{watM99}.

\begin{figure}
\epsscale{1.0} \plotone{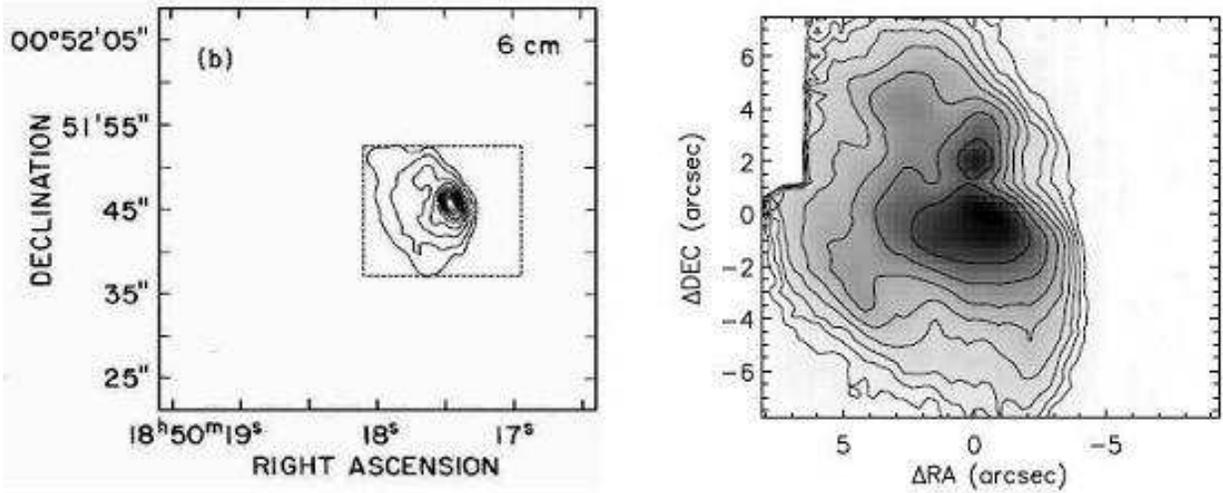} \caption{6~cm continuum map
\citep[left,][restoring beam: $1.''6 \times 1.''4$, positions in
B1950 coordinates]{feyCGNJ92} and integrated [Ne~II] line flux map
(right) of G33.92~+0.11. The box drawn with dashed lines in the
continuum map indicates the approximate area covered by the [Ne~II]
line observations. The (0,0) position in the line map is the
location of the peak [Ne~II] line emission. Contours were drawn at
$100\%$, $90\%$, $80\%$, $70\%$, $60\%$, $50\%$, $40\%$, $30\%$,
$20\%$, $10\%$ and $5\%$ of the peak value for the 6 cm continuum
map and at $70\%$, $50\%$, $35\%$, $25\%$, $17.5\%$, $12.5\%$, $9\%$
and $6\%$ of the peak value for the [Ne~II] line map. The peak
[Ne~II] surface brightness is
0.24~ergs~cm$^{-2}$~s$^{-1}$~sr$^{-1}$. The total [Ne~II] flux is
2.7$\times$10$^{-10}$~ergs~cm$^{-2}$~s$^{-1}$. \label{g3392map}}
\end{figure}

The integrated [Ne~II] line map shows an irregular peak surrounded
by extended emission (Figure~\ref{g3392map}). Mid-infrared continuum
observations show similar structure to that seen in the [Ne~II] line
observations \citep{givRBW07}. The major structure is a resolved
ellipsoid oriented northeast-southwest with an additional extension
to the east along the minor axis. The [Ne~II] and radio continuum
maps look very similar, but the central peak splits into two
components in the neon line, possibly as a result of foreground
extinction.

\begin{figure}
\epsscale{1.0} \plotone{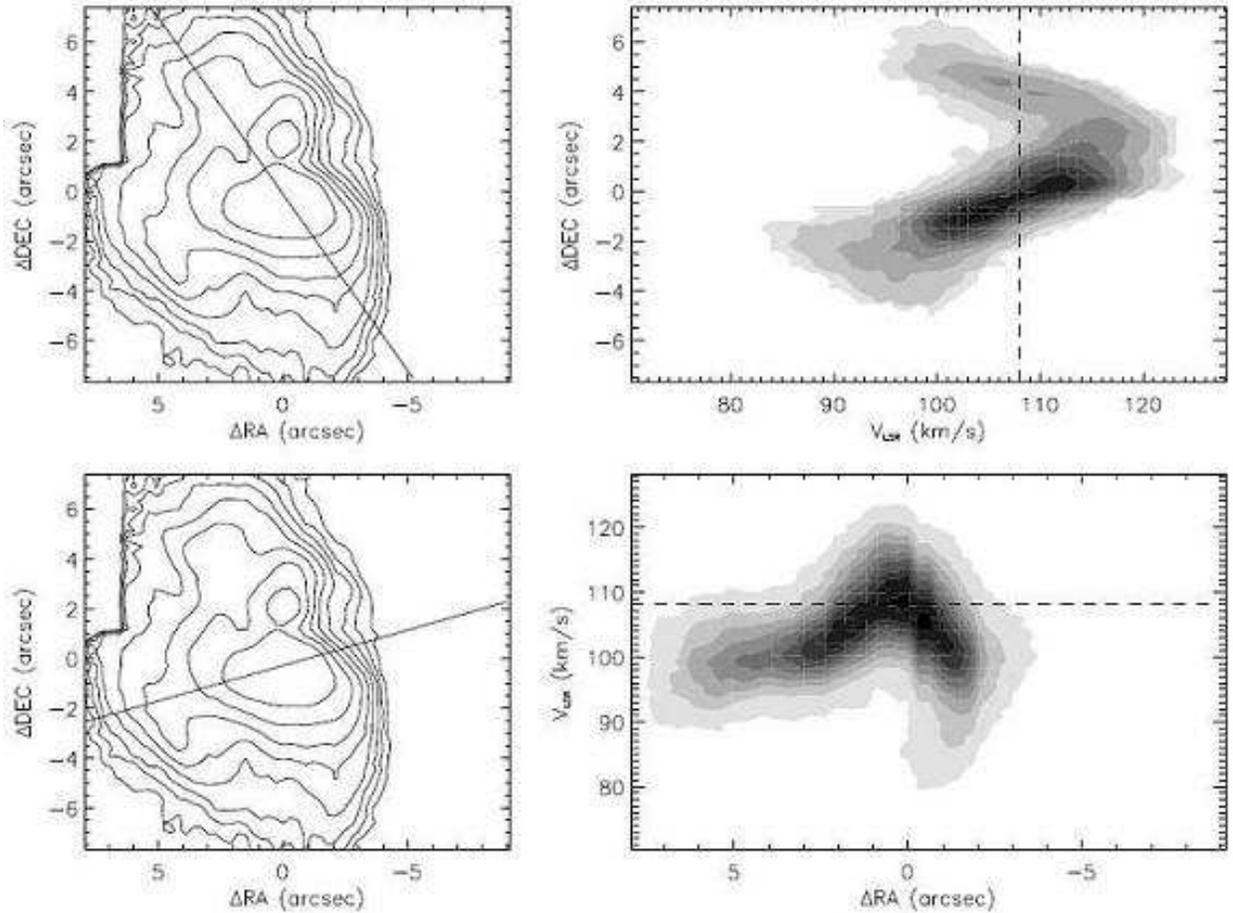} \caption{Position-velocity diagrams
of [Ne~II] observations toward G33.92~+0.11. Contours in line flux
maps are drawn at $70\%$, $50\%$, $35\%$, $25\%$, $17.5\%$,
$12.5\%$, $9\%$ and $6\%$ of the peak values
(0.24~ergs~cm$^{-2}$~s$^{-1}$~sr$^{-1}$) of the maps. Dashed lines
in p-v diagrams show the ambient molecular material velocity.
\label{g3392pv}}
\end{figure}

\begin{figure}
\epsscale{1.0} \plotone{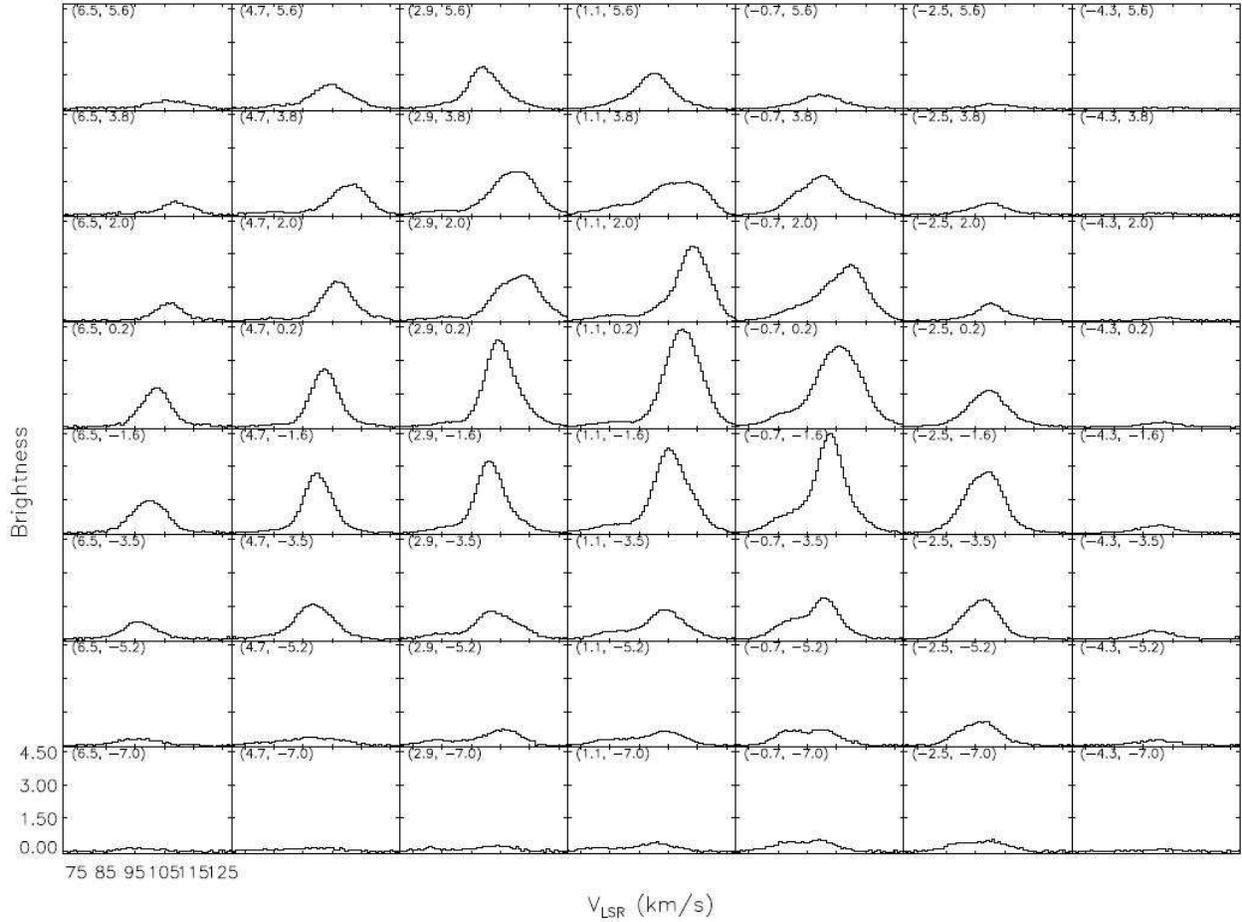} \caption{[Ne~II] line profiles for
G33.92~+0.11. Spectra are made by averaging spectra over $1.8''
\times 1.8''$. Central coordinates of the averaging areas are shown
in parentheses. The brightness has units of
erg~cm$^{-2}$~s$^{-1}$~sr$^{-1}$~(cm$^{-1}$)$^{-1}$. The pixel size
along the spectral direction is 0.0025~cm$^{-1}$.
\label{g3392multispec}}
\end{figure}

\begin{figure}
\epsscale{1.0} \plotone{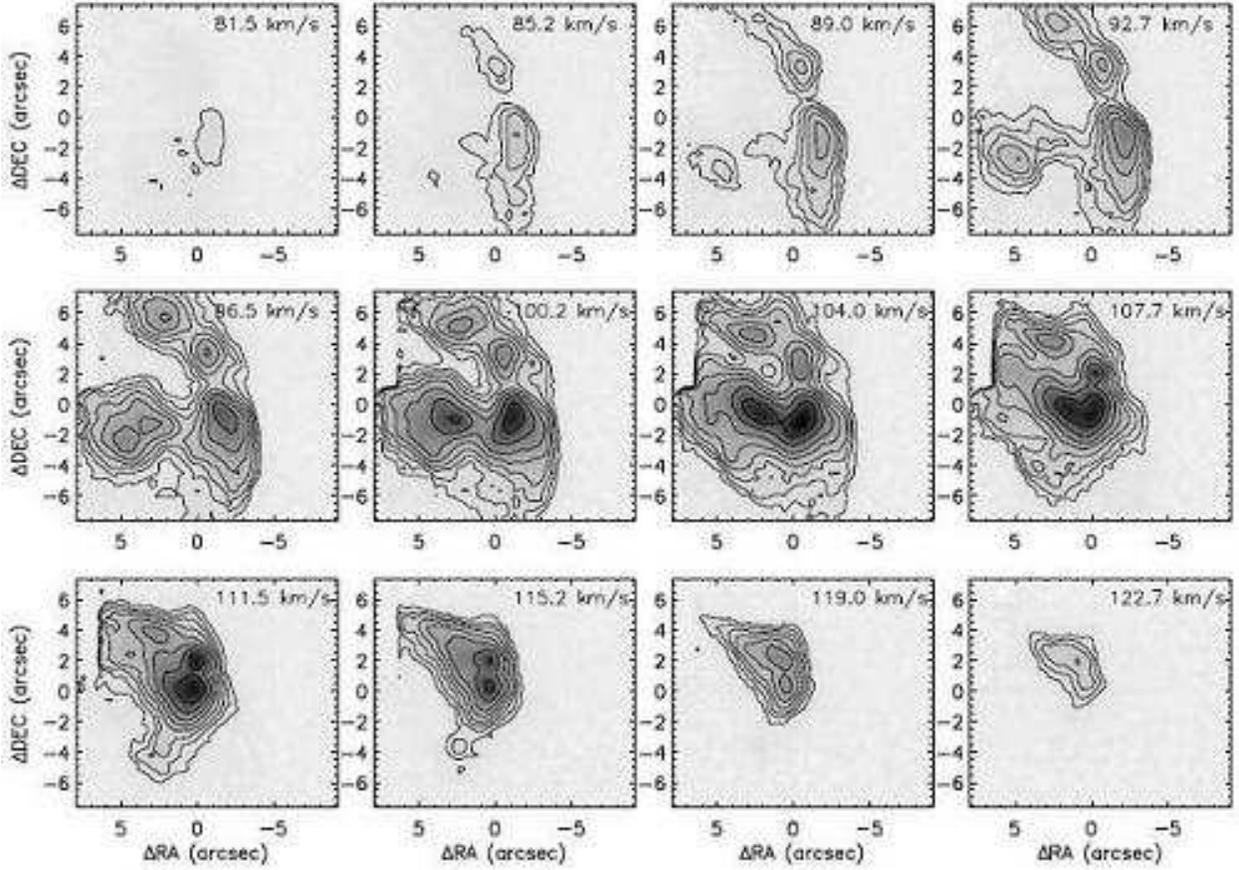} \caption{Channel maps of [Ne~II]
line observations for G33.92~+0.11. Contours are drawn at $70\%$,
$50\%$, $35\%$, $25\%$, $17.5\%$, $12.5\%$, $9\%$ and $6\%$ of the
peak value
(5.9~ergs~cm$^{-2}$~s$^{-1}$~sr$^{-1}$~(cm$^{-1}$)$^{-1}$). The
molecular cloud velocity is V$_{LSR}$ = 108 km s$^{-1}$.
\label{g3392chanmapneii}}
\end{figure}

Position-velocity diagrams along a few cross cuts through G33.92
(Figure~\ref{g3392pv}) show kinematics identifiable with the
kinematics typical of a flow along a surface similar to that seen in
G29.96 including the characteristic ``7'' and ``$\Lambda$''
patterns, despite its more nondescript morphology. Profiles at
individual positions are broad and frequently asymmetric
(Figure~\ref{g3392multispec}). Channel maps
(Figure~\ref{g3392chanmapneii}) show that three prominent components
form an arc at blue velocities and persist to the ambient material
velocity, V$_{LSR} \simeq $108~km~s$^{-1}$. This arc has a sharp
falloff to the west. A fourth emission peak and extended emission
can be seen to the east of the arc. The positions of these emission
peaks shift from channel to channel. This position shift is
especially prominent for the peak to the east of the arc. It moves
almost 5$''$ southeast from a position close to the arc in the
channel at V$_{LSR} {\simeq} $107~km~s$^{-1}$ to (+6,-3) in the
channel at V$_{LSR} {\simeq} $89~km~s$^{-1}$. The three components
along the arc move a smaller distance to the northwest. Towards the
red end of the spectrum, the emission components merge together and
fade away. We note that the overall shape of the region changes from
a thin crescent at blue-shifted velocities to a fan at red-shifted
velocities. Bow shock models, like that for G29.96 predict a similar
morphological change.

\subsection{G43.89~-0.78}
G43.89~-0.78 was classified as a cometary UCHII
region by \cite{wooC89b}.
\cite{macVW91} used G43.89 as one of their bow shock
model tests. They concluded, based on the continuum maps
that they needed a viewing angle between 45$^{\circ}$ and 60$^{\circ}$
from head-on in order to fit the morphology well.
\cite{sheC96} found
evidence for high-velocity molecular material in the region with a CO
line width over 39~km~s$^{-1}$. Low-resolution 21~$\mu$m imaging
observations showed the region as a point source with extended
emission to the southeast \citep{croC03}, while the radio continuum
observations showed extended emission northwest of an emission peak.

\begin{figure}
\epsscale{1.0} \plotone{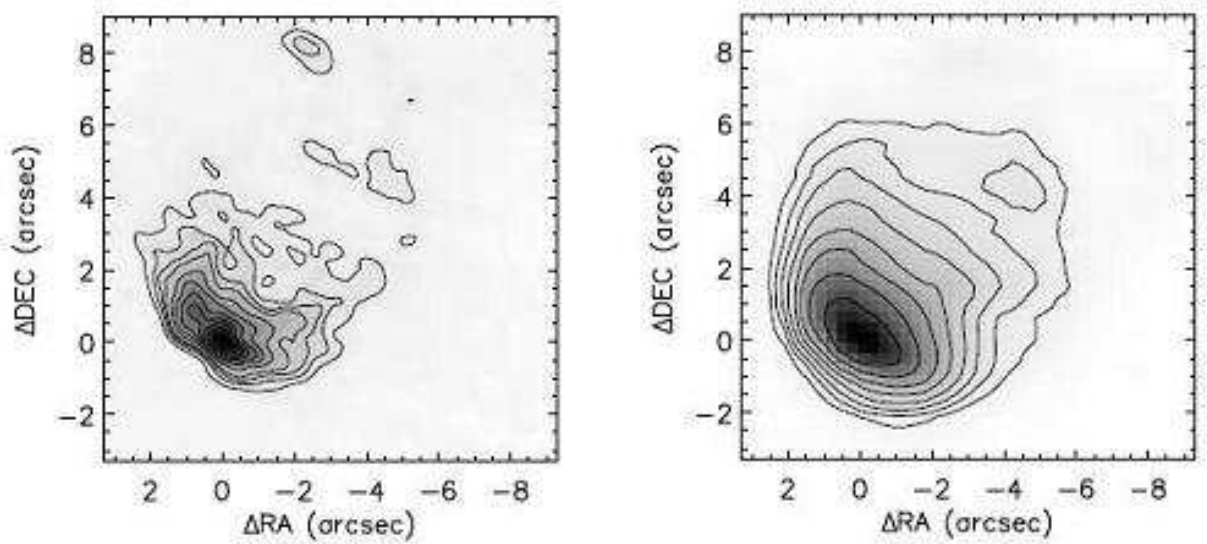} \caption{2~cm continuum contour map
\citep[left,][restoring beam: $0.''41 \times 0.''41$]{wooC89b} and
integrated [Ne~II] line flux map (right) of G43.89~-0.78. The line
map has been cross-correlated with the continuum map and shifted to
match. The (0,0) position is the location of peak emission in the
line map. Contours are drawn at $95\%$, $90\%$, $80\%$, $70\%$,
$60\%$, $50\%$, $40\%$, $30\%$, $20\%$, $15\%$, $10\%$ and $5\%$ of
the peak value for the radio map and at $70\%$, $50\%$, $35\%$,
$25\%$, $17.5\%$, $12.5\%$, $9\%$ and $6\%$ of the peak value for
the line map. The peak [Ne~II] surface brightness is
0.22~ergs~cm$^{-2}$~s$^{-1}$~sr$^{-1}$. The total [Ne~II] flux is
7.3$\times$10$^{-11}$~ergs~cm$^{-2}$~s$^{-1}$. \label{g4389map}}
\end{figure}

\begin{figure}
\epsscale{1.0} \plotone{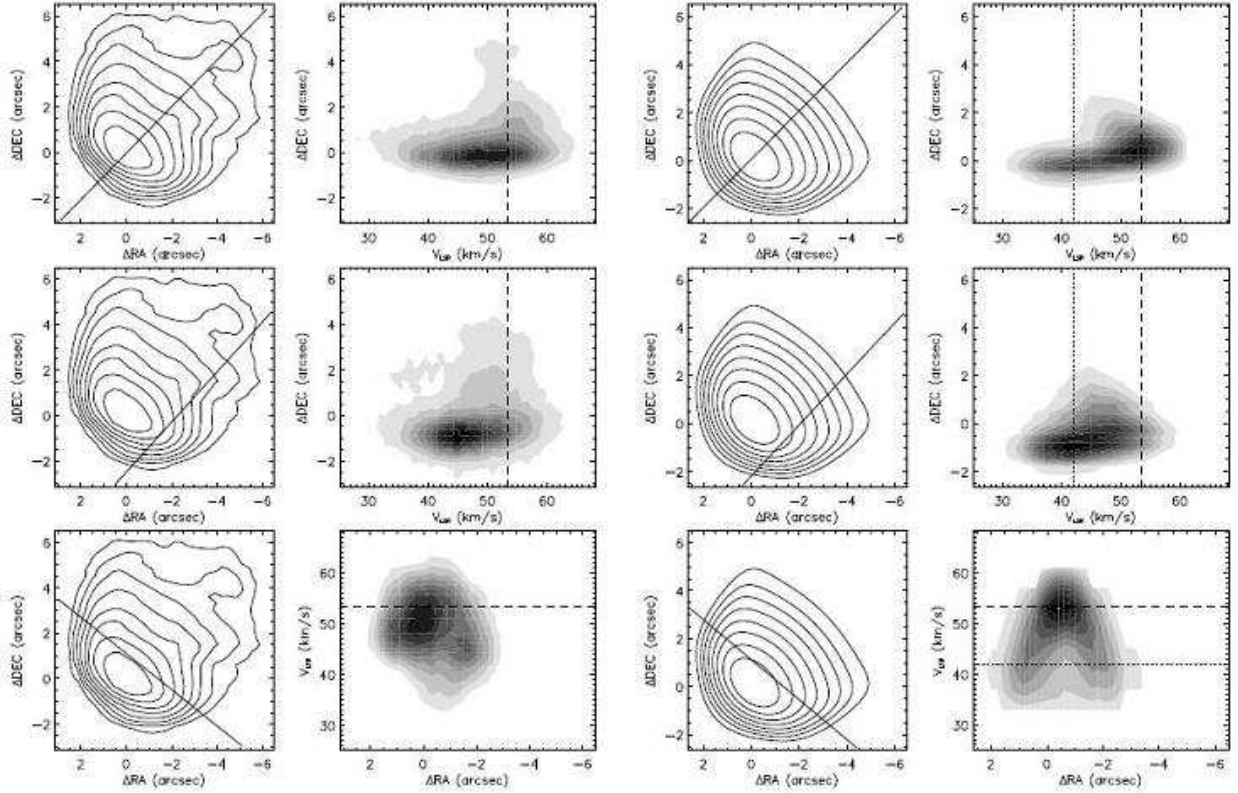} \caption{Position-velocity diagrams
of [Ne~II] line observations (two columns on the left) and a bow
shock model (two columns on the right) of G43.89~-0.78. Contours in
line flux maps are drawn at $70\%$, $50\%$, $35\%$, $25\%$,
$17.5\%$, $12.5\%$, $9\%$ and $6\%$ of the peak values
(0.22~ergs~cm$^{-2}$~s$^{-1}$~sr$^{-1}$) in the maps. See text for a
description of the model. The dashed lines in the p-v diagrams show
the ambient molecular cloud velocity. Dotted lines in the model p-v
diagrams show the ambient material velocity in the model.
\label{g4389pv}}
\end{figure}

The cometary morphology and kinematics of the [Ne~II] line emission
(Figure~\ref{g4389map}-\ref{g4389pv}) are consistent with flows
along the surface of the region like those seen in G29.96. We fitted
the p-v diagrams with a model like that used for G29.96
\citep{zhuLJRG05}. The best fit was with a stellar motion of
15~km~s$^{-1}$ and 140$^{\circ}$ away from the direction to the
observer. However, as with G29.96, there is a discrepancy between
the model, which predicts that the ionic emission should be
redshifted relative to the molecular emission, and the observed
emission, which is predominantly blueshifted.

\subsection{G45.07~+0.13}
G45.07~+0.13 was classified as an UCHII region with a shell
morphology by \citet{turM84}. At a distance of $\sim$6.0~kpc
\citep{chuWC90}, it has a diameter $\sim$0.9$''$ at $\lambda$=2~cm
\citep{wooC89b}. The slight velocity gradient in the H76$\alpha$
line was interpreted as the result of either an expanding ring
\citep{garRV86} or ionized bipolar outflows \citep{limW99}. A group
of water masers lie 2$''$ to the north of the radio continuum peak
\citep{hofC96,debRTP03}.

\begin{figure}
\epsscale{1.0} \plotone{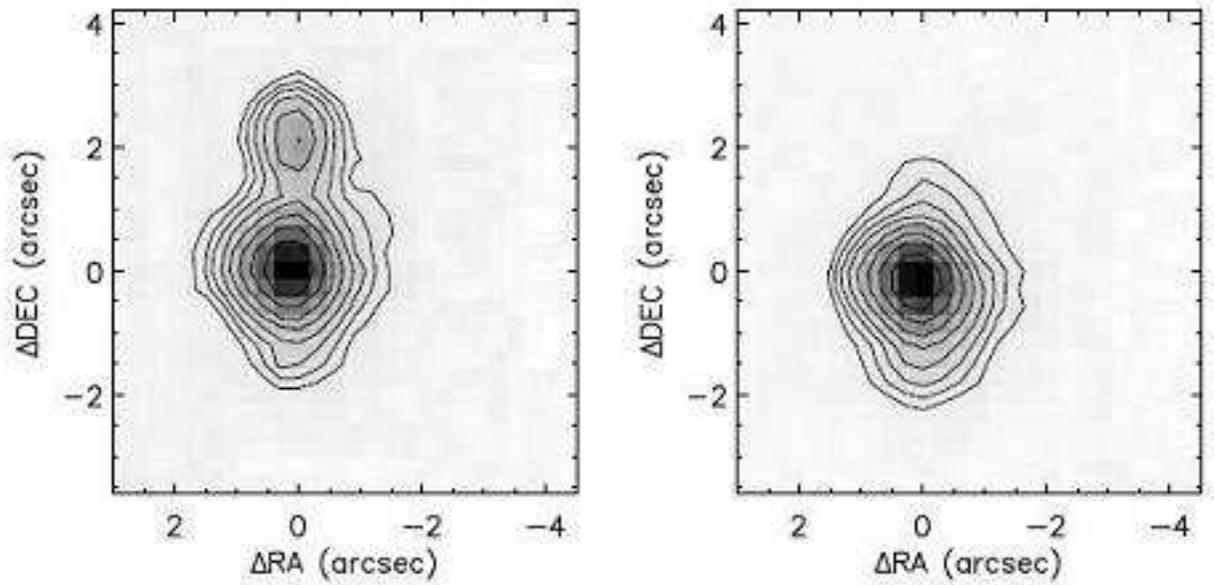} \caption{Dust 12.8~$\mu$m continuum
map (left) and integrated [Ne~II] line flux map (right) of
G45.07~+0.13. The fainter northern source in the left figure is a
hot molecular core candidate. The (0,0) position marks the peak
[Ne~II] line emission. Contours are drawn at $70\%$, $50\%$, $35\%$,
$25\%$, $17.5\%$, $12.5\%$, $9\%$ and $6\%$ of peak values in both
maps. The peak continuum flux density is
0.23~ergs~cm$^{-2}$~s$^{-1}$~sr$^{-1}$~(cm$^{-1}$)$^{-1}$. The peak
[Ne~II] surface brightness is
0.14~ergs~cm$^{-2}$~s$^{-1}$~sr$^{-1}$. The total [Ne~II] flux is
7.8$\times$10$^{-12}$~ergs~cm$^{-2}$~s$^{-1}$. \label{g4507mapneii}}
\end{figure}

G45.07 is unresolved in our 12.8~$\mu$m [Ne~II] map at 2$''$ spatial
resolution (Figure~\ref{g4507mapneii}, right). The [Ne~II] line
toward G45.07 is much narrower (V$_{FWHM} \simeq$ 20~km~s$^{-1}$)
than is the H76$\alpha$ radio recombination line
\citep[$\sim$48~km~s$^{-1}$,][]{garRV86}. The difference in widths
can arise either from pressure broadening of the H76$\alpha$ line or
from the presence of high velocity material at very high density
where the recombination line emissivity is proportional to n$_e^2$
while the [Ne~II] emissivity only scales with n$_e$. Dust continuum
emission is observed at G45.07 and at a second source $\sim$2$''$
north of G45.07 (Figure~\ref{g4507mapneii}, left). The location of
the second source matches the positions of the H$_2$O masers. We did
not detect any line emission from the second source, supporting the
suggestion that this source is at an early evolutionary stage
\citep{hofC96,debRTP03}.

\subsection{G45.12~+0.13}
G45.12~+0.13 (G45.12 hereafter) is located at a distance of
6.9~kpc \citep{chuWC90}.
It was observed by \cite{wooC89b} with the VLA at 2 and 6~cm and revealed
at least 3 peaks with a roughly arc-like overall morphology.

\begin{figure}
\epsscale{1.0}
\plotone{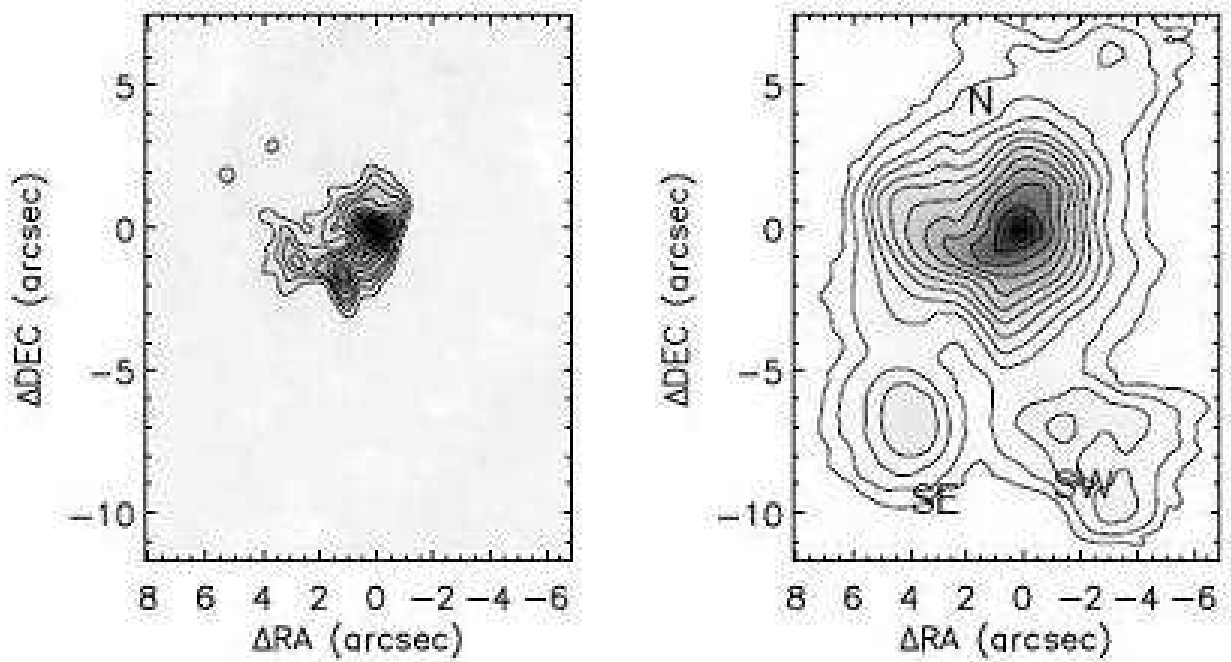}%{g4512_contimap_ne.eps}
\caption{2~cm continuum map \citep[left,][restoring beam:
$0.''41 \times 0.''41$]{wooC89b}
and integrated [Ne~II] line map (right) of G45.12~+0.13.
The (0,0) position is the location of the [Ne~II] line emission peak.
The contours are drawn at
$95\%$, $90\%$, $80\%$, $70\%$, $60\%$,
$50\%$, $40\%$ ,$30\%$, $20\%$,$15\%$, $10\%$ and $5\%$
of peak value for the radio continuum map and at
$70\%$, $50\%$, $35\%$, $25\%$, $17.5\%$,
$12.5\%$, $9\%$, $6\%$, $4\%$, $2\%$ and $1\%$ of the
peak value for the [Ne~II] map.
The peak [Ne~II] surface brightness is 1.2~ergs~cm$^{-2}$~s$^{-1}$~sr$^{-1}$.
The total [Ne~II] flux is 4.5$\times$10$^{-10}$~ergs~cm$^{-2}$~s$^{-1}$.
\label{g4512map}}
\end{figure}

\begin{figure}
\epsscale{1.0} \plotone{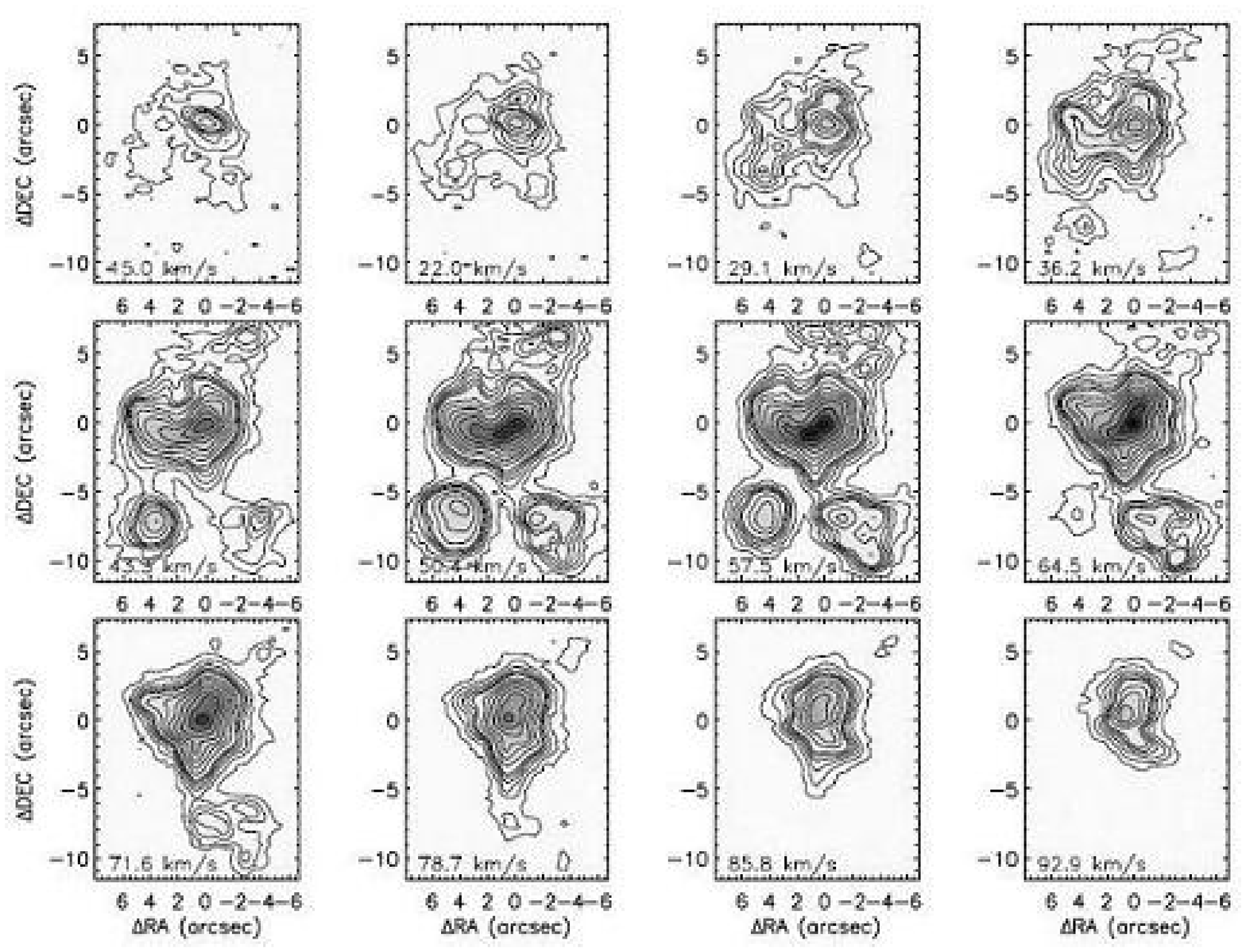} \caption{Channel maps for
G45.12~+0.13. Contours are drawn at $70\%$, $50\%$, $35\%$, $25\%$,
$17.5\%$, $12.5\%$, $9\%$, $6\%$, $4\%$, $2\%$ and $1\%$ of the peak
value of all channels. The peak value is
11.1~ergs~cm$^{-2}$~s$^{-1}$~sr$^{-1}$~(cm$^{-1}$)$^{-1}$. The
molecular cloud velocity is V$_{LSR}$ = 59 km s$^{-1}$.
\label{g4512chanmap}}
\end{figure}

\begin{figure}
\epsscale{1.0} \plotone{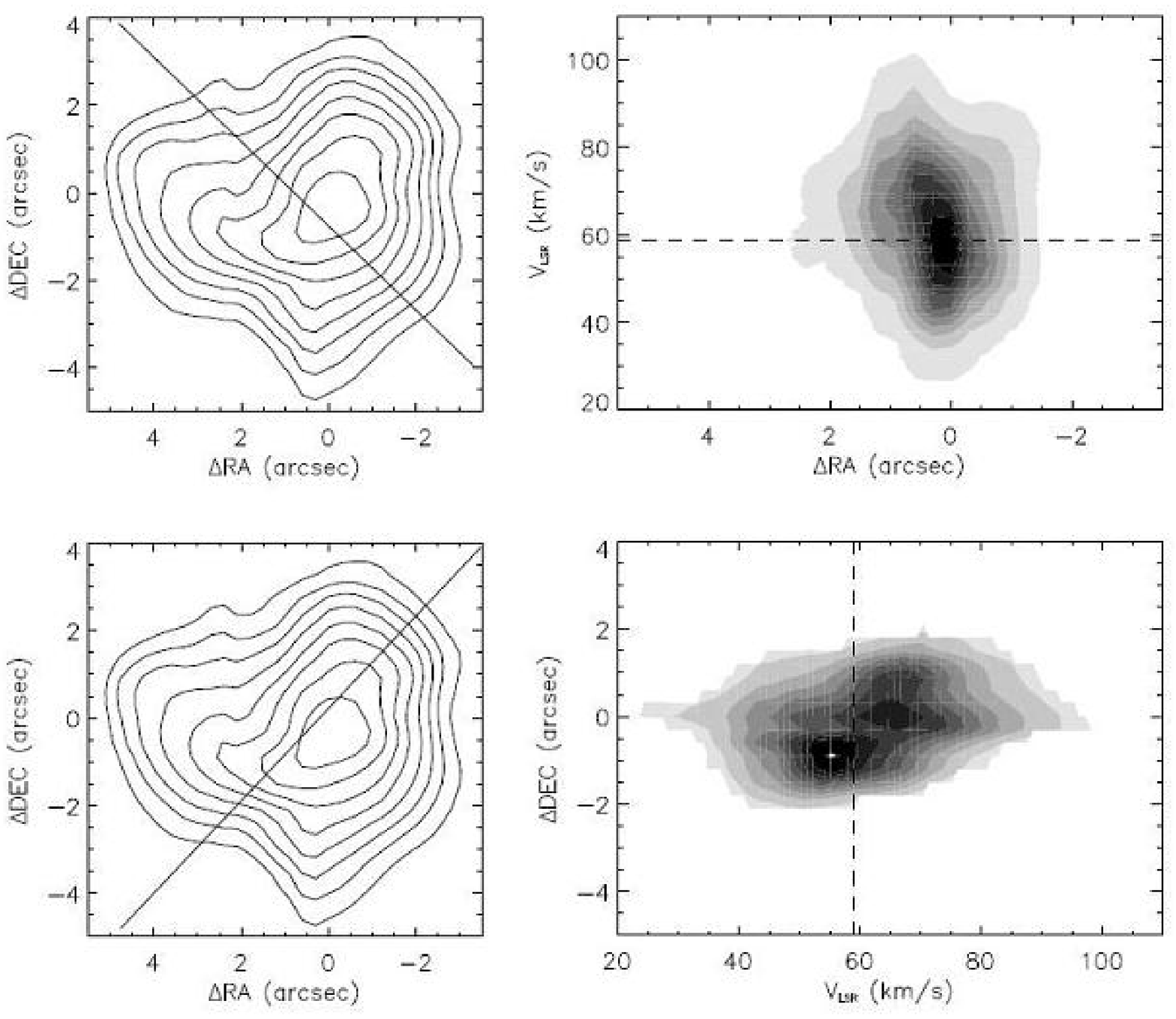} \caption{Position-velocity diagrams
of G45.12~+0.13N. Contours in line flux maps are drawn at $70\%$,
$50\%$, $35\%$, $25\%$, $17.5\%$, $12.5\%$, $9\%$ and $6\%$ of the
peak values (1.23~ergs~cm$^{-2}$~s$^{-1}$~sr$^{-1}$) in the maps.
Dashed lines in p-v diagrams show the ambient molecular material
velocity. \label{g4512Npv}}
\end{figure}

Our [Ne~II] line observations do not completely resolve the three
components along the radio emission arc, which is only $\sim$3''
long (referred to as Source ``N'' in Figure~\ref{g4512map}). The
channel maps (Figure~\ref{g4512chanmap}) show that the ``horns'' of
the arc-like structure are more prominent at lower redshifts while
the bright spot at the middle of the arc dominates at higher
velocities. The p-v diagrams of a cut parallel to the symmetry axis
of the source N do show ``7'' patterns of cometary H~II regions
(Figure~\ref{g4512Npv}, top panels), but a cut perpendicular to the
axis does not show the expected ``$\Lambda$'' shaped pattern
(Figure~\ref{g4512Npv}, bottom panels). It is not clear how much the
kinematics in G45.12N resembles that in a cometary HII region,
because its ``7'' type p-v diagram is not dramatic compared to that
in G29.96 (Figure~\ref{g2996pv1}).

\begin{figure}
\epsscale{1.0} \plotone{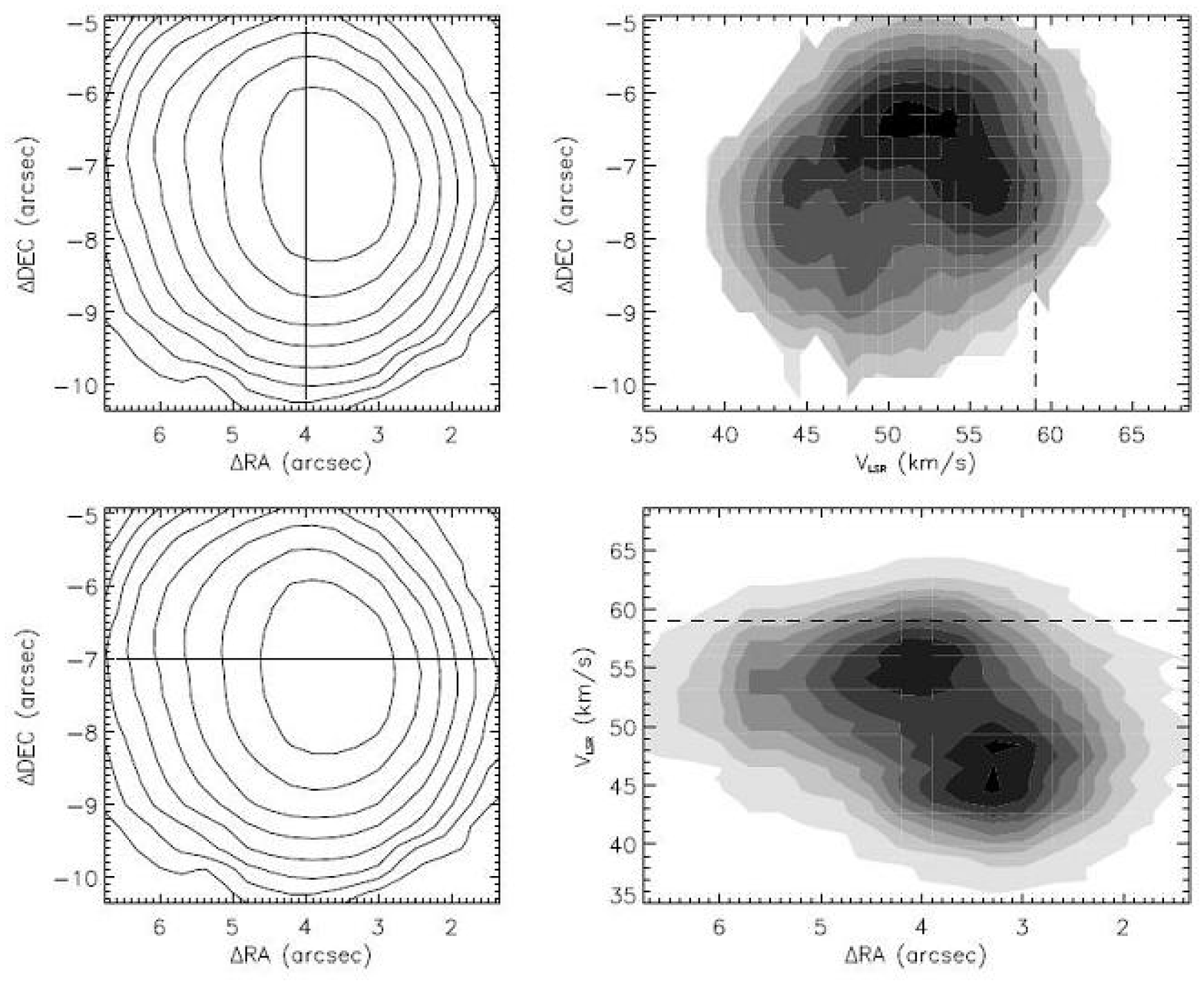} \caption{Position-velocity diagrams
of G45.12~+0.13SE. Contours in line flux maps are drawn at $70\%$,
$50\%$, $35\%$, $25\%$, $17.5\%$, $12.5\%$, $9\%$ and $6\%$ of the
peak values (0.10~ergs~cm$^{-2}$~s$^{-1}$~sr$^{-1}$) in the maps.
Dashed lines in p-v diagrams show the ambient molecular material
velocity. \label{g4512SEpv}}
\end{figure}

We also find two fainter sources to the south of the arc which were
not seen in the sparsely sampled VLA snapshot image. To show these
two faint sources clearly, we extend contour levels down to 1$\%$ of
the peak value in our line map. The southeastern source (SE) is
extended and almost round-shaped. It is hard to determine its
symmetry axes. Its p-v diagrams along RA and DEC cuts
(Figure~\ref{g4512SEpv}) show evidence of gas bulk motion and most
of emission is blue-shifted relative to the ambient molecular gas.
Tentative ``7'' shaped emission distribution can be seen in these
diagrams, suggesting that surface flows may be present in the
source. The southwestern source (SW) is irregular and contains
multiple peaks. These emission peaks surround an emission minimum at
the same location $\sim$(2.$''$7, -8$''$) in almost all channels
(Figure~\ref{g4512chanmap}). A broken shell-like morphology with an
opening to the west suggests an expanding shell structure for SW,
but the ring of the line emission does not collapse at the ends of
the Doppler shift-range as would be expected for an expanding shell.
The velocity distributions in both N and SW are more or less
symmetric about the ambient material velocity, which is
59~km~s$^{-1}$. Line emission from SE is relatively blue-shifted,
and the width of the [Ne~II] line is narrower in SE than in the
other two sources.

\subsection{G45.45~+0.06}
\begin{figure}
\epsscale{1.0} \plotone{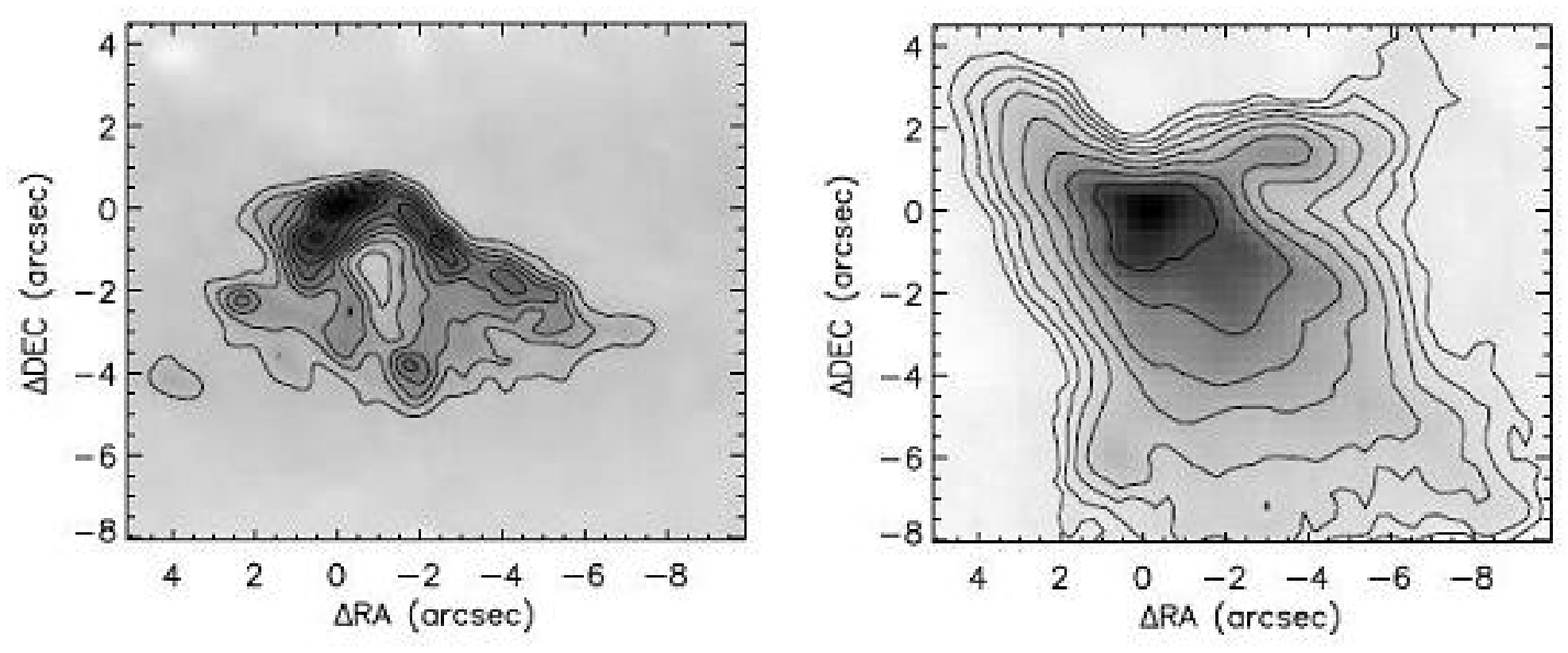} \caption{6~cm continuum map
\citep[left,][restoring beam: $0.''38 \times 0.''38$]{wooC89b} and
integrated [Ne~II] line map (right) of G45.45~+0.06. The [Ne~II]
line map has been cross-correlated with the radio map and the (0,0)
position has been shifted to the best match. The (0,0) position is
the location of the emission peak in the line map. Contours are
drawn at $95\%$, $90\%$, $80\%$, $70\%$, $60\%$, $50\%$, $40\%$
,$30\%$, $20\%$ and $10\%$ of peak value for the continuum map and
at $70\%$, $50\%$, $35\%$, $25\%$, $17.5\%$, $12.5\%$, $9\%$ and
$6\%$ of the peak value for the line map. The peak [Ne~II] surface
brightness is 0.16~ergs~cm$^{-2}$~s$^{-1}$~sr$^{-1}$. The total
[Ne~II] flux is 1.2$\times$10$^{-10}$~ergs~cm$^{-2}$~s$^{-1}$.
\label{g4545map}}
\end{figure}

G45.45~+0.06 was categorized by \cite{wooC89b} as a cometary UCHII
region. In the 6~cm radio continuum image, the source extends
$\sim9''$~EW and $6''$~NS and has a sharp and bright ionization
front to the north and extended emission to the south
(Figure~\ref{g4545map}, left panel). Multiple emission peaks
arranged along the ionization front form a distorted ring around a
cavity. At a distance 6.6~kpc, the integrated continuum flux density
implies an O7.5 main sequence star as the ionizing star
\citep{chuWC90}. \cite{felSHH98} acquired images at H and K$'$ bands
and three mid-infrared bands centered at 3.8$\mu$m, 10.5$\mu$m and
11.7$\mu$m and found that the region contained a young OB cluster.

\begin{figure}
\epsscale{1.0} \plotone{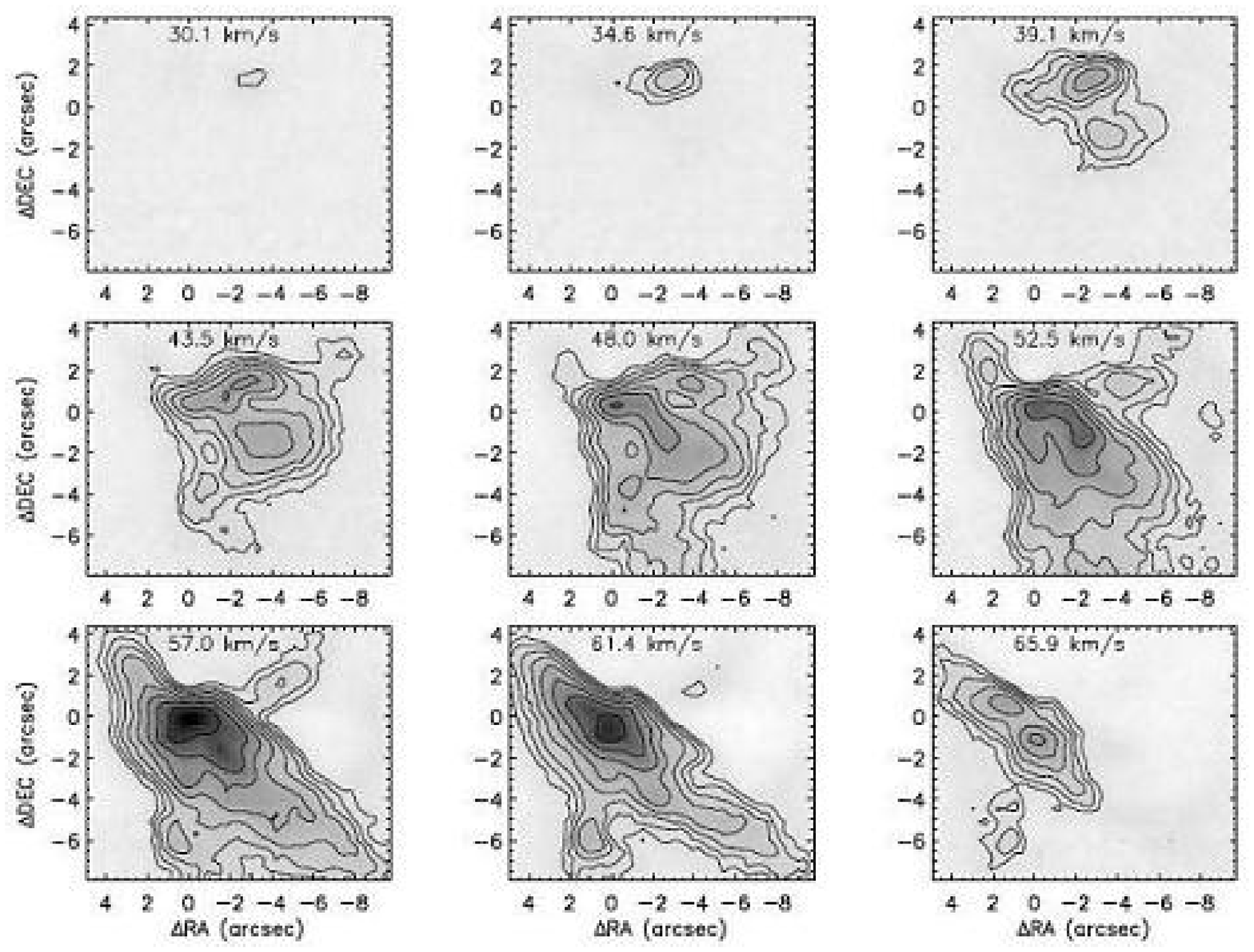} \caption{Channel maps for
G45.45~+0.06 [Ne~II] line observations. Contours are drawn at
$70\%$, $50\%$, $35\%$, $25\%$, $17.5\%$, $12.5\%$, $9\%$ and $6\%$
of the peak value of all channels. The peak value is
4.2~ergs~cm$^{-2}$~s$^{-1}$~sr$^{-1}$~(cm$^{-1}$)$^{-1}$. The
molecular cloud velocity is V$_{LSR}$ = 59 km s$^{-1}$.
\label{g4545chanmapneii}}
\end{figure}

The overall morphology of the [Ne~II] line emission
(Figure~\ref{g4545map}, right) does not look like that of the radio
continuum map. Most of the line emission comes from an elongated
structure aligned northeast to southwest with the extended emission
to the southwest roughly matching that seen in Br$\gamma$ by
\cite{felSHH98}. The intermediate velocity (43-53~km~s$^{-1}$)
channel maps (Figure~\ref{g4545chanmapneii}) most nearly match the
radio emission. The NE-SW elongated structure is primarily present
in red-shifted velocity channels. Extended emission north of the
radio arc and a overlying compact 11.7~$\mu$m source
\citep{felSHH98} dominate the blue-shifted channels. The differences
between the [Ne~II] and radio distributions could be due to
incomplete uv coverage of the snap shot radio observations or/and to
greater sensitivity of [Ne II] line observations.

\subsection{W51 IRS2}
W51 IRS2 (W51d) is a luminous \citep[L = 2-4 $\times $10$^6$
L$_{\odot}$,][]{eriT80,jafHG87} ultracompact HII region/molecular
cloud core within a more extensive star forming complex (Carpenter
\& Sanders 1998). Proper motion studies of H$_2$O maser features
place W51 IRS2 at a distance of $\sim$ 7~kpc
\citep{genDSR81,schMG81,imaW02}. Elsewhere, we have presented
0.5$''$ resolution maps of the [S~IV] 10.5~$\mu$m and [Ne~II] 12.8
$\mu$m lines toward this source \citep{lacJZRBG07}. These maps
reveal a number of embedded UCHII regions, which dominate the
[Ne~II] emission, a more extensive surface flow, seen mostly in
[S~IV], and a prominent $\sim$ 100~km~s$^{-1}$ jet, which we
attribute to ionization of one lobe of a protostellar jet. The
surface flow has a position-velocity pattern indicating that the gas
is moving along a shell with its vertex pointed away from the
observer. At the shell edges, the emission is blue shifted by
10-15~km~s$^{-1}$ with respect to the ambient cloud. The flow is
less evident in [Ne~II] where several compact sources are much
brighter, either because of the lower extinction at 12.8~$\mu$m or
because of the lower ionization potential of neon than of S$^{++}$.

\begin{figure}
\epsscale{1.0} \plotone{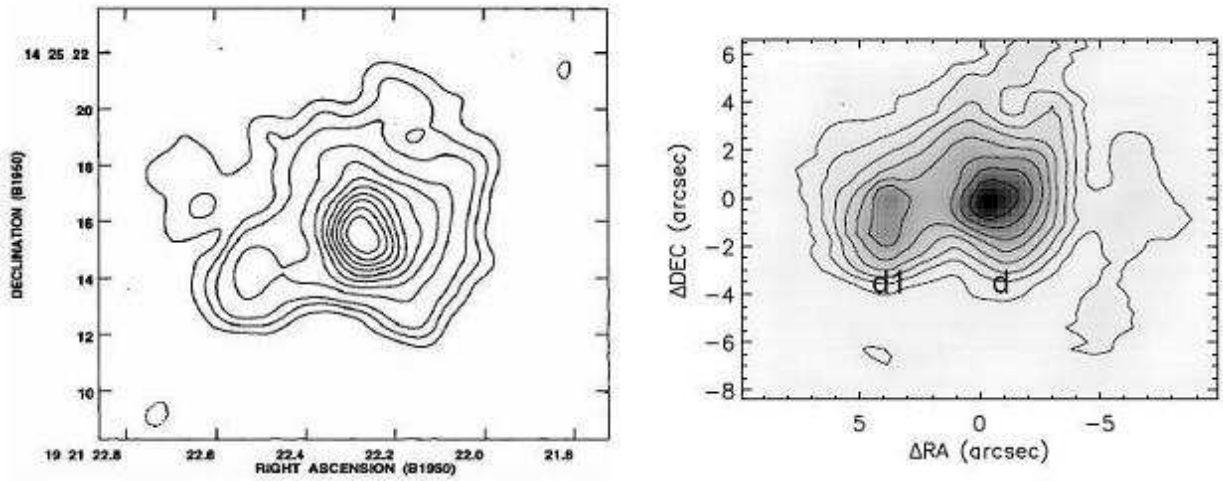} \caption{2~cm continuum map
\citep[left,][resolution: $1.''2$]{gauM87} and integrated [Ne~II]
line flux map (right) of W51 IRS2. Contours in the radio map were
drawn at $90\%$, $80\%$, $70\%$, $60\%$, $50\%$, $40\%$, $30\%$
,$20\%$, $10\%$ and $-10\%$ of the peak value. Contours in the
[Ne~II] map are drawn at $70\%$, $50\%$, $35\%$, $25\%$, $17.5\%$,
$12.5\%$, $9\%$ and $6\%$ of peak line emission at (0,0). The peak
[Ne~II] surface brightness is
0.66~ergs~cm$^{-2}$~s$^{-1}$~sr$^{-1}$. The total [Ne~II] flux is
4.2$\times$10$^{-10}$~ergs~cm$^{-2}$~s$^{-1}$. \label{w51dmapneii}}
\end{figure}

\begin{figure}
\epsscale{1.0} \plotone{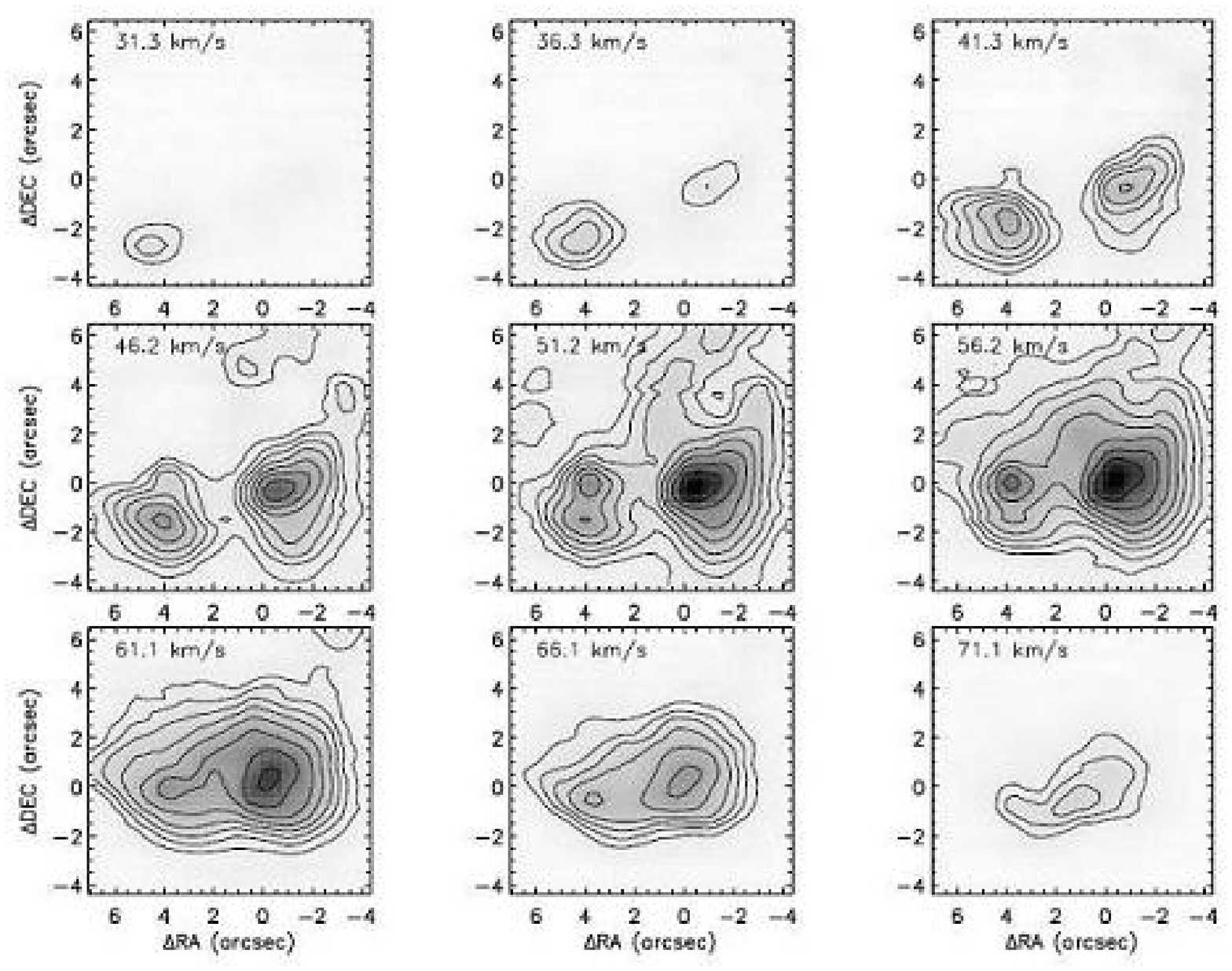} \caption{Channel maps of W51 IRS2
[Ne~II] line observations. Contours are drawn at $70\%$, $50\%$,
$35\%$, $25\%$, $17.5\%$, $12.5\%$, $9\%$ and $6\%$ of the peak
value of all channels. The peak value is
13.2~ergs~cm$^{-2}$~s$^{-1}$~sr$^{-1}$~(cm$^{-1}$)$^{-1}$. The
molecular cloud velocity is V$_{LSR}$ = 58 km s$^{-1}$.
\label{w51dchanmapneii}}
\end{figure}

\begin{figure}
\epsscale{0.6} \plotone{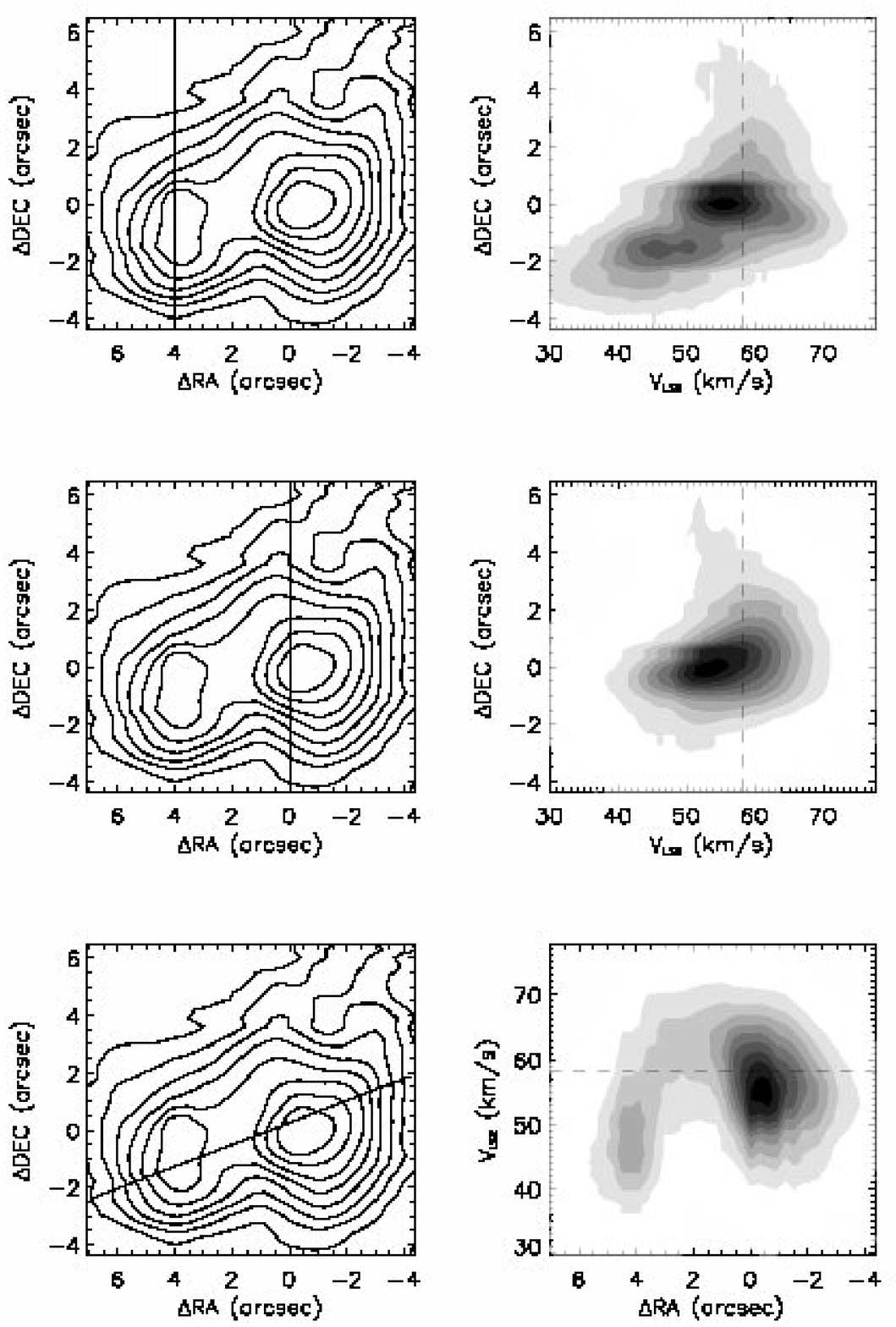} \caption{Position-velocity diagrams
of W51 IRS2 [Ne~II] line observations. Contours in line flux maps
are drawn at $70\%$, $50\%$, $35\%$, $25\%$, $17.5\%$, $12.5\%$,
$9\%$ and $6\%$ of peak values
(0.66~ergs~cm$^{-2}$~s$^{-1}$~sr$^{-1}$) of the maps. Dashed lines
in p-v diagrams show the ambient molecular material velocity.
\label{w51dpv}}
\end{figure}

We include our somewhat more extensive 2$''$ resolution IRTF map of
W51 IRS2 in the [Ne~II] line here for completeness.
Figure~\ref{w51dmapneii} compares the distribution of radio
continuum and [Ne~II] emission. The channel maps in
Figure~\ref{w51dchanmapneii}, as well as the top p-v diagram in
Figure~\ref{w51dpv} show that the two components of the eastern
source d1 have velocities that differ by $\sim$ 10~km~s$^{-1}$, with
the southern component blueshifted by $\sim$ 12~km~s$^{-1}$ from the
molecular cloud velocity. The molecular emission spectra of CS show
a second feature close to the [Ne~II] velocity in the southern
component (49.5~km~s$^{-1}$, Plume et al. 1997). The northern
component of W51d1 and W51d have [Ne~II] velocities close to the
velocity of the main component of the molecular cloud. We also
notice that the ionized gas toward the center of the region is
somewhat redshifted with respect to ambient material, which is
likely because the ionized gas is pushed into the cloud by
stellar/cluster winds.

\subsection{G61.48~+0.09B}
\begin{figure}
\epsscale{0.5} \plotone{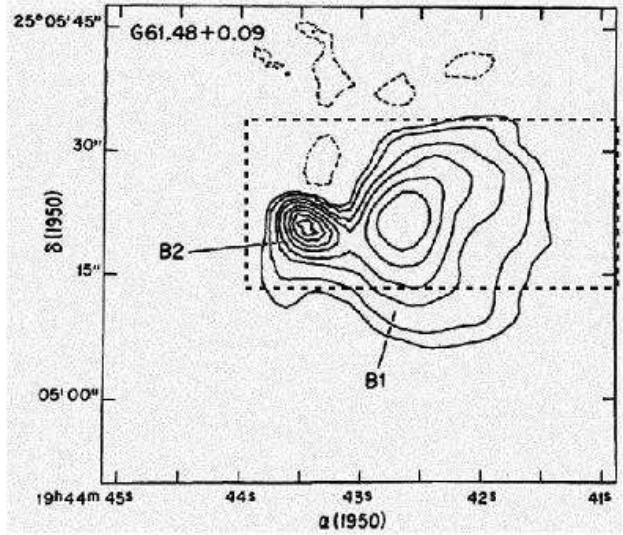} \caption{1.3~cm continuum map of
G61.48~+0.09 \citep[][resolution $\sim 4''$]{gomGL95}. Contours were
drawn at $90\%$, $75\%$, $60\%$, $45\%$, $30\%$ ,$20\%$, $10\%$,
$5\%$ and $-5\%$ of the peak value. The box drawn with dashed lines
indicates the approximate area covered by our [Ne~II] observations.
\label{g6148mapradio}}
\end{figure}

\begin{figure}
\epsscale{0.5} \plotone{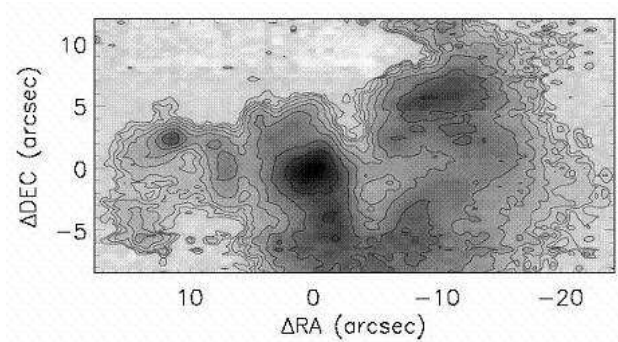} \caption{Integrated [Ne~II] line
flux map of G61.48~+0.09. The (0,0) position is the location of the
[Ne~II] line emission peak. Contours are drawn at $70\%$, $50\%$,
$35\%$, $25\%$, $17.5\%$, $12.5\%$, $9\%$ and $6\%$ of the peak
value. The peak [Ne~II] surface brightness is
0.052~ergs~cm$^{-2}$~s$^{-1}$~sr$^{-1}$. The total [Ne~II] flux is
2.2$\times$10$^{-10}$~ergs~cm$^{-2}$~s$^{-1}$. \label{g6148mapneii}}
\end{figure}

G61.48~+0.09B  belongs to the emission nebula complex Sh2-88B at a
distance of 5.4~kpc \citep{chuWC90}. \cite{garLG94} compared
kinematics from hydrogen recombination line and molecular line
observations \citep{macLBRMF81,chuWC90} and concluded that the
region was undergoing a blister flow to the southwest. They argued
that the hydrogen recombination line kinematics and the radio
continuum morphology of the region were best explained by expansion
of ionized gas in a non-uniform medium. Radio continuum observations
show that the region consists of two components, designated as B1
and B2 in Figure~\ref{g6148mapradio}. The more compact and spherical
B2 is located $\sim$11$''$ east of the peak of B1, which is extended
and has slightly curved contours that hint toward a cometary
morphology. Our [Ne~II] line emission distribution
(Figure~\ref{g6148mapneii}), which is very similar to the
distribution of Br$\gamma$ emission \citep{pugAFHW04}, shows
significantly different structure than the radio continuum map. The
morphological difference between the [Ne~II] line emission and the
radio continuum emission is likely due to dust extinction. The
[Ne~II] emission has complex morphology and kinematics. Multiple
components are present over a 35$''$ by 20$''$ area. A bright ridge
of emitting gas runs diagonally from the northeast to south central
in the map. A less bright irregular structure extends to the
northwest from the south.

\begin{figure}
\epsscale{1.0} \plotone{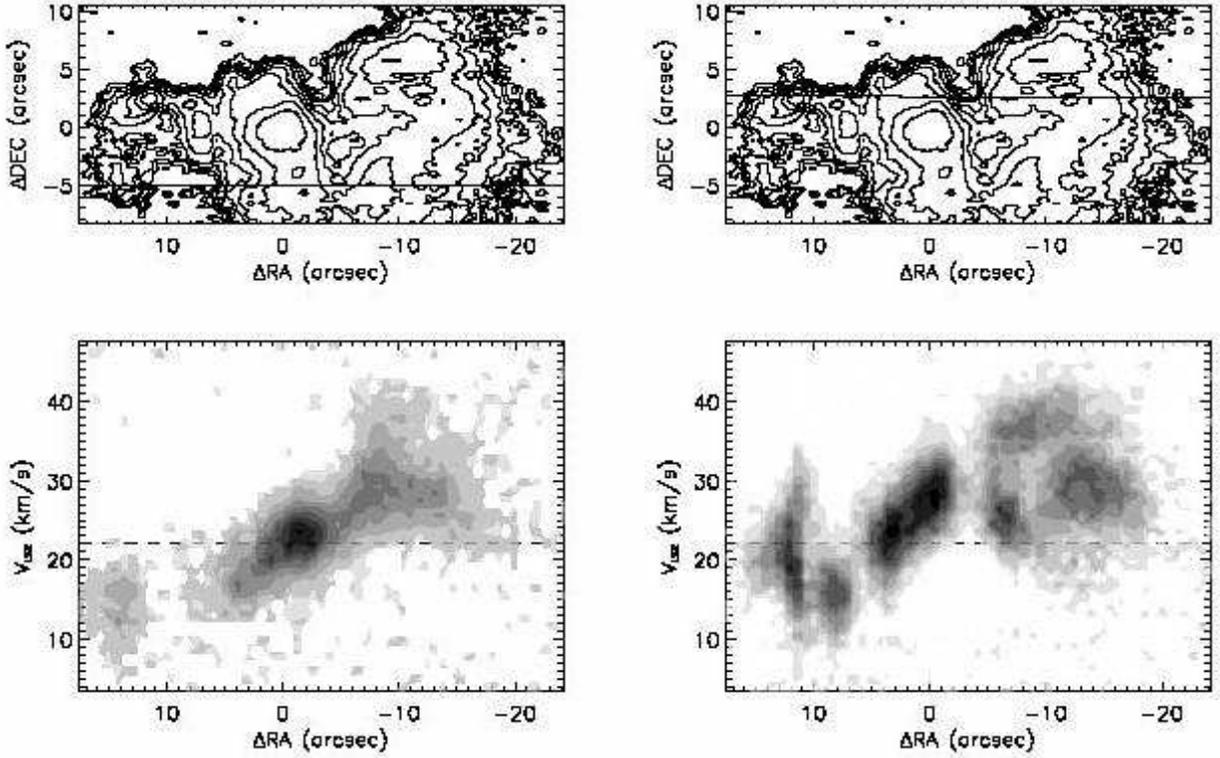} \caption{Position-velocity diagrams
of G61.48~+0.09 [Ne~II] line observations. Contours in line flux
maps are drawn at $70\%$, $50\%$, $35\%$, $25\%$, $17.5\%$,
$12.5\%$, $9\%$ and $6\%$ of peak values
(0.052~ergs~cm$^{-2}$~s$^{-1}$~sr$^{-1}$) in the maps. The spatial
resolution of the map is $\sim 4 ''$. Dashed lines in p-v diagrams
show the ambient molecular material velocity. \label{g6148pv}}
\end{figure}

The p-v diagrams (Figure~\ref{g6148pv}) show that the gas observable
in the [Ne~II] line has a steep velocity gradient. The [Ne~II]
emission near B2 is blueshifted by a few km~s$^{-1}$ with respect to
the neutral cloud velocity. Over much of B1, the bulk of the neon
emission arises in redshifted material with the westernmost gas
ranging from 0-20 km~s$^{-1}$ to the red of the molecular line
center. The morphological differences between the radio and [Ne~II]
maps make it difficult to associate the kinematic and morphological
structures. The RRL observations of \cite{garLG94}, which are not
affected by extinction, show that the bulk of the ionized material
is redshifted with respect to the molecular cloud.

\subsection{K3-50A (G70.3~+1.6)}
\begin{figure}
\epsscale{1.0} \plotone{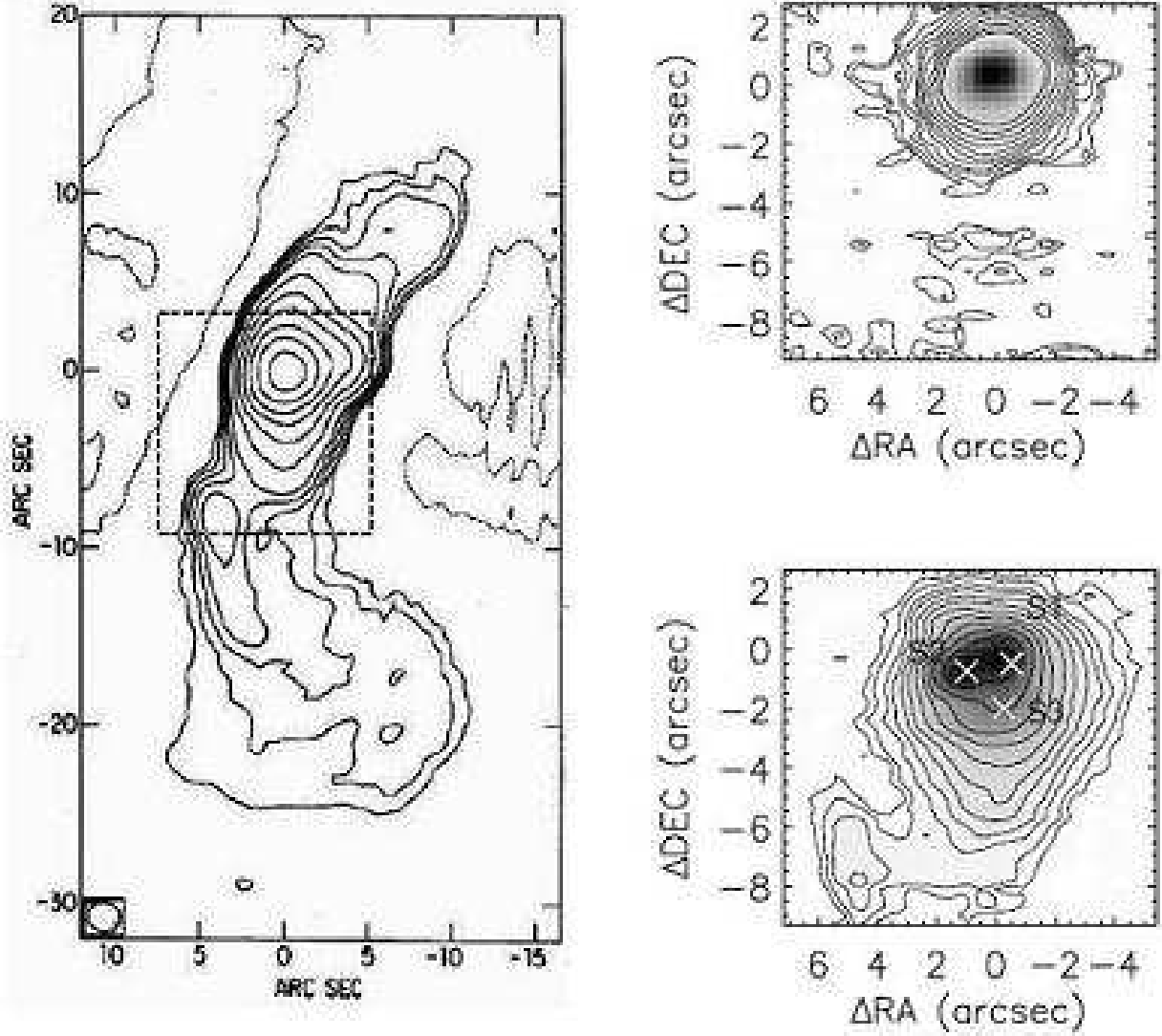} \caption{The 2.0~cm continuum map
of K3-50A (left) \citep[][resolution$\sim 1.''3$]{depGPR94}. The
(0,0) position marks the peak continuum emission. First positive and
negative contours were drawn at the 3$\sigma$ flux level. Other
contours were drawn at 1.4, 2, 2.8, 4, 8, 16, 32, 64, 128 and 256
times the 3$\sigma$ flux level. The box drawn with dashed lines
indicates the approximate area covered by our [Ne~II] observations.
The 12.8 $\mu$m dust continuum map (top right) and integrated
[Ne~II] line flux map (bottom right). For both maps, the (0,0)
position is the location of the [Ne~II] emission peak.  Three
crosses show the approximate positions of S1, S2, and S3 which are
[Ne~II] emission concentrations identified in the channel maps
(Figure~\ref{k350chanmapneii}). Contours are drawn at $70\%$,
$50\%$, $35\%$, $25\%$, $17.5\%$, $12.5\%$, $9\%$, $6\%$, $4\%$,
$2\%$ and $1\%$ of the peak values in both maps. The peak dust
continuum flux density is
0.82~ergs~cm$^{-2}$~s$^{-1}$~sr$^{-1}$~(cm$^{-1}$)$^{-1}$. The peak
[Ne~II] surface brightness is
0.34~ergs~cm$^{-2}$~s$^{-1}$~sr$^{-1}$. The total [Ne~II] flux is
1.0$\times$10$^{-10}$~ergs~cm$^{-2}$~s$^{-1}$. \label{k350map}}
\end{figure}

K3-50 is a four-component \citep[A-D,][]{wyn69} HII region complex
at a distance of 8.7~kpc \citep{har75}. The kinematic distance was
computed using the galactic rotational curve of \cite{sch65} with
the assumed Sun-to-GC distance of 10~kpc and a 250~km~s$^{-1}$
circular velocity at the position of the Sun \citep{rubT69,rub65}.
More recent estimates of the distance to the Galactic Center
($sim$7.94~kpc) imply that this distance is too large
\citep{eisSGOT03}.
%radial velocity $\sim$-22~km~s$^{-1}$.
The UCHII region K3-50A appears to
be the youngest and dominates the emission at infrared wavelengths.
\cite{hofBPW04} presented K$'$ band bispectrum speckle
interferometric data of K3-50A and their observations suggested that
K3-50A is excited by a small cluster of massive to intermediate-mass
stars. Radio continuum and hydrogen recombination line observations
found that the ionized gas in K3-50A was undergoing a high-velocity
bipolar outflow \citep{depGPR94}. The left panel of
Figure~\ref{k350map} shows the radio continuum image
\citep{depGPR94}. There is extended emission both north and south of
the dashed box that shows the extent of the [Ne~II] map.

The [Ne~II] line emission in K3-50A peaks at the radio peak and has
a remarkably similar distribution to that of the radio continuum
within the dashed box (Figure~\ref{k350map}). This is consistent
with its relatively low extinction \citep[$A_K=1.6$,][]{marPMTCR02}.
Most of the line emission comes from the core of the region. The
integrated line map shows that the core of K3-50A consists of two
sources, source 1 (S1) at the line emission peak and source 2 (S2)
$\sim 1.5''$ southeast of S1. Line emission extends north and south
from the central region. In the 12.8~$\mu$m continuum
(Figure~\ref{k350map}), a more compact emission peak is slightly
offset ($\sim$1$''$) to the north of S1.

\begin{figure}
\epsscale{1.0} \plotone{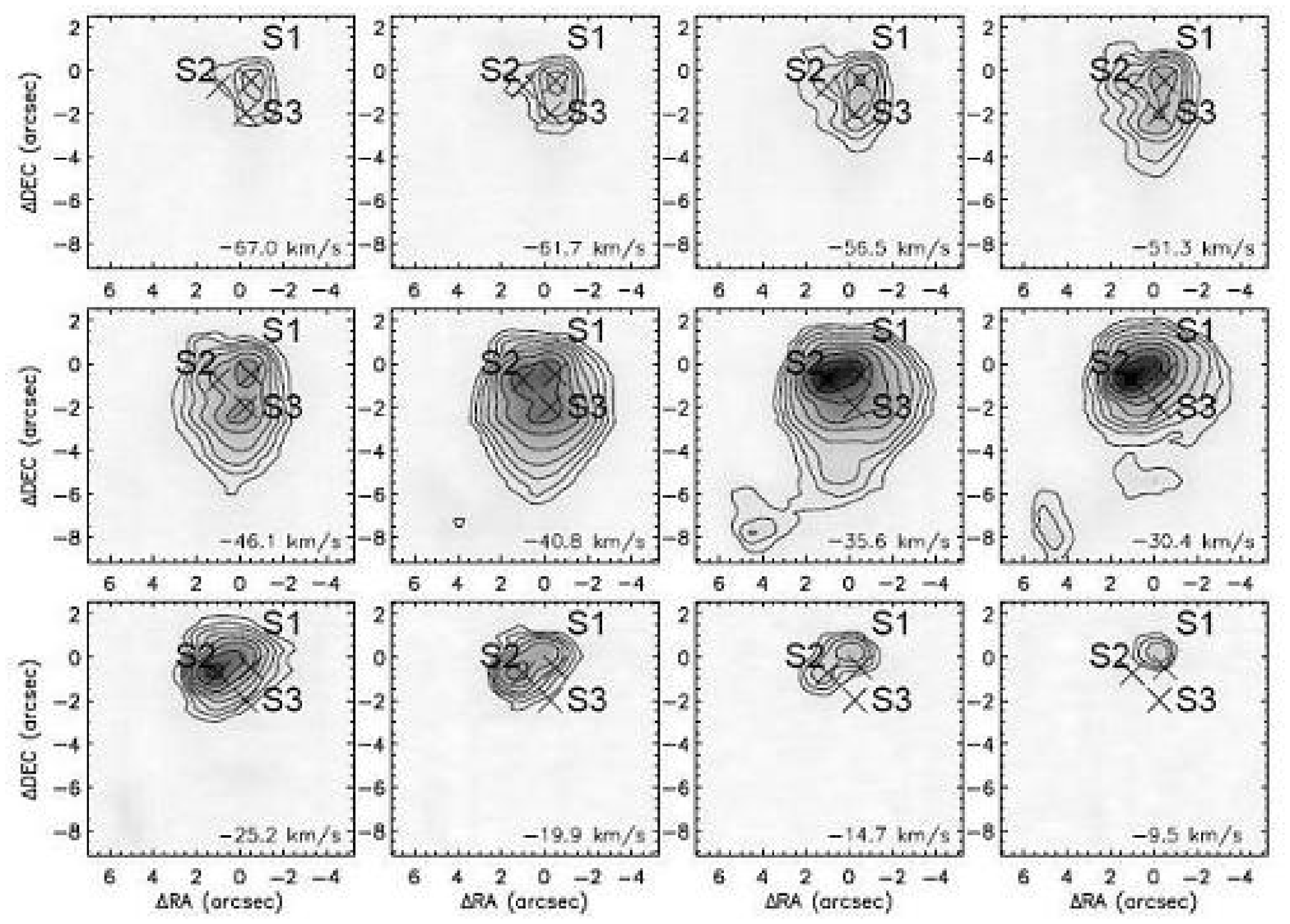} \caption{Channel maps of K3-50A
[Ne~II] line observations. Contours are drawn at $70\%$, $50\%$,
$35\%$, $25\%$, $17.5\%$, $12.5\%$, $9\%$ and $6\%$ of the peak
value of all channels. The peak value is
5.1~ergs~cm$^{-2}$~s$^{-1}$~sr$^{-1}$~(cm$^{-1}$)$^{-1}$. The
molecular cloud velocity is V$_{LSR}$ = -24 km s$^{-1}$.
\label{k350chanmapneii}}
\end{figure}

\begin{figure}
\epsscale{0.8} \plotone{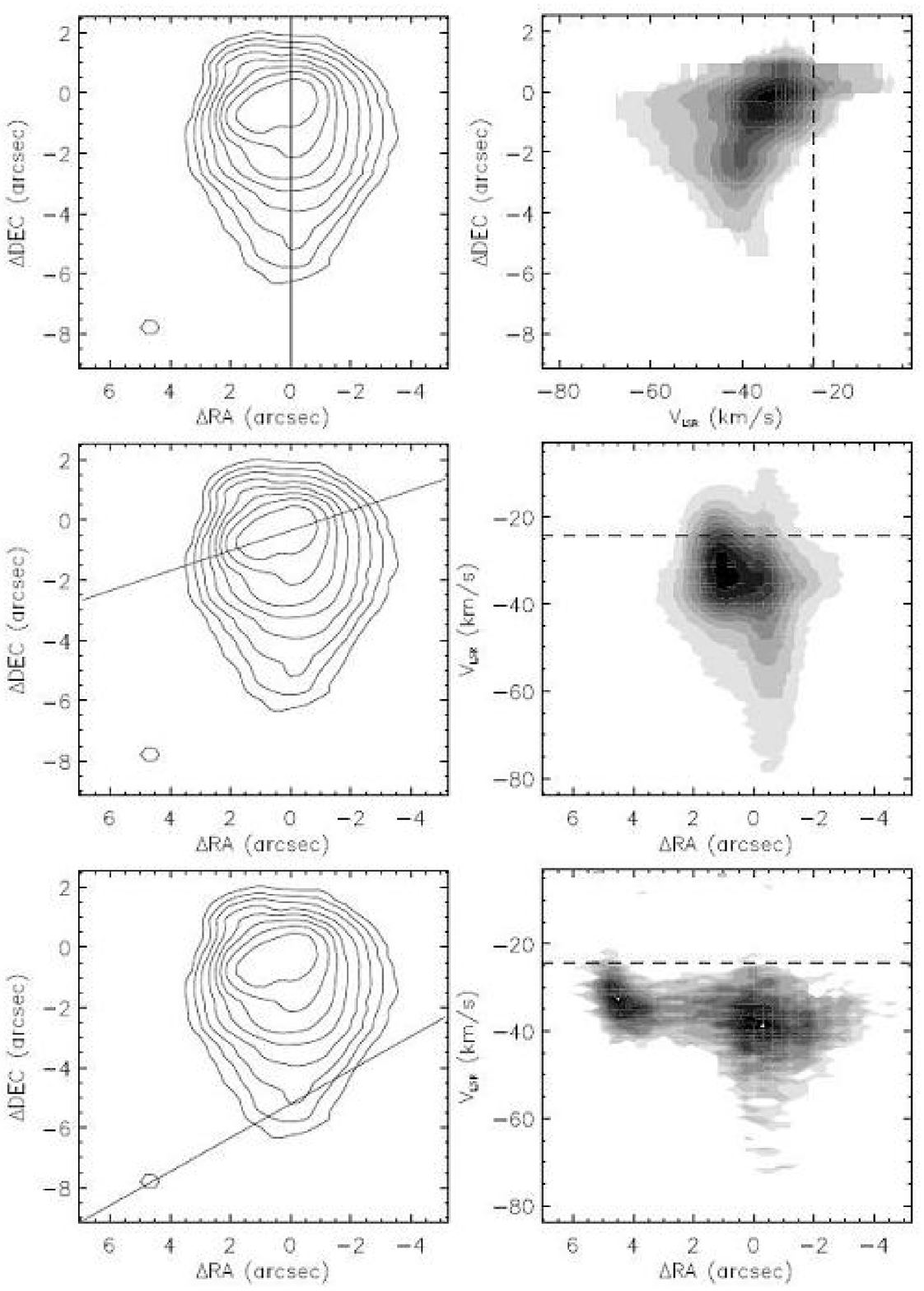} \caption{Position-velocity diagrams
of K3-50A [Ne~II] line observations. Contours in line flux maps are
drawn at $70\%$, $50\%$, $35\%$, $25\%$, $17.5\%$, $12.5\%$, $9\%$
and $6\%$ of peak values (0.34~ergs~cm$^{-2}$~s$^{-1}$~sr$^{-1}$) of
the maps. Dashed lines in p-v diagrams show the ambient molecular
material velocity. \label{k350pv}}
\end{figure}

The [Ne~II] emission has a very broad velocity range, from
-67~km~s$^{-1}$ to -8~km~s$^{-1}$ (Figure~\ref{k350chanmapneii}).
S1, at (0,0), is present in all velocity channels. S2 is bright in
the channels at the red end of the spectrum. A third component (S3)
$\sim2''$ to the south of S1 is present in the channels at central
and blue-shifted velocities. All three components lie significantly
to the blue of the molecular cloud which has V$_{LSR}$ =
-24~km~s$^{-1}$ \citep{sheC96}. S3 has the biggest velocity shift
with a line center at $\sim$-39~km~s$^{-1}$, with S1 and S2 at
$\sim$-35~km~s$^{-1}$. Position-velocity diagrams of K3-50A show the
velocity differences between different sources
(Figure~\ref{k350pv}). The position-velocity diagram for a
north-south cut (top panel of Figure~\ref{k350pv}) shows a velocity
gradient while the region right around S1 shows evidence for a
barely resolved bipolar structure with a total velocity extent of 60
km s$^{-1}$ (middle panel of Figure~\ref{k350pv}). Because the
region is very extended north-south, line emission from the region
fills our slit from end to end and emission from the northern and
southern lobes are very faint. The emission indicated by the last
four contours in Figure~\ref{k350map} is at least 100 times fainter
than emission at the peak. Our map did not cover enough area to show
the two more extended emission lobes.

\subsection{S106}
\begin{figure}
\epsscale{1.0} \plotone{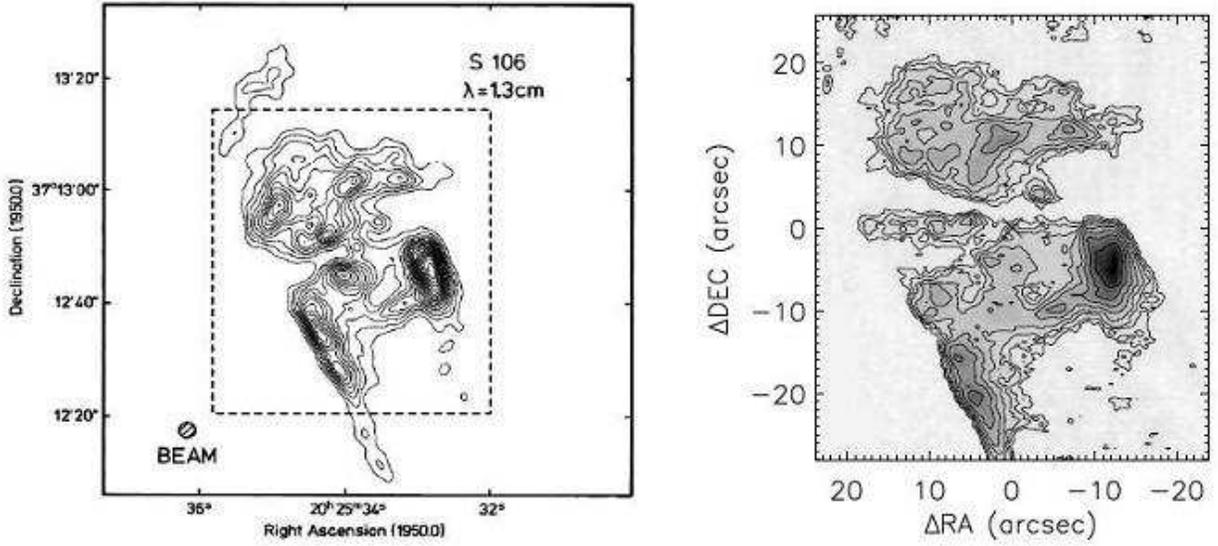} \caption{1.3~cm continuum map
\citep[left,][resolution$\sim 2.''3$]{felMSRE84} and integrated
[Ne~II] line flux map (right) of S106. Contours in the continuum map
were drawn at each $5\%$ step from $5\%$ to $95\%$ of the peak
value. The box drawn with dashed lines indicates the approximate
area covered by the [Ne~II] observations. Contours in the [Ne~II]
line map are drawn at $70\%$, $50\%$, $35\%$, $25\%$, $17.5\%$,
$12.5\%$, $9\%$ and $6\%$ of the peak line emission. The spatial
resolution of the radio map is $\sim 2.''3$. The (0,0) position is
the location of a weak continuum source seen in the line
observations. The peak [Ne~II] surface brightness is
0.23~ergs~cm$^{-2}$~s$^{-1}$~sr$^{-1}$. The total [Ne~II] flux is
1.1$\times$10$^{-9}$~ergs~cm$^{-2}$~s$^{-1}$.
\label{s106mapradioneii}}
\end{figure}

S106 (G76.4~-0.6) is a well-known bipolar HII region consisting of
two lobes with the symmetry axis oriented at a position angle of
$\sim$30$^{\circ}$.
An equatorial gap is seen in all wavelengths from optical to radio.
Earlier investigators suggest that a dense, circumstellar disk surrounds a
star and blocks the ionizing photons in the disk plane
forming a biconical nebula \citep{balS82}.
A point source, presumably the excitation source of the nebula,
is located at the center of the equatorial gap
\citep{pipSKKSM76,felMSRE84}.
\cite{felMSRE84} noticed that the bright radio continuum peaks in
S106 tended to be distributed at the edges of the radio lobes and
inferred that the inner volume was filled with gas with lower density.
A radio continuum map from their observations is shown in
Figure~\ref{s106mapradioneii}.
Dense molecular clumps were identified on both sides of the central
source of the HII region \citep{schSKSB02} but there is
no evidence of a smooth molecular disk.
\cite{smiJGK01} imaged the region at 3 to 20~$\mu$m.
Some of their images (11.7~$\mu$m) show two dark lanes.
One crosses the center of the region from the east to the west and
divides the region into the north and south lobes.
The other originates east of the region,
crosses the south lobe and forms a small angle ($\sim$30$^{\circ}$)
with the first dark lane.

The integrated [Ne~II] line map and the radio continuum map are
stunningly similar (Figure~\ref{s106mapradioneii}).
The bipolar morphology and individual emission concentrations are
clearly seen in both maps.
The south lobe is characterized by
a few bright clumps along the edge, while the north lobe contains
more diffuse emission.
The axis of the bipolar structure, which is indicated by the sharp edges of
the south lobe, runs NNE-SSW.
The main difference between the radio and [Ne~II] maps is that
the second dark lane seen in the infrared continuum is also seen in
[Ne~II].

\begin{figure}
\epsscale{0.8} \plotone{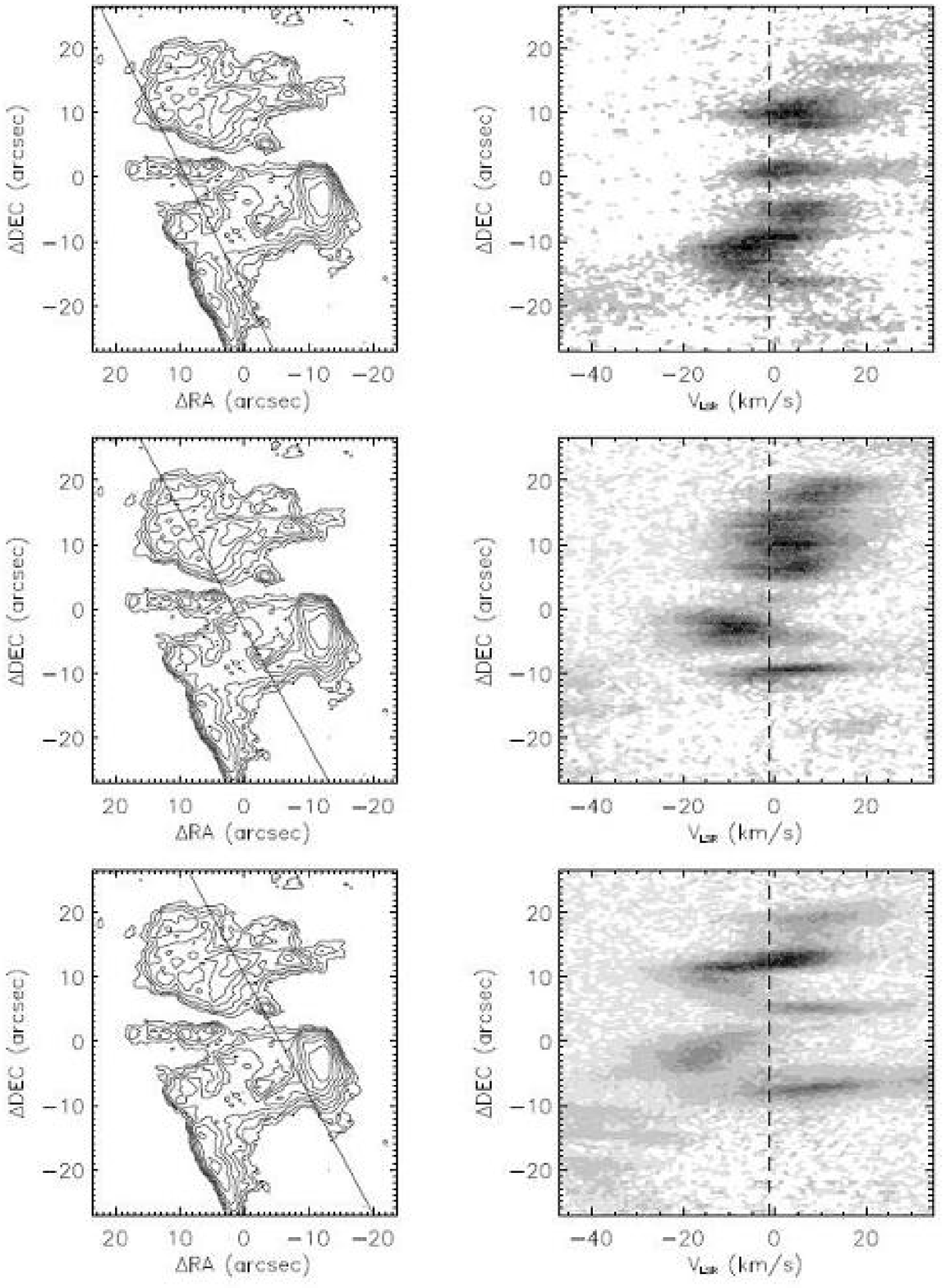} \caption{Position-velocity diagrams
of S106 [Ne~II] line observations. Contours in [Ne~II] line flux
maps are drawn at $70\%$, $50\%$, $35\%$, $25\%$, $17.5\%$,
$12.5\%$, $9\%$ and $6\%$ of peak values
(0.23~ergs~cm$^{-2}$~s$^{-1}$~sr$^{-1}$) in the maps. Dashed lines
in p-v diagrams show the ambient molecular material velocity.
\label{s106pv}}
\end{figure}

\begin{figure}
\epsscale{1.0} \plotone{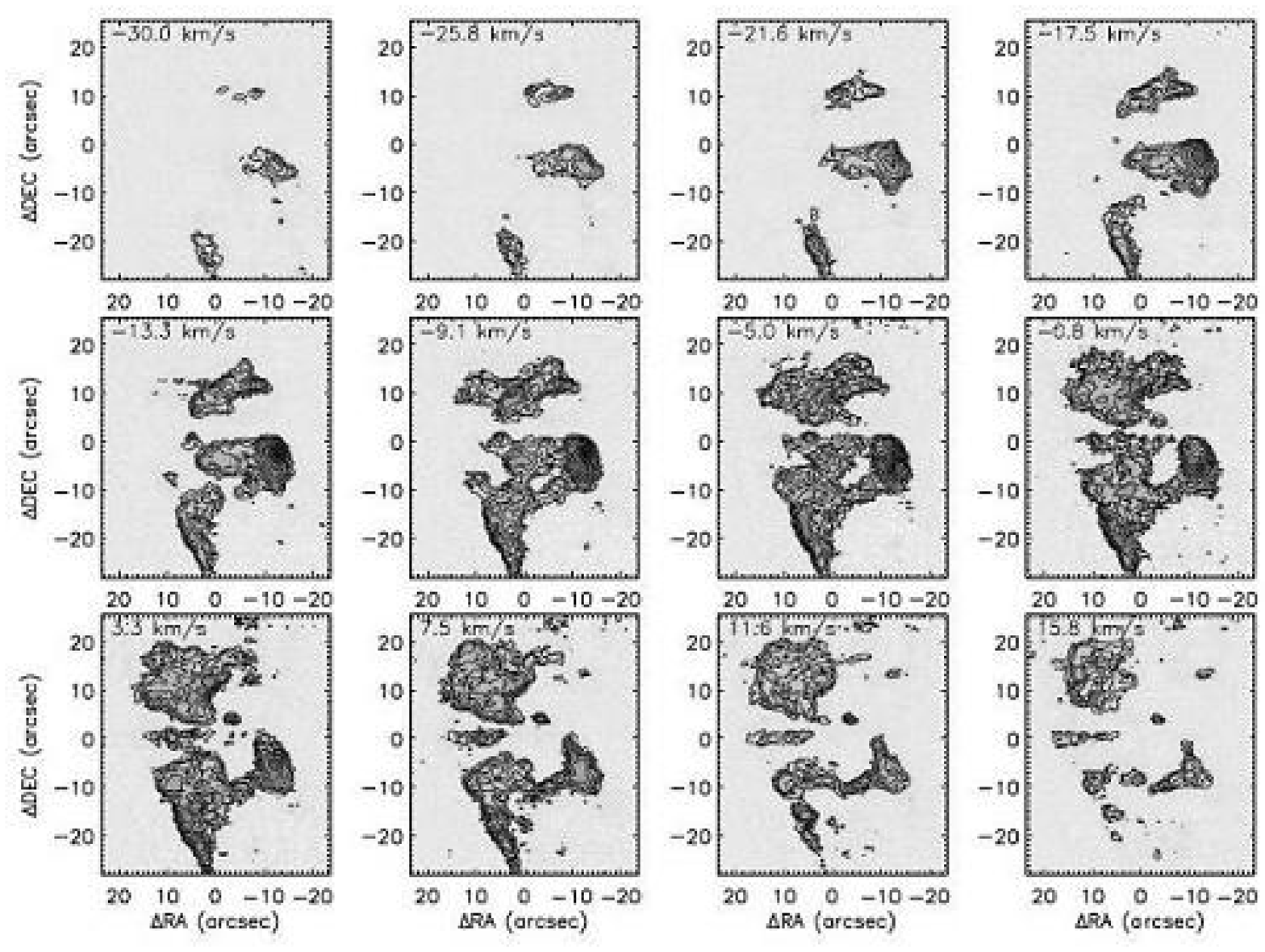} \caption{Channel maps of S106
[Ne~II] line observations. Contours are drawn at $70\%$, $50\%$,
$35\%$, $25\%$, $17.5\%$, $12.5\%$, $9\%$ and $6\%$ of the peak
value of all channels. The peak value is
3.6~ergs~cm$^{-2}$~s$^{-1}$~sr$^{-1}$~(cm$^{-1}$)$^{-1}$. The
molecular cloud velocity is V$_{LSR}$ = -1 km s$^{-1}$.
\label{s106chanmapneii}}
\end{figure}

The velocity structure shown in the p-v diagrams
(Figure~\ref{s106pv}) and the channel maps
(Figure~\ref{s106chanmapneii}) is consistent with a picture of S106
as a nearly edge-on bipolar nebula. Both figures indicate that the
south lobe is slightly blue shifted ($\sim$3 km s$^{-1}$) with
respect to the molecular cloud velocity while the material in the
northern lobe is on-average red-shifted by $\sim$4 km s$^{-1}$. At
the same time, individual intensity peaks show moderately narrow
($\sim$10 km s$^{-1}$ lines. Projection effects may lead to velocity
shifts from one emission peak to its neighbors.

\subsection{NGC7538A}
\begin{figure}
\epsscale{1.0} \plotone{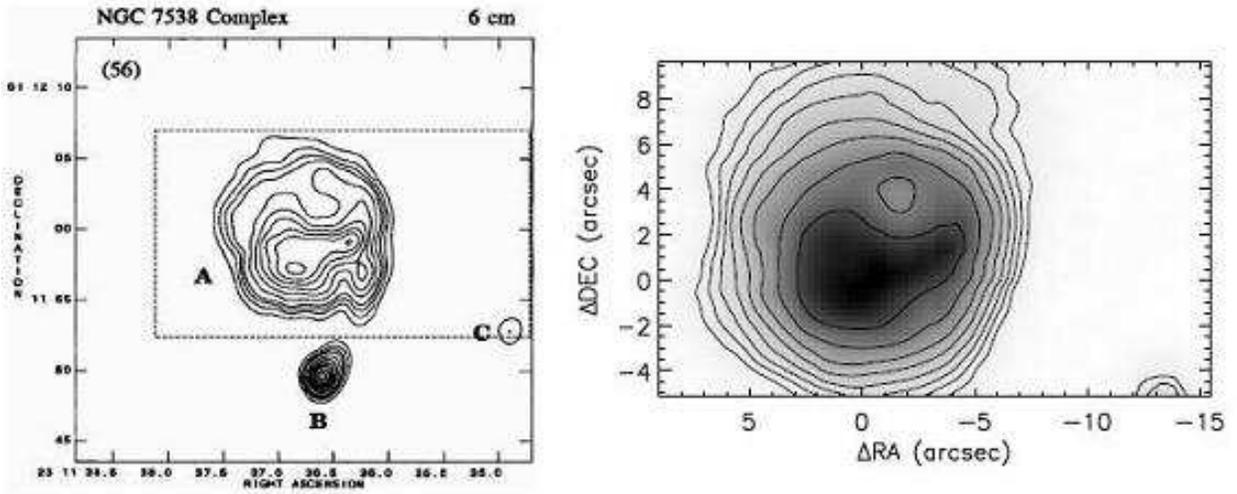} \caption{6~cm continuum map
\citep[left,][restoring beam: $0.''38 \times 0.''38$]{wooC89b} and
integrated [Ne~II] line flux map (right) of NGC7538A. Contours in
the continuum map were drawn at $95\%$, $90\%$, $80\%$, $70\%$,
$60\%$, $50\%$, $40\%$ ,$30\%$, $20\%$, $15\%$, $10\%$, $5\%$,
$-5\%$, $-10\%$, $-15\%$ and $-20\%$ of the peak value. The box
drawn with dashed lines indicates approximately the area covered by
the [Ne~II] observations. Contours in the [Ne~II] line map are drawn
at $70\%$, $50\%$, $35\%$, $25\%$, $17.5\%$, $12.5\%$, $9\%$ and
$6\%$ of the peak value in the map. The radio map has a synthesized
beam with FWHM$\sim 0.''38 \times 0.''38$. The (0,0) position is the
location of the [Ne~II] line emission peak. The peak [Ne~II] surface
brightness is 0.16~ergs~cm$^{-2}$~s$^{-1}$~sr$^{-1}$. The total
[Ne~II] flux is 2.7$\times$10$^{-10}$~ergs~cm$^{-2}$~s$^{-1}$.
\label{n7538map}}
\end{figure}

NGC7538A belongs to an HII region-molecular cloud complex at a
distance of ${\sim}3.5$~kpc \citep{hanLR02}. In the 6 cm radio
continuum image, it is a compact, spherical HII region with a radius
$\sim5''$ \citep{wooC89b} (Figure~\ref{n7538map}, left),
corresponding to $\sim$19,000~AU. Molecular line observations show
that molecular gas toward NGC7538 has a LSR velocity
$-56.9$~km~s$^{-1}$ \citep{dicDW81}. \cite{bloWPFAG98} suggested
that NGC7538A is formed by a stellar wind bow shock with a star
moving from the northwest to the southeast away from the Earth into
the molecular cloud.

\begin{figure}
\epsscale{1.0} \plotone{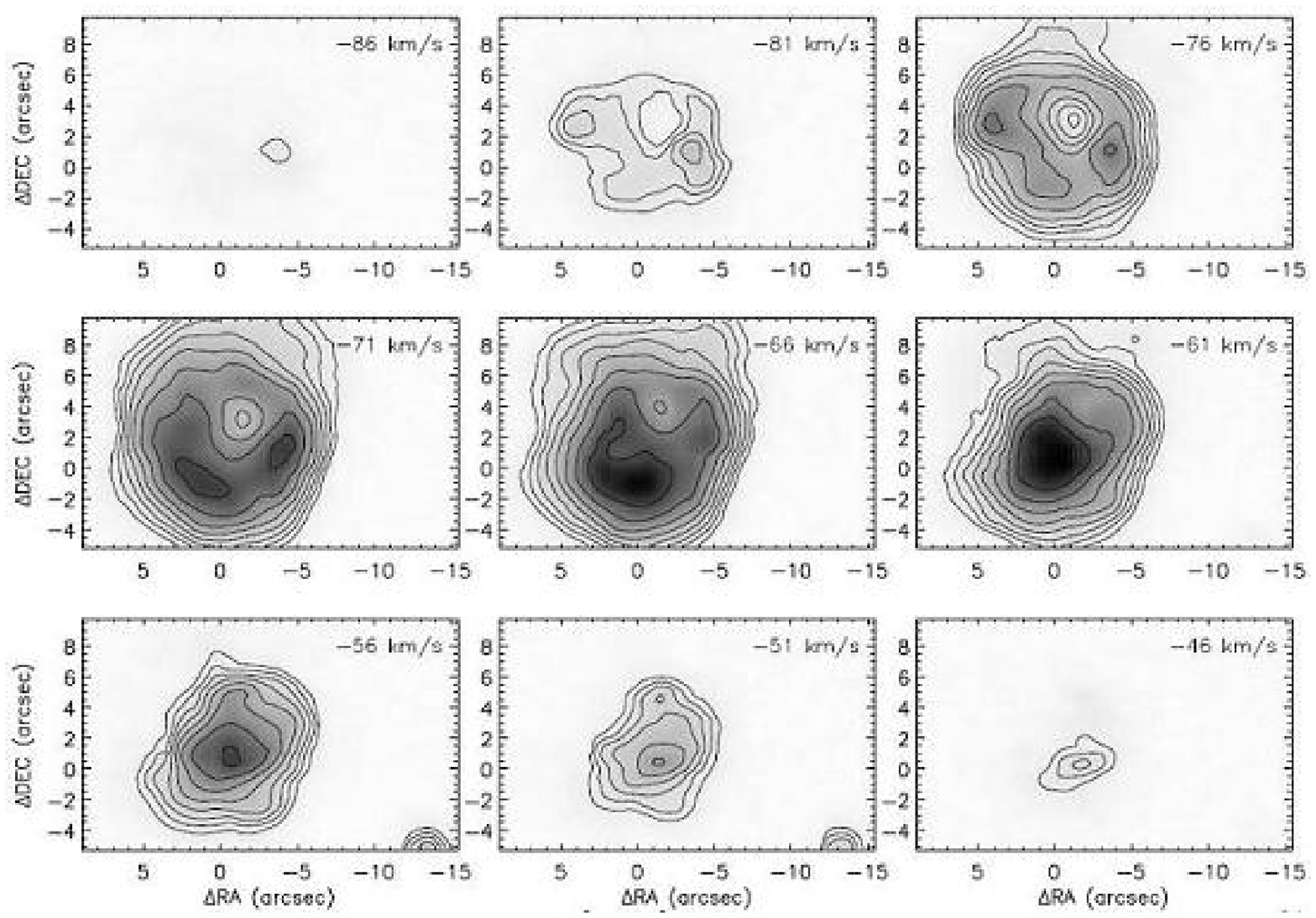} \caption{Channel maps of N7538~A
[Ne~II] line observations. Contours are drawn at $70\%$, $50\%$,
$35\%$, $25\%$, $17.5\%$, $12.5\%$, $9\%$ and $6\%$ of the peak
value of all channels. The peak value is
3.1~ergs~cm$^{-2}$~s$^{-1}$~sr$^{-1}$~(cm$^{-1}$)$^{-1}$. The
molecular cloud velocity is V$_{LSR}$ = -57 km s$^{-1}$.
\label{n7538chanmapneii}}
\end{figure}

\begin{figure}
\epsscale{0.7} \plotone{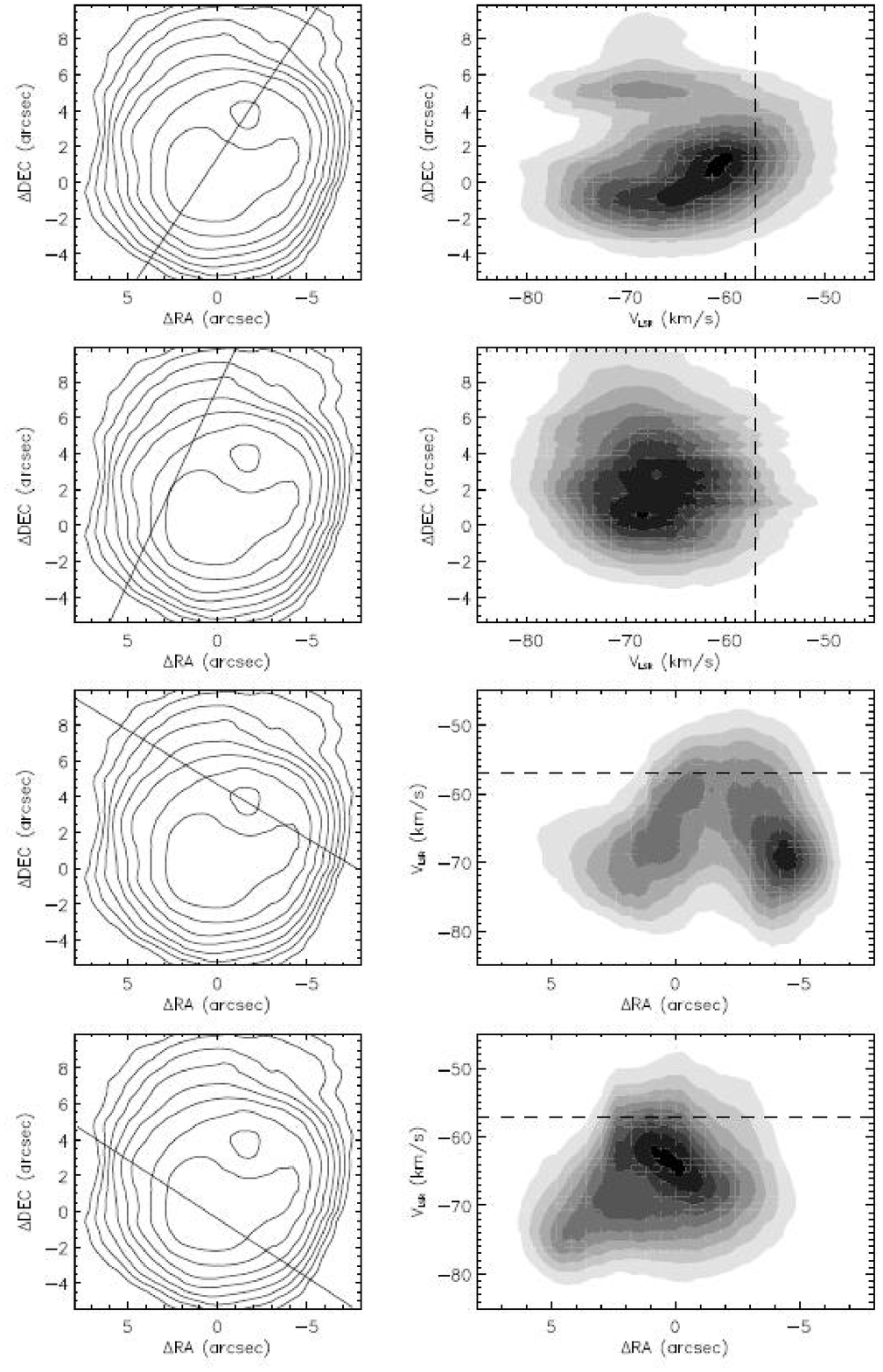} \caption{Position-velocity diagrams
of NGC7538~A [Ne~II] line observations. Contours in line flux maps
are drawn at $70\%$, $50\%$, $35\%$, $25\%$, $17.5\%$, $12.5\%$,
$9\%$ and $6\%$ of the peak values
(0.16~ergs~cm$^{-2}$~s$^{-1}$~sr$^{-1}$) of the maps. Dashed lines
in p-v diagrams show the ambient molecular material velocity.
\label{n7538apv}}
\end{figure}

Since dust extinction toward the region is not huge
\citep[$A_K\sim1.3$,][]{werBGMNW79,herHFMHWP81,camT84}, our [Ne~II]
line map of NGC7538A (Figure~\ref{n7538map}, right) resembles the
radio continuum map closely except that the radio component which
should be at the southwestern corner of the source A, $\sim$(-5,-2),
is not present in our [Ne~II] map. This is probably due to our poor
spatial resolution. [Ne~II] line channel maps, spanning a velocity
range from -82 to -51~km~s$^{-1}$, are shown in
Figure~\ref{n7538chanmapneii}. A cavity can be seen in the most
blue-shifted channels and in the integrated line map. The shape of
the region changes from ring-like at the blue-shifted channels to
core-like at the red-shifted channels. The channel maps in
Figure~\ref{n7538chanmapneii} show that most of the ionized gas in
NGC7538A is blue-shifted with respect to the molecular cloud. The
limb-brightened appearance of the source in the integrated line
strength and radio continuum maps (Figure~\ref{n7538map}) might lead
to the conclusion that the source is a limb-brightened spherical
shell. However, the pattern of a small central source at the
molecular cloud velocity and a larger, shell-like structure at bluer
velocities, combined with the ``$\Lambda$'' and ``7'' structures
seen in the pv cuts through the source (Figure~\ref{n7538apv}) argue
against this picture. The kinematics and morphology are more
consistent with a tangential flow along a shell with the symmetry
axis close to the line of sight and with the vertex pointing away
from the observer.

\subsection{W3A+B}
W3 belongs to an extensive HII region-molecular cloud complex
located in the Perseus spiral arm \citep[D = 2.0~kpc,][]{hacBMRI06}.
Aperture synthesis observations identified four compact/ultracompact
radio continuum sources in a 3 arc minute diameter region
\cite[A-D,][]{wyn71}. Infrared imaging at 1.65 to 20$\mu$m revealed
nine objects (IRS~1-9) in W3 \citep{wynBN72}, four of which are
associated with the compact radio continuum sources: IRS1 and IRS2
with the circular source W3A; IRS3 with W3B and IRS4 with W3C.
Earlier authors have used the ``broken-shell'' morphology and the
small velocity gradients seen in hydrogen recombination lines toward
W3A to argue for a picture of this region as a blister flow
\citep{dicHG83,tieGCWJ97}. W3B has also been characterized as an
emerging blister flow \citep{tieGCWJ97}.

\begin{figure}
\epsscale{1.0} \plotone{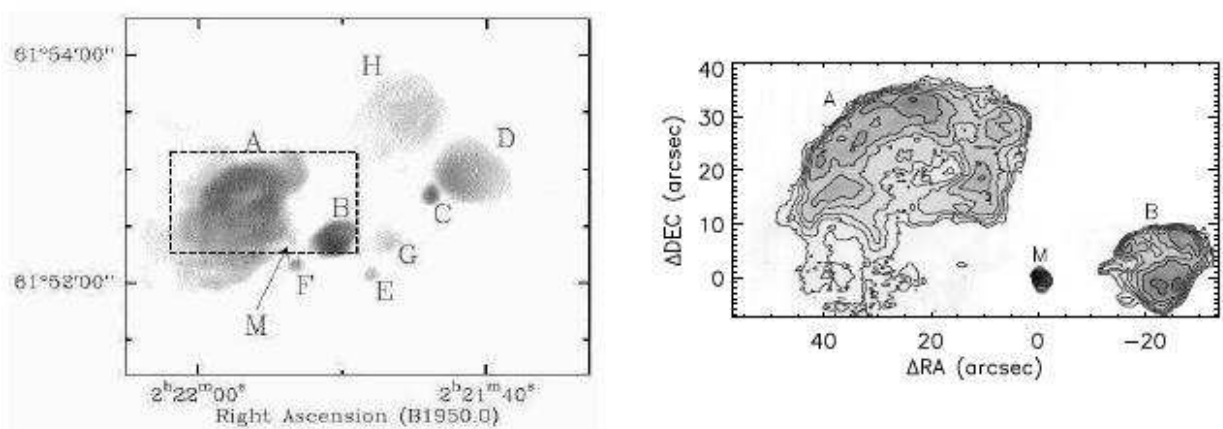} \caption{Grey-scale map of the
radio 6 cm free-free continuum of W3 core (left) \citep{tieGCWJ97}.
The observations have a FWHM beam with a size of $0.''71 \times
0.''56$. The flux in the region ranges from 0.1~mJy to
40~mJy~beam$^{-1}$. The box drawn with dashed lines indicates the
approximate area covered by our [Ne~II] observations. Integrated
[Ne~II] line flux map of W3 A and B region (right). The (0,0)
position is the peak location of integrated emission map. Contours
are drawn at $70\%$, $50\%$, $35\%$, $25\%$, $17.5\%$, $12.5\%$,
$9\%$ and $6\%$ of the peak value. The peak [Ne~II] surface
brightness is 0.26~ergs~cm$^{-2}$~s$^{-1}$~sr$^{-1}$. The total
[Ne~II] flux from W3A is
1.3$\times$10$^{-9}$~ergs~cm$^{-2}$~s$^{-1}$. The total [Ne~II] flux
from W3B is 3.7$\times$10$^{-10}$~ergs~cm$^{-2}$~s$^{-1}$.
\label{w3map}}
\end{figure}

\begin{figure}
\epsscale{1.0} \plotone{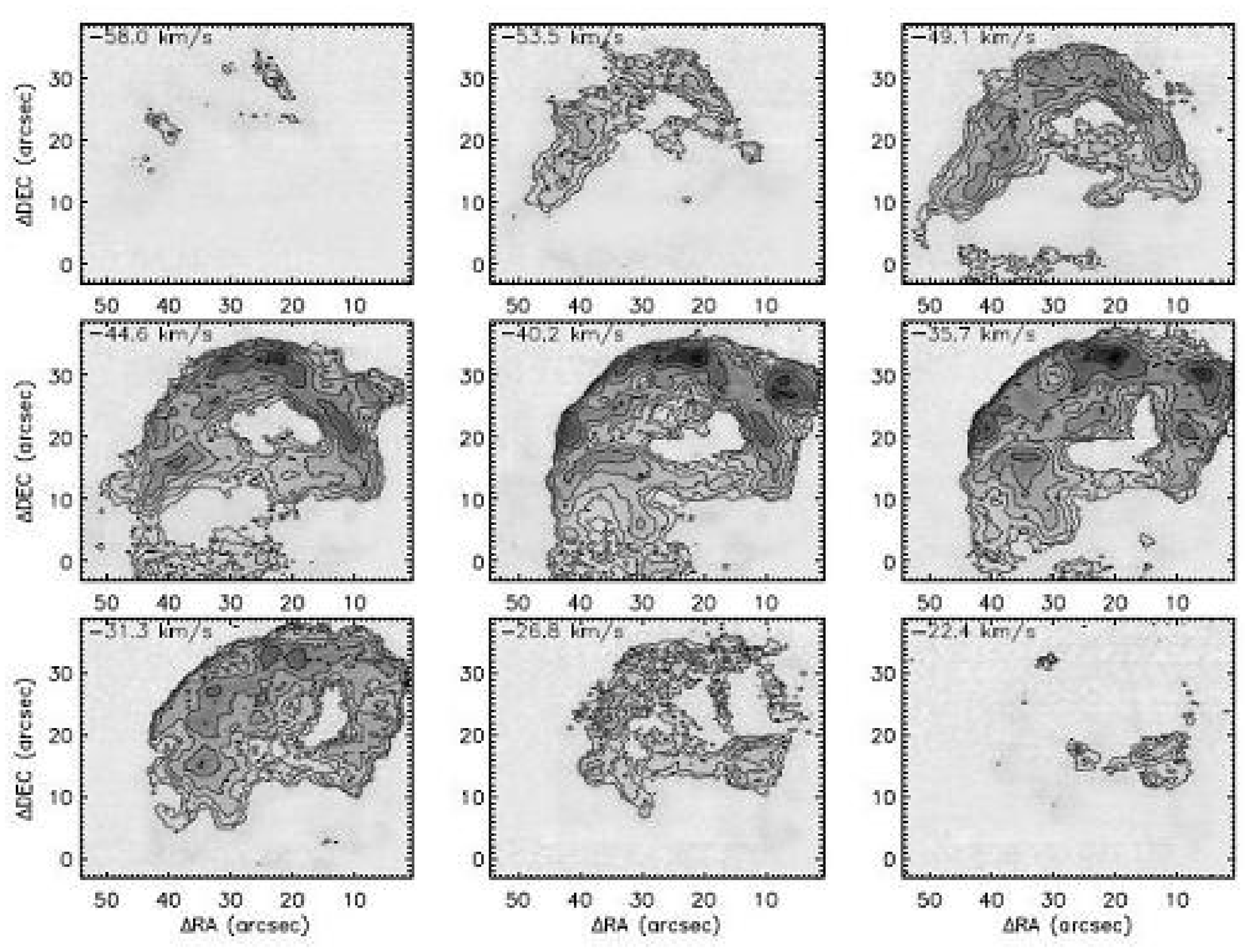} \caption{Channel maps of W3~A
[Ne~II] line observations. Contours are drawn at $70\%$, $50\%$,
$35\%$, $25\%$, $17.5\%$, $12.5\%$, $9\%$ and $6\%$ of the peak
value of all channels. The peak value is
2.8~ergs~cm$^{-2}$~s$^{-1}$~sr$^{-1}$~(cm$^{-1}$)$^{-1}$. The
molecular cloud velocity is V$_{LSR}$ = -40 km s$^{-1}$.
\label{w3achanmapneii}}
\end{figure}

W3A has the morphology of a limb-brightened broken shell in our
[Ne~II] map (Figure~\ref{w3map}), in close agreement with the
distribution of the radio continuum emission. Most of the line
emission forms a crescent open to the south with a horizontal
structure crossing the center of the incomplete shell. An irregular
emission cavity is present inside the crescent. Two finger-like
structures extend from the southeast corner of the ring to the
south. If we ignore these finger-like structures and another small
lobe of emission on the northwest edge of the shell, the region
closely resembles a half-ring. Such an image becomes clear in the
channel maps of the region (Figure~\ref{w3achanmapneii}). The
half-ring morphology is present in channels over a broad velocity
range. The sharp outer edge of the ring is evident. The emission
level drops rapidly to the background level within a few arcseconds.

The shell-like morphology of W3A might lead one to expect the source
to have the kinematics of an expanding shell. However, p-v diagrams
(Figure~\ref{w3apv}) of the region do not support this picture. We
do not see the two components toward the center of the source which,
in an expanding shell, would arise from the front and back sides of
the shell and would gradually merge into one component at an
intermediate velocity at the edge of the shell. Instead, we see a
very broad line ($\geq$25 km~s$^{-1}$) along the bright arc-like rim
of the region becoming narrow blue-shifted and red-shifted lines
toward the center of the region. The change of the line width and
the center velocities of the lines are abrupt. A surface flow,
rather than an expanding shell, can better explain the observed p-v
structure. The fact that both p-v cuts show the same pattern of a
broad feature at the bright limb followed by a double line within
the cavity implies that the vertex of the flow lies between the two
cuts. An examination of the -53.5 km s$^{-1}$ and -26.8 km s$^{-1}$
channel maps in Figure~\ref{w3achanmapneii} supports this picture.
The emission in the southern part of W3A comes from a narrower
velocity range and therefore is not part of the surface flow seen in
the north.

The compact HII regions W3B and W3A have similar central velocities
and line widths. The overall [Ne~II] line emission distribution in
W3B (see left panels in Figure~\ref{w3bpv}) looks like an irregular
shell. The line profile at most positions in the region can be fit
by a single Gaussian. W3B contains a low surface brightness cavity,
which can be seen in both the integrated line map and the channel
maps (Figure~\ref{w3bchanmapneii}) of the region. The total shift in
velocity centroid in W3B is relatively modest, from $\sim$-42 km
s$^{-1}$ to -34 km s$^{-1}$ and there is an apparently random
element to the velocity variations at the 2-3 km s$^{-1}$ level.
Nonetheless, both the channel maps (Figure~\ref{w3bchanmapneii}) and
the p-v cuts (Figure~\ref{w3bpv}) are consistent with tangential
flow along the ionized walls of a cavity. The channel maps show
emission from the central region primarily at the least negative
velocities, while the shell is most prominent in the bluer channels.
The p-v diagrams also show the reddest velocities at positions
crossing the center of the cavity and the bluest velocities along
the edge.

\section{Discussion}
\label{secdiscussion}

We have presented high spatial and spectral resolution mapping of
the 12.8 $\mu$m [Ne~II] line toward a sample of 18 compact and
ultracompact HII regions, where we count distinct HII regions within
clusters as separate sources. In our discussion, we make use of
previously published results for two of the 18 regions: G29.96-0.02
\citep{zhuLJRG05} and W51 IRS2 \citep{lacJZRBG07}. We also include
in the discussion one source \citep[Mon~R2,][]{jafZLR03}, for which
no results are presented in the current paper, bringing our sample
to 19 sources.

In order to discuss possible interpretations of the observed source
morphologies and kinematics, we first sketch p-v diagrams of several
types of models that have been suggested for UCHII regions. We then
compare the observed morphologies and kinematics to these sketches.
We emphasize that the sketches are only meant to be qualitative
descriptions of generic models. They are not derived from
hydrodynamic simulations.

\subsection{Qualitative Models}

To facilitate the discussion of the observed gas motions, we first
consider qualitative kinematic models of UCHII regions.
Sketches of several types of models of UCHII regions are shown in
Figures~\ref{ssspv}-\ref{sshpv}.

A. (Figure~\ref{ssspv})  An uniformly expanding shell source appears
as a limb-brightened disk. Its p-v diagram is ring-like, centered on
the molecular cloud velocity, V$_{amb}$.

B. (Figure~\ref{ssspv})  An expanding filled sphere appears as a
limb-darkened disk. Its p-v diagram is a filled ellipse centered at
V$_{amb}$.

C. (Figure~\ref{sbspv})  The appearance of a bow-shock source, in
which a wind from a star moving supersonically through a molecular
cloud sweeps up a paraboloidal shell, with the motion of the gas in
the shell resulting from the combined momenta of the stellar wind
and the swept-up material \citep{macVW91,vanM92,wil00,zhuLJRG05,
artH06} depends on the viewing angle. If viewed from the side
($\theta = 90^\circ$, left panel), two velocities are seen at each
offset, on opposite sides of V$_{amb}$. The line is broad in the
head, and the two velocities approach each other in the tail. If
viewed from the tail ($\theta = 180^\circ$, right panel), the source
would not be obviously cometary in appearance, and the velocity
would vary from V$_{\star}$ at the center to V$_{amb}$ at the edges.
From an intermediate angle ($\theta = 135^\circ$, middle panel), we
see a p-v pattern like those seen in G29.96, but offset in velocity.
A broad line, centered between V$_{amb}$ and V$_{\star}$ is seen at
the head of the comet. Of the two velocities seen at $\theta =
90^\circ$, the V $<$ V$_{amb}$ branch is foreshortened and may blend
into the head (see Figure~\ref{g2996pv2}). The V $>$ V$_{amb}$
branch rises toward V$_{\star}$ then falls back toward V =
V$_{amb}$. The line centroid is at V $>$ V$_{amb}$. The p-v diagrams
would be mirrored about V$_{amb}$ if the star were moving toward the
observer ($\theta < 90^{\circ}$).

D. (Figure~\ref{sppv})  A paraboloidal shell can also occur in which
the motion of the gas in the shell is predominantly due to the
pressure gradient along the shell, which causes an acceleration of
the gas toward the tail \citep{zhuLJRG05,artH06}. In this case, the
paraboloidal shape of the shell could result either from a
relatively slow motion of the star through the molecular cloud or
from a density gradient in the cloud. The resulting appearance and
p-v diagrams are similar to those of the bow-shock model, but the
p-v diagrams are offset toward negative velocities, with the
velocity centroid at V $<$ V$_{amb}$ (for $\theta > 90^{\circ}$).
This is the offset seen in G29.96, indicating that it belongs to
this category.

E. (Figure~\ref{sshpv})  A filled blister flow could result if the wind
from the ionizing star is too weak to prevent the ionized gas from
flowing away from the surface of the molecular cloud, past the star
\citep{yorTB83,com97,henAG05,artH06}. Such a region is formed when a
density gradient is present in the natal molecular cloud. In this
case, the source morphology and p-v diagrams would depend on the
shape of the blister cavity in the molecular cloud, but blister
flows are largely axial and a broad line would be likely to be seen
at all positions. For $\theta > 90^{\circ}$, the line centroid is at
V $<$ V$_{amb}$. The largest velocities would be seen toward the
tail, or at the center for $\theta \approx 180^{\circ}$.

A number of sources in our sample either have confused kinematics or
have kinematics for which we have no simple qualitative models:
Several sources show ring, horseshoe, or bipolar morphologies,
without the kinematic signatures of shells or paraboloidal flows.
Several regions appear to be clusters of overlapping sources, making
the interpretation of their kinematics difficult. Several sources
have distinctly different morphologies in [Ne~II] and the radio
continuum, suggesting that variable extinction is distorting their
appearance in the infrared. These sources also have confusing
kinematics. And several sources are too small for their [Ne~II]
kinematics to have been resolved.

\subsection{Observed Source Morphologies}

The extraordinarily good match between the radio and [Ne~II]
morphologies in the majority (12 of 16) of the resolved sources
indicates that most neon is in Ne$^+$, that the gas density is
almost everywhere below the critical density of the 12.8 $\mu$m
line, and that the centimeter-wave radio continuum emission is for
the most part optically thin. Extinction is the most likely cause of
the strong differences between radio continuum and [Ne II]
morphologies in the other four sources. Some of the differences
could also be due to incomplete UV coverage in the radio snapshot
observations. And, differences in sensitivity between [NeII] and
radio continuum observations may also play a role.

In many cases, radio interferometer observations show that UCHII
regions are accompanied by substantial radio continuum flux from
extended, low-surface brightness ionized gas
\citep{kurWHO99,kimK01}. These results imply that UCHII regions may
exist in asymmetric regions where the ionizing radiation reaches
dense gas in one direction and more extended low density gas in the
other direction. Our scanning technique does not permit us to find
these extended haloes in [Ne~II] due to limited sky area we can
cover and the sky background subtraction method we use, but the
frequent presence of emission halos may have implications for the
interpretation of the source kinematics.

\subsection{Observed Source Kinematics}

Of the 12 sources that are resolved and not badly confused by
extinction, nine show clear evidence of bulk motions, which are
usually not symmetric about the molecular cloud velocities. These
nine sources all show strong evidence of tangential flows along the
surfaces of shells. Three of these sources, G29.96
(Figures~\ref{g2996pv1} and \ref{g2996pv2}), G33.92
(Figure~\ref{g3392pv}), and G43.89 (Figure~\ref{g4389pv}), show,
with varying degrees of clarity, the ``$\Lambda$'' and ``7''
patterns in p-v diagrams that are similar to the patterns
established in the surface flow models (types C and D). In these
three cases, the patterns indicate that the terminal velocities are
blueshifted with respect to the molecular cloud velocities. Along
the symmetry axis, the velocities cross the molecular cloud velocity
from blue to red and back to blue going from the vertex into the
source for G29.96 and G33.92. In G43.89, the velocity approaches the
cloud velocity from the blue side but does not cross it before
moving back to the blue. The kinematic behavior we observe in all
three sources is very similar to what we see in the models of G29.96
(Figures~\ref{g2996pv1} and \ref{g2996pv2}), which are tilted with
the vertex away from the observer and have an inclination of the
symmetry axis of about 45$^{\circ}$ with respect to the line of
sight. With their vertices pointed away from the observer, the
blueshift relative to V$_{amb}$ means that these sources are of
kinematic type D (pressure-driven surface flows; Figure~\ref{sppv}).

The three sources in our sample whose [Ne~II] and radio continuum
morphologies most closely resemble limb-brightened spherical shells
do not have kinematic patterns consistent with a closed-shell
geometry. The kinematic patterns for three of the sources, NGC~7538A
(Figure~\ref{n7538apv}), W3B (Figure~\ref{w3bpv}), and W51 IRS2,
\citep[see][]{lacJZRBG07}, are consistent with motion of the ionized
gas along roughly parabolic shells with vertices pointing almost
directly away from the observer. In the position-velocity plots, the
most positive velocities, usually close to the velocity of the
dense molecular gas, are toward the vertex while the velocities
along the bright rim are shifted to the blue. The presence of this
pattern is largely independent of the position angle of the
position-velocity cut through the center of the source.
These sources are of kinematic type D with $\theta \approx 180^{\circ}$.

The remaining sources with understandable kinematics also have a
strongly limb-brightened appearance. The kinematics revealed by the
[Ne~II] spectroscopy show that G30.54 (Figure~\ref{g3054pv}), W3A
(Figure~\ref{w3apv}), and Mon~R2 \citep{jafZLR03} are also likely to
be limb-brightened parabolas with significant flow of the ionized
gas along the ionized shell. For these three sources, however, the
kinematics indicate that the symmetry axes lie near the plane of the
sky. The sources show broad lines at the source vertex and then
double lines farther along the symmetry axis with the two components
appearing at the extreme velocities of the line seen at the vertex.
Along the edges of each source, the lines are single and
significantly narrower. These sources are of either kinematic type D
or type C, with $\theta \approx 90^{\circ}$.

Four sources have very similar [Ne~II] and radio morphologies, but
do not fit into any of the kinematic categories we have discussed.
S106 (Figure~\ref{s106mapradioneii}) has a clear bipolar morphology,
but confused kinematics. G45.12N (Figure~\ref{g4512Npv}) has a
cometary morphology, but does not have obvious cometary kinematics,
of either type C or type D. G45.12SW (Figure~\ref{g4512SEpv}) has a
ring or shell morphology, but does not have the kinematic signature
of a shell. K3-50A (Figure~\ref{k350map}) appears to be a cluster of
sources, which are too blended to allow us to sort out their
kinematics. The peaks in the IRTF map of [Ne~II] in W51~IRS2
(Figure~\ref{w51dmapneii}) are also blended. The Gemini map of
\cite{lacJZRBG07} separates these sources but does not resolve their
kinematics. However, the kinematics of the extended ionized gas in
W51~IRS2 shown in \cite{lacJZRBG07} look much like a type D surface
flow with $\theta \approx 180^{\circ}$.

Complex morphologies and kinematics in the four sources with
substantially different [Ne~II] and radio continuum morphologies
prevent a coherent explanation for gas motion in these sources.
However, it is clear that systematic gas motions are present. In
G11.94, there is a component of the [Ne~II] emission at the same
velocity as the ambient cloud (Figure~\ref{g1194ABCspec}), but there
is also a substantial amount of emission at many positions in a red
wing or in an identifiable red component. In G45.45
(Figure~\ref{g4545map}), the kinematic pattern is confusing, and the
most prominent feature in [Ne~II] is not apparent in the radio
continuum. G61.48 (Figure~\ref{g6148mapneii}) is the source where
the [Ne~II] morphology least resembles the distribution of the radio
continuum emission. The [Ne~II] emission we see may therefore not be
sampling the bulk of the gas, but the dominance of red-shifted gas
in the RRL observations \citep{garLG94} is consistent with the
conclusion from the [Ne~II]-radio continuum morphology difference,
namely that the source is on the back side of the molecular cloud
and the ionized material is flowing away from the neutral core.

\subsection{Comparison of Observed Sources with Qualitative Models}

The kinematic types that we have assigned to observed sources are
given in Table~\ref{kintypes}. Examination of this list shows that
none of the observed sources fits into category A or B, the
classical expanding shell and sphere models of HII regions. None has
a ring-like p-v diagram of an expanding shell, and the sources with
broad lines are either too small for their morphologies to be
apparent or have confused morphologies. In addition, none of the
sources fits into category E, the classical filled blister flow with
the ionized gas flowing away from the ionization front.

Even more remarkably, in spite of the frequency of bow-shock
morphologies, none of the observed sources clearly fits into
category C, a bow shock moving with a star through a molecular
cloud. Because they are nearly edge-on, Mon R2, G30.54, and W3A
could fit into either category C or category D. The other sources
with cometary kinematic signatures all have p-v patterns
characteristic of sources tipped away from the observer (${\theta}
> 90^{\circ}$) with their velocity centroids blueshifted relative to
the neighboring molecular material, the signature of a
pressure-driven surface flow.

The preponderance of sources with their tails tipped toward the
observer ($\theta \geq 90^{\circ}$) probably results from a
combination of extinction and selection effects. The extinction at
12.8~$\mu$m is relatively low
\citep[A$_{12.8}$/A$_V$$\sim$0.05,][]{rieL85,rosBD00}, but the
column density through the kind of massive core that harbors massive
young stars and UCHII regions can be substantial \citep[A$_V$
$\approx$ a few hundred,][]{beuKDS00}. The density in such cores
drops rather steeply, with power-law exponents of 1.6-1.8
\citep{mueSEJ02,beuSMMSW02}. If UCHII regions are not in the centers
of cores, they should therefore be visible in [Ne~II] when the
exciting stars lie between the front side and $\theta \approx
90^{\circ}$. Our survey was somewhat biased in favor of sources
lying on the front sides of molecular clouds by including sources
known from optical and near-infrared observations (NGC 7538, W3, and
Mon R2). In addition, we were forced to abandon efforts to observe
several UCHII regions (G8.67-0.36, G10.10+0.74, G10.62-0.38,
G12.208-0.10, G33.50+0.20, G42.42-0.27, G45.47+0.05, G50.23+0.33)
because they were undetectable or too faint to map.
%g34.30, g35.19, w49
These selection effects are likely to explain the lack of sources
with kinematics indicating that they lie on the back sides of their
clouds.

\begin{figure}
\epsscale{1.0} \plotone{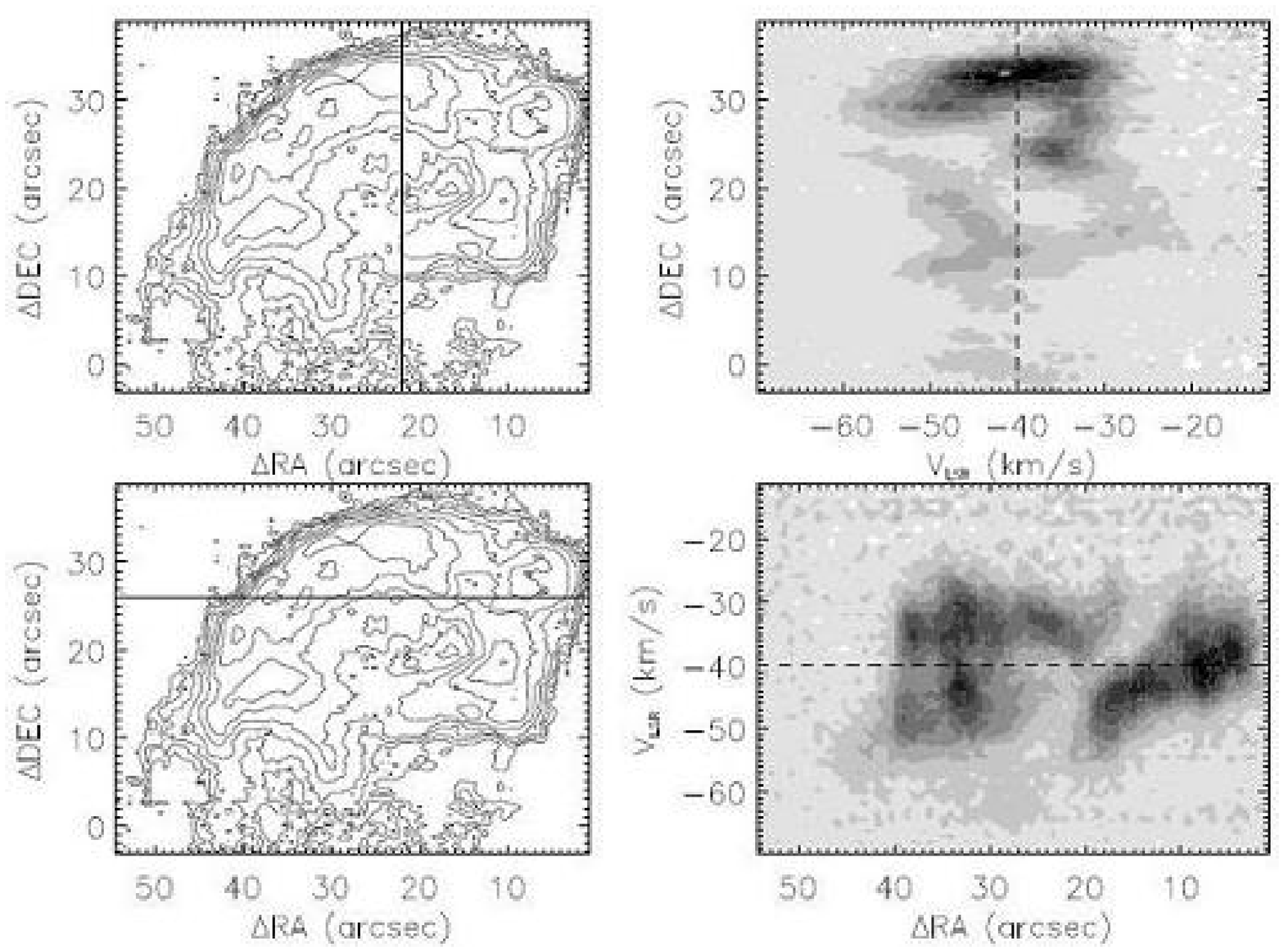} \caption{Position-velocity diagrams
of W3~A line observations. Contours in line flux maps are drawn at
$70\%$, $50\%$, $35\%$, $25\%$, $17.5\%$, $12.5\%$, $9\%$ and $6\%$
of peak values (0.13~ergs~cm$^{-2}$~s$^{-1}$~sr$^{-1}$) of the maps.
Dashed lines in p-v diagrams show the ambient molecular material
velocity. \label{w3apv}}
\end{figure}

\begin{figure}
\epsscale{0.8} \plotone{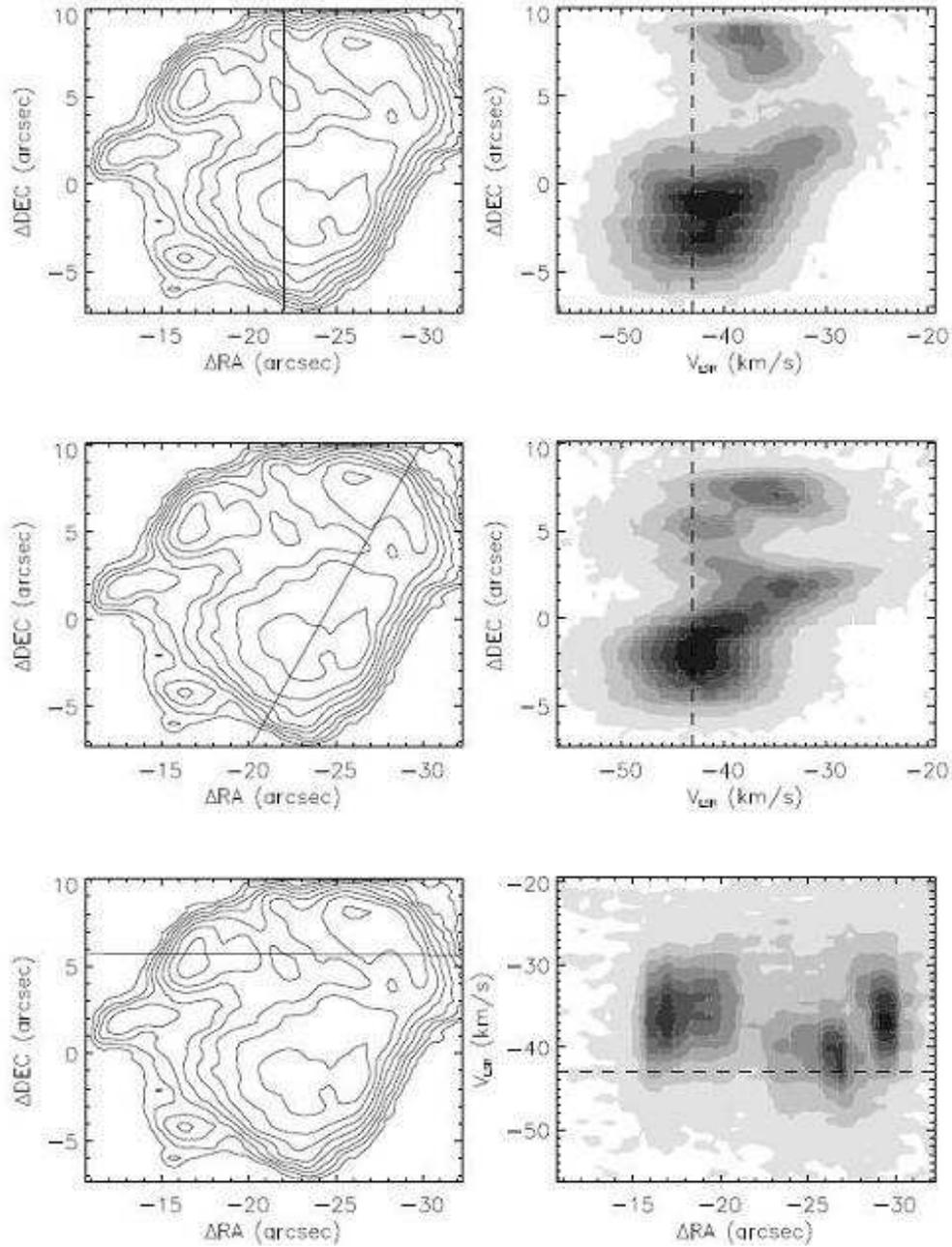} \caption{Position-velocity diagrams
of W3~B line observations. Contours in line flux maps are drawn at
$70\%$, $50\%$, $35\%$, $25\%$, $17.5\%$, $12.5\%$, $9\%$ and $6\%$
of peak values (0.17~ergs~cm$^{-2}$~s$^{-1}$~sr$^{-1}$) of the maps.
\label{w3bpv}}
\end{figure}

\begin{figure}
\epsscale{1.0} \plotone{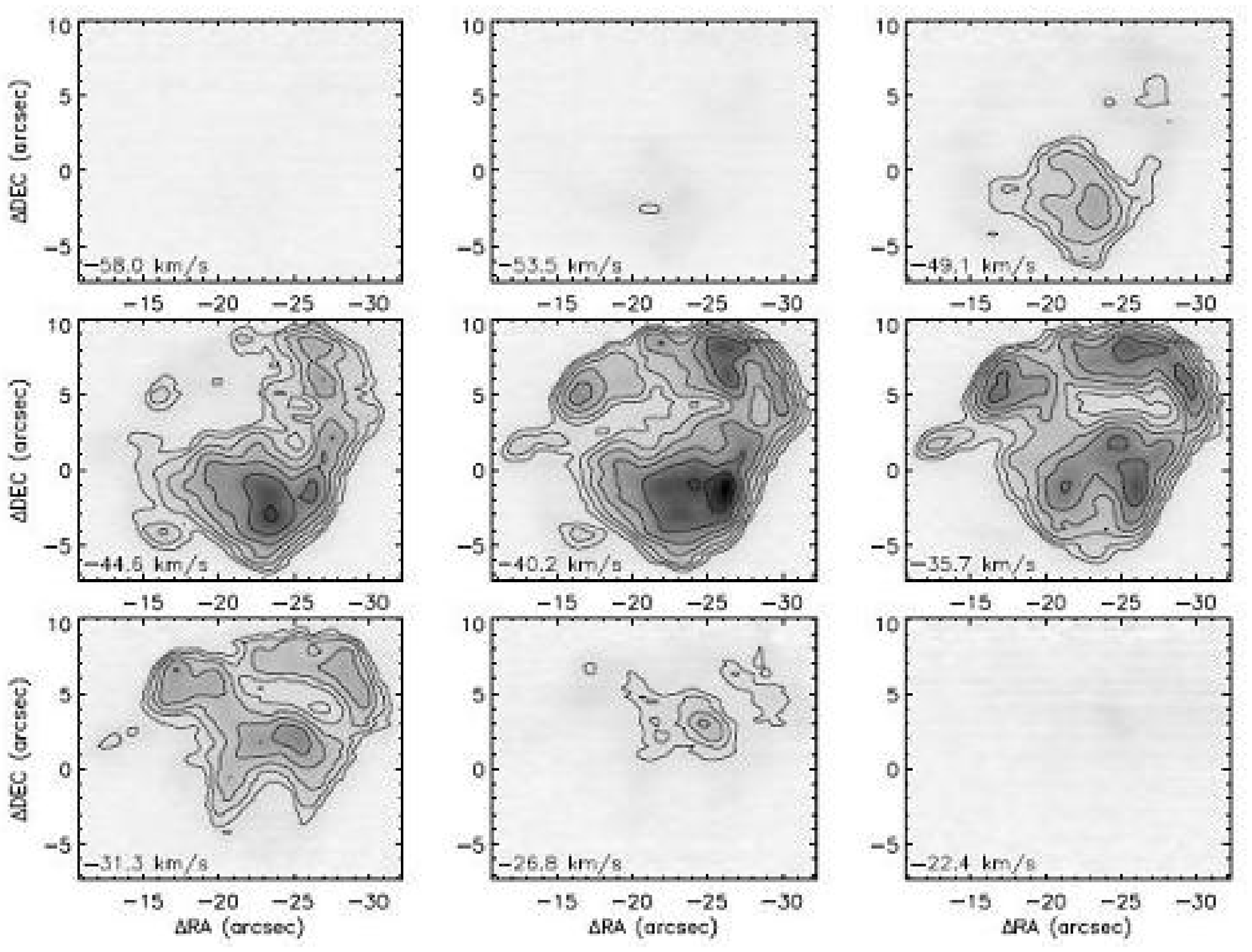} \caption{Channel maps of W3~B
[Ne~II] line observations. Contours are drawn at $70\%$, $50\%$,
$35\%$, $25\%$, $17.5\%$, $12.5\%$, $9\%$ and $6\%$ of the peak
value of all channels. The peak value is
5.2~ergs~cm$^{-2}$~s$^{-1}$~sr$^{-1}$~(cm$^{-1}$)$^{-1}$. The
molecular cloud velocity is V$_{LSR}$ = -43 km s$^{-1}$.
\label{w3bchanmapneii}}
\end{figure}

\begin{figure}
\epsscale{1.0} \plotone{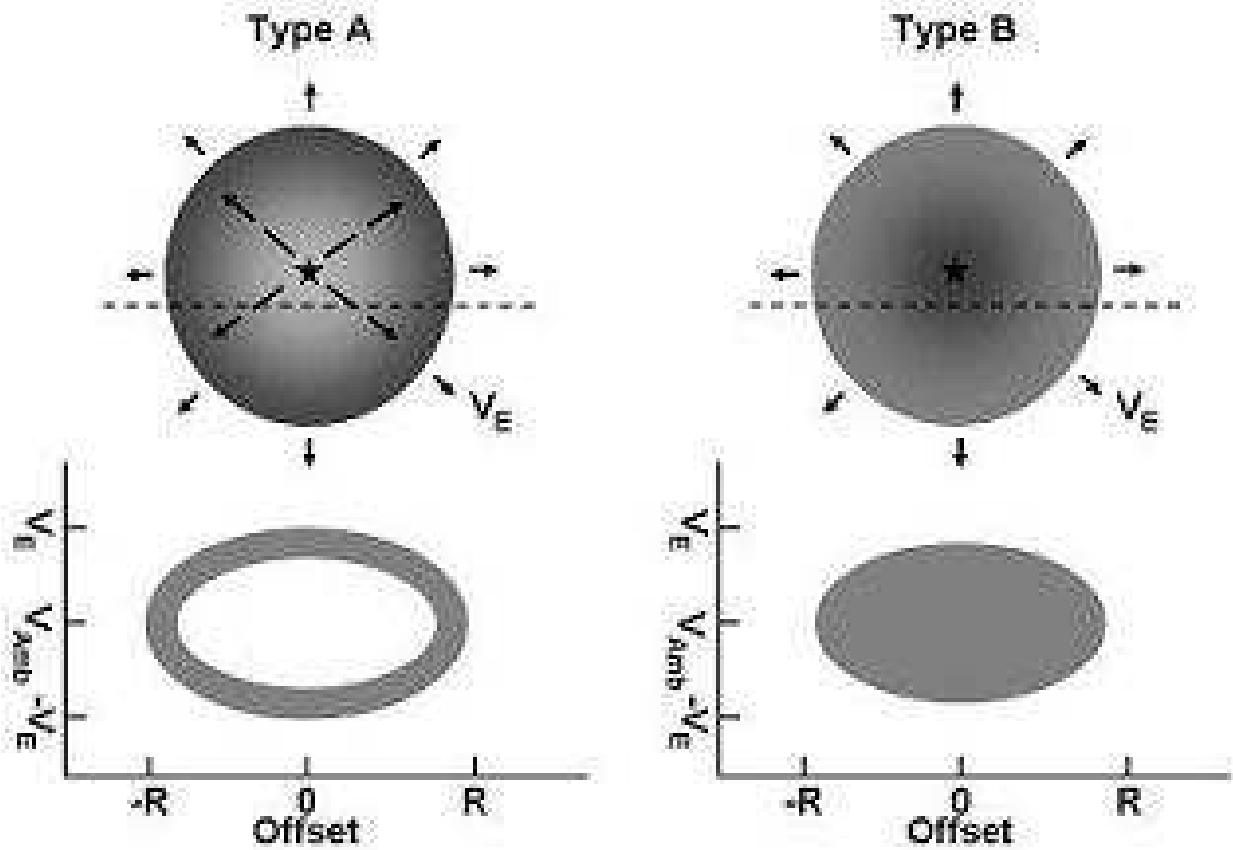} \caption{Sketches of
position-velocity diagrams of an expanding shell-like HII region
(left) and an expanding spherical HII region (right). The expansion
speed is V$_{E}$. The molecular cloud velocity is V$_{amb}$. The
dashed lines show the locations of the p-v cuts. The solid arrows in
the morphology diagrams indicate the moving directions of ionized
gas relative to ambient molecular material and the dashed arrows
show the moving directions of stellar wind material with respect to
the central stars. \label{ssspv}}
\end{figure}

\begin{figure}
\epsscale{1.0} \plotone{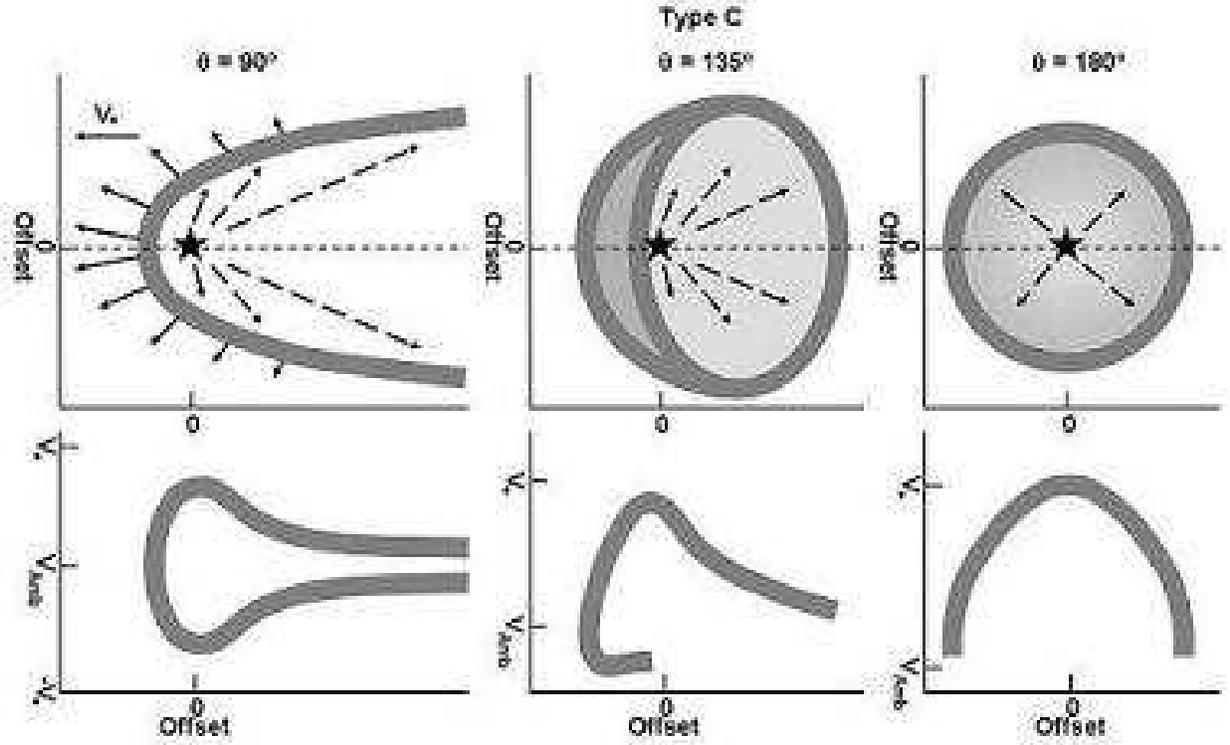} \caption{Sketches of
position-velocity diagrams of a stellar wind bow shock HII region at
three different viewing angles. Note that the velocity vectors are
shown in the frame of reference of the ambient molecular cloud; the
shell is not expanding in the moving frame of reference of the star.
The molecular cloud velocity is V$_{amb}$ and the ionizing star is
moving supersonically at a speed of V$_{\star}$ relative to the
molecular cloud. The ram pressures of incoming cloud material and
stellar wind material confine the ionized material in a compressed
shell which moves with the star in the cloud. The dashed lines show
the locations of the p-v cuts. The p-v diagrams in the middle and
right panels are drawn for stars moving away from the observer
($\theta > 90^\circ$). The p-v diagrams would be mirrored about
V$_{amb}$ for stellar motions toward the observer ($\theta <
90^\circ$).
%The solid arrows in the morphology diagrams
%indicate the moving directions of ionized gas relative to ambient
%molecular material and the dashed arrows show the moving directions
%of stellar wind material with respect to the central
%stars.
\label{sbspv}}
\end{figure}

\begin{figure}
\epsscale{1.0} \plotone{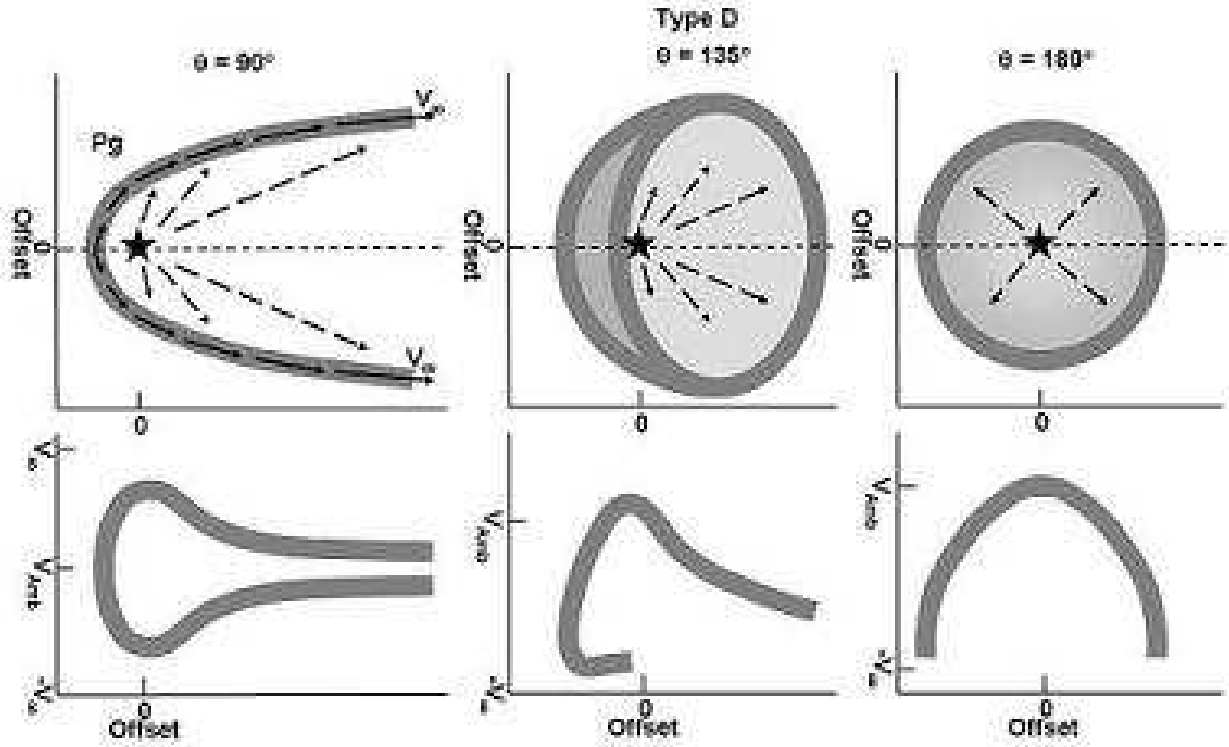} \caption{Sketches of
position-velocity diagrams of a stellar wind pressure-driven HII
region at three different viewing angles. The molecular cloud
velocity is V$_{amb}$ and the star with a stellar wind is stationary
with the cloud. The ionized gas is compressed by the stellar wind
into a shell and moves along the shell under the influences of the
density gradient and the stellar wind pressure. V$_{\infty}$ is the
exiting speed of the ionized gas. The dashed lines show the
locations of the p-v cuts. The p-v diagrams are similar to those in
Fig.~\ref{sbspv}, but are offset toward negative V$_{LSR}$ for the
head of the shell tipped away from the observer ($\theta >
90^\circ$). The p-v diagrams would be mirrored about V$_{amb}$ for
stellar motions toward the observer ($\theta < 90^\circ$).
\label{sppv}}
\end{figure}

\begin{figure}
\epsscale{1.0} \plotone{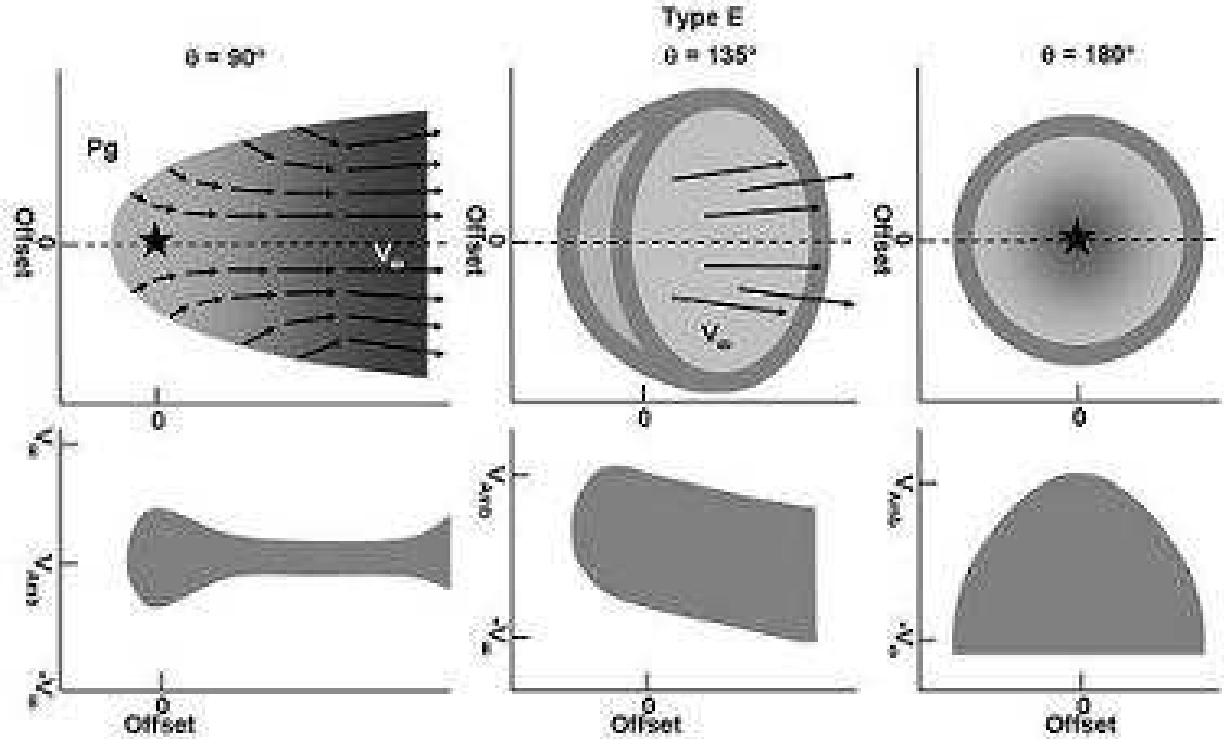} \caption{Sketches of
position-velocity diagrams of a blister flow HII region at three
different viewing angles. The molecular cloud velocity is V$_{amb}$
and the star is stationary. No stellar wind is present. The cloud
material is continually evaporated by the stellar radiation and
forced by pressures to flow along the density gradient. The
V$_{\infty}$ is the exiting speed of the ionized gas. The dashed
lines show the locations of the p-v cuts. The figures show p-v
diagrams of the region when the head of the region is tipped away
from the observer ($\theta > 90^\circ$). The p-v diagrams would be
mirrored about V$_{amb}$ for stellar motions toward the observer
($\theta < 90^\circ$). \label{sshpv}}
\end{figure}

More than half of the observed sources fit into category D, a flow along
the surface of a paraboloidal shell, which accelerates relative to
molecular cloud as it moves away from the apex of the paraboloid.
Although we have not made a hydrodynamic model of such a flow, we
know of no driving mechanism for the acceleration other than the
pressure gradient in the ionized gas. \cite{zhuLJRG05} found that
the pressure gradient acceleration would be small if the ionized and
swept-up molecular gas were coupled so both had to be accelerated.
But we have estimated the acceleration if the ionized gas is allowed
to slip past the molecular shell, and we find that speeds
$\sim$20-30~km~s$^{-1}$ can occur. The hydrodynamic models of
\cite{artH06} include the essential effects present in the
combination of blister and bow shock models: density gradients in
the neutral medium, stellar winds, and stellar motion. The
\cite{artH06} models do an excellent job of matching the range of
source morphologies and reproduce the most significant kinematic
signatures using reasonable values for density gradients, mass loss
rates, and stellar velocities. One critical area where they fall
short is in explaining the width of the lines at the source vertex.
The observed effect is most evident in W3A (Figure~\ref{w3apv}) and
in Mon~R2 \citep{jafZLR03} where the symmetry axes are near the
plane of the sky. The splitting of the broad line into two narrow
lines at the velocity extremes of the broad line as one crosses the
shell to the inside implies that the gas accelerates to
$\sim~\pm$10~km~s$^{-1}$ within a small solid angle (less than a
steradian, as seen from the point of view of the ionizing star) at
the vertex. This, in turn, implies a significant pressure gradient.

The [Ne~II] kinematics of our sample unite a large and diverse group
of UCHII and compact HII regions into a single empirical picture:
The HII regions coexist with dense and massive molecular cores
\citep{beuSMMSW02,shiEYKJ03,sriBSMW02}. The ionized regions have the
form of partial, roughly parabolic shells with their open sides
usually pointing away from the centers of the dense neutral
condensations and gas flows tangentially along these shells.

The evidence that very many UCHII regions are dense shells with
tangential flows away from the parent molecular cores has
implications for the survivability of the molecular cores. Earlier
examinations of the erosion of dense cores by external O stars did
not take the compression of neutral material by stellar wind shocks
into account. These efforts led to estimates that when as little as
4\% of the core material formed into stars with a Salpeter IMF,
these stars could erode the remaining core entirely away
\citep{whi79}. It has already been pointed out, however that the
dense shells formed by stellar wind pressure would greatly inhibit
the process of erosion \citep{chu91,lumH99,hoa06}. Our results
indicate that most if not all OB stars breaking out of dense cores
form such dense shells at their boundaries.

\acknowledgments

The development of TEXES was made possible with support from the NSF
and the SOFIA program through grant USRA-8500-98-008. Observations
with TEXES are supported by NSF grant AST-0607312. Part of this
research was performed in the Rochester Imaging Detector Laboratory
with support from a NYSTAR Faculty Development Program grant. MJR
acknowledges the support of NSF grant NSF-0708074 and NASA grant
NNG04GG92G. TKG was supported by the Lunar and Planetary Institute,
which is operated by the Universities Space Research Association
under NASA CAN-NCC5-679. DTJ would like to thank the
Max-Planck-Institut fuer extraterrestrische Physik for its
hospitality. We thank Alan Fey and Ed Churchwell for letting us to
have their data on G29.96~-0.02 and other objects. We also want to
acknowledge Achim Tieftrunk, Yolanda Gomez, Marcello Felli, Chris
DePree, Ralph Gaume for their permission to use their published
figures in this work.

%\bibliography{master}
%\end{thebibliography}

\clearpage

\clearpage

\clearpage

\clearpage

\clearpage

\clearpage

\clearpage

\clearpage

\clearpage

\clearpage

\clearpage

\clearpage

\clearpage

\clearpage

\clearpage

\clearpage

\clearpage

\clearpage

\clearpage

\clearpage

\clearpage

\clearpage

\clearpage

\clearpage

\clearpage

\clearpage

\clearpage

\clearpage

\clearpage

\clearpage

\clearpage

\clearpage

\clearpage

\clearpage

\clearpage

\clearpage

\clearpage

\clearpage

\clearpage

\clearpage

\clearpage

\begin{deluxetable}{lccccccccccc}
\tabletypesize{\scriptsize}
\tablewidth{0pt}
\tablecaption{OBSERVATIONAL DATA \label{obspara}}
\tablehead{
Object& Date& RA(J2000.0) &DEC &d &V$_{LSR}$  &SW & SL             &STS& PS   &ME                 & IT \\
    &mm/dd/yy&hh mm ss.sss &dd mm ss.sss & (kpc) & (km~s$^{-1}$) & ($''$) & ($''$) & ($''$) &  ($''$)  &($''$ $\times$ $''$)&  (sec)
}
\startdata
$^2$G5.89~-0.39  &06/11/03&18 00 30.4 &-24 04 00&2.6~\tablenotemark{a}&9~\tablenotemark{b} &0.9 &10.8 &0.4& 0.360&6${\times}$6  &180 \\
$^1$G11.94~-0.62 &07/02/03&18 14 02.2 &-18 53 26&4.2~\tablenotemark{c}&39~\tablenotemark{c} &1.4 &10.8 &0.7& 0.360&18${\times}$16  &12 \\
$^2$G29.96~-0.02 &06/11/01&18 46 03.9 &-02 39 22&7.4~\tablenotemark{c}&98~\tablenotemark{c} &0.9 &10.8 &0.4& 0.360&20.4${\times}$13 & 16 \\
$^2$G30.54~+0.02 &06/30/03&18 46 59.5 &-02 07 26&13.8~\tablenotemark{c}&48~\tablenotemark{c}&1.4 &10.8 &0.7& 0.360&12${\times}$12  &16 \\
$^2$G33.92~+0.11 &06/11/01&18 52 50.2 & 00 55 29&8.3~\tablenotemark{c}&108~\tablenotemark{c}&0.9 &10.8 &0.4& 0.360&17${\times}$15  &50 \\
$^2$G43.89~-0.78 &07/01/03&19 14 26.2 & 09 22 34&4.2~\tablenotemark{c}&53.5~\tablenotemark{c}&1.4 &10.8 &0.7& 0.360&12${\times}$12  &16 \\
$^2$G45.07~+0.13 &07/01/03&19 13 22.1 & 10 50 53&6.0~\tablenotemark{c}&59~\tablenotemark{c}&1.4 &10.8 &0.7& 0.360& 7${\times}$8   &12 \\
$^2$G45.12~+0.13 &07/01/03&19 13 27.8 & 10 53 36&6.9~\tablenotemark{c}&59~\tablenotemark{c}&1.4 &10.8 &0.7& 0.360&15${\times}$18  &16 \\
$^2$G45.45~+0.06 &06/11/01&19 14 21.4 & 11 09 14&6.6~\tablenotemark{c}&59.5~\tablenotemark{c}&0.9 &10.8 &0.4& 0.360&15${\times}$12  &52 \\
$^1$W51 IRS2     &06/30/03&19 23 39.9 & 14 31 09&6.6~\tablenotemark{c}&58.3~\tablenotemark{e}&1.4 &10.8 &0.7& 0.360&30${\times}$14  &12 \\
$^1$G61.48~+0.09B&07/01/03&19 46 49.2 & 25 12 43&5.4~\tablenotemark{c}&22~\tablenotemark{c}&1.4 &10.8 &0.7& 0.360&60${\times}$19  &16 \\
$^1$K3-50A       &06/30/03&20 01 45.7 & 33 32 43&8.7~\tablenotemark{f}&-24.4~\tablenotemark{e}&1.4 &10.8 &0.7& 0.360&12${\times}$12  &24 \\
$^1$S106&06/29/03&20 27 26.8 & 37 22 48&0.6~\tablenotemark{g}&-1.0~\tablenotemark{h}&1.4 &10.8 &0.7& 0.360&47${\times}$53  &16 \\
$^1$NGC7538A     &09/13/02&23 13 45.6 & 61 28 18&3.5~\tablenotemark{i}&-56.9~\tablenotemark{j} &1.4 &10.1 &1.0& 0.360&29.4${\times}$16.2&$\geq$20\\
$^1$W3A &07/01/03&02 25 40.8 & 62 05 53&2.0~\tablenotemark{k}&-40~\tablenotemark{l}&1.4 &10.8 &0.7& 0.360&86${\times}$37  &$>$8 \\
$^1$W3B &07/03/03&02 25 36.9 & 62 05 45& 2.0~\tablenotemark{k}&
-43~\tablenotemark{m}& 1.4 & 10.8 & 0.7& 0.360&61${\times}$44  &$>$8
\enddata
\tablecomments{The observational parameters (SW: slit width, SL:
slit length, STS: step size, PS: plate scale, ME: map extent, IT:
pixel integration time) of observations. The shown coordinates are
for the (0,0) positions in [Ne~II] line maps. Two methods are used
in determining these coordinates: $^1$ Read from published figures.
Both qualities of the figures and the reading process can affect
uncertainties of the resulting coordinates, which can be over 1$''$.
$^2$ Cross-correlate [Ne~II] line maps resampled on fine grids
(pixel size $\leq$0.1$''$) and available radio continuum maps to
find the best morphological fit. Uncertainties of these coordinates
should be better than 0.5$''$. Nasa/IPAC Extragalactic Database
(NED) coordinate calculator is used to convert obtained B1950.0
coordinates to J2000.0 coordinates. $^a$~\cite{kimK03};
$^b$~\cite{hatTM98}; $^c$~\cite{chuWC90}; $^d$~\cite{wooC89b};
$^e$~\cite{sheC96}; $^f$~\cite{har75}; $^g$~\cite{staLDS82};
$^h$~\cite{schSKSB02}; $^i$~\cite{hanLR02} $^j$~\cite{dicDW81};
$^k$~\cite{hacBMRI06}; $^l$~only for W3A, \cite{kanAG98}; $^m$~only
for W3B, \cite{tieWSGJC95}}
\end{deluxetable}

\begin{deluxetable}{lcl}
\tabletypesize{\scriptsize}
\tablewidth{0pt}
\tablecaption{OBSERVED KINEMATIC TYPES \label{kintypes}}
\tablehead{Object & Type & Comments}
\startdata
Mon R2         & C/D & 70$^{\circ}$ viewing angle, see \cite{zhuLJRG05}\\
G5.89~-0.39    & ? & barely resolved\\
G11.94~-0.62   & ? & multiple peaks, extinction lanes, complex kinematics, $\bar{V} \approx V_{amb}$\\
G29.96~-0.02   & D & 140$^{\circ}$ viewing angle, see \cite{zhuLJRG05}\\
G30.54~+0.02   & C/D? & horseshoe morphology, weak type D signature on axis\\
G33.92~+0.11   & D & confused morphology, but clear kinematic signature\\
G43.89~-0.78   & D & clear type D morphology and kinematics\\
G45.07~+0.13   & ? & unresolved, 1$''$ diameter\\
G45.12~+0.13N  & ? & appears cometary but without obvious cometary kinematics\\
G45.12~+0.13SE & ? & barely resolved\\
G45.12~+0.13SW & ? & broken ring, multiple peaks \\
G45.45~+0.06   & ? & confused line morphology, with different [Ne~II] and ff morphologies\\
W51 IRS2       & D & [S~IV] has $\theta \approx 180^{\circ}$, type D kinematics \citep{lacJZRBG07},\\
 & & [Ne~II] shows superimposed sources\\
G61.48~+0.09B  & ? & extinction lanes, complex line morphology and kinematics, $\bar{V} > V_{molec}$\\
K3-50A         & ? & three blended peaks, extended line emission\\
S106           & ? & bipolar, with unclear kinematics\\
NGC7538A       & D & shell-like appearance, tail-on type D kinematics\\
W3A            & C/D & cometary with $\theta \approx 90^{\circ}$\\
W3B            & D? & shell-like appearance, confusing kinematics
\enddata
\tablecomments{Kinematic types: A. expanding shell, B. expanding
sphere, C. bow shock from moving star,  D. pressure-driven surface
flow, E. blister. }
\end{deluxetable}

\end{document}